\documentclass[12pt,letterpaper]{article}

\usepackage[title]{appendix}
\usepackage{lscape}
\usepackage{amsfonts}
\usepackage{amsmath}
\usepackage{amssymb}
\usepackage{amsthm}
\usepackage{color}
\usepackage{dsfont}
\usepackage{epstopdf}
\usepackage{eurosym}
\usepackage[bottom,multiple]{footmisc}
\usepackage{float}
\usepackage{natbib}
\usepackage[margin=.75in]{geometry}
\usepackage[hang]{caption}
\usepackage[pdftex]{graphicx}
\usepackage{rotating}
\usepackage{mathrsfs}

\newcommand{\abs}[1]{\left|#1\right|}

\renewcommand{\qed}{\hfill $\blacksquare$}
\numberwithin{equation}{section}
\renewcommand{\baselinestretch}{1.2}
\allowdisplaybreaks
\newtheorem{assumption}{Assumption}
\newtheorem{corollary}{Corollary}

\newtheorem{lemma}{Lemma}

\newtheorem{theorem}{Theorem}
\newtheorem{remark}{Remark}

\begin{document}

\title{An unbounded intensity model for point processes\thanks{We are indebted to Phillip Heiler, Roberto Ren\`{o}, Yoann Potiron, and participants at the ''Volatility, Jumps, and Bursts`` workshop at Lancaster University, the 14th annual SoFiE meeting at University of Cambridge, the 33rd annual (EC)$^{2}$ at ESSEC Business School, the Financial Econometrics conference to mark Stephen Taylor's Retirement at Lancaster University, the 10th Italian Congress of Econometrics and Empirical Economics (ICEEE) at University of Cagliari, and the 16th annual SoFiE meeting at Pontifical Catholic University of Rio de Janeiro for constructive comments and insightful suggestions on our work. At the Journal of Econometrics, we appreciate a stimulating interaction with an associated editor and two anonymous referees, whose reports helped to significantly improve several aspects of the submitted version of the article. Christensen is grateful for support from the Independent Research Fund Denmark (DFF 1028–00030B).}}
\author{Kim Christensen\thanks{Aarhus University, Department of Economics and Business Economics, Denmark. \textsc{kim@econ.au.dk}.} \and Aleksey Kolokolov\thanks{University of Manchester, Alliance Manchester Business School, United Kingdom. \textsc{aleksey.kolokolov@manchester.ac.uk}.}}
\date{May, 2024}

\maketitle

\begin{abstract}
We develop a model for point processes on the real line, where the intensity can be locally unbounded without inducing an explosion. In contrast to an orderly point process, for which the probability of observing more than one event over a short time interval is negligible, the bursting intensity causes an extreme clustering of events around the singularity. We propose a nonparametric approach to detect such bursts in the intensity. It relies on a heavy traffic condition, which admits inference for point processes over a finite time interval. With Monte Carlo evidence, we show that our testing procedure exhibits size control under the null, whereas it has high rejection rates under the alternative. We implement our approach on high-frequency data for the EUR/USD spot exchange rate, where the test statistic captures abnormal surges in trading activity. We detect a nontrivial amount of intensity bursts in these data and describe their basic properties. Trading activity during an intensity burst is positively related to volatility, illiquidity, and the probability of observing a drift burst. The latter effect is reinforced if the order flow is imbalanced or the price elasticity of the limit order book is large.


\end{abstract}

\vspace*{-0.5cm}

\setlength{\baselineskip}{18pt}\setlength{\abovedisplayskip}{10pt} \belowdisplayskip \abovedisplayskip \setlength{\abovedisplayshortskip }{5pt} \abovedisplayshortskip \belowdisplayshortskip \setlength{\abovedisplayskip}{8pt} \belowdisplayskip \abovedisplayskip \setlength{\abovedisplayshortskip }{4pt}

\vfill

\thispagestyle{empty}

\pagebreak

\section{Introduction} \setcounter{page}{1}

Trading activity is an important variable in financial economics.\footnote{Trading activity is a measure of how much a given financial asset has traded over a period of time. It can be taken as the either the number of transactions, the number of shares traded, or the dollar volume (price times quantity). In this paper, we mainly employ the transaction count.} In the mixture of distribution hypothesis \citep*[e.g.][]{clark:73a, epps-epps:76a, tauchen-pitts:83a}, trades are driven by a latent stochastic process, interpreted as the flow of news. If the unobservable arrival of news is random and persistent, then so is the observable trading activity. In market microstructure, a number of theories suggest that optimal execution strategies induce time-varying trading activity \citep*[e.g.][]{kyle:85a, admati-pfleiderer:88a, almgren-chriss:01a}. Moreover, the increased volatility typically observed during distressed market conditions often coincide with abnormal increases in trading activity.

The literature has therefore devoted a lot of attention to build models of trading activity \citep*[e.g.][and references therein and thereto]{engle-russell:98a}. In practice, the sequence of trades constitute a time series of irregularly spaced high-frequency data, so point processes are a natural starting point. Trading activity can be described by a point process with random---and possibly persistent---intensity, such as an inhomogeneous Poisson process \citep*[e.g.][]{cox:55a}. The rate of trade arrivals can also depend on other observable characteristics, even the history of the process itself as in \citet*{hawkes:71a}. In the latter class of self-exciting processes, each event increases the likelihood of a new event for a short while. However, the assumption in all this literature is that the intensity process remains locally bounded.

In this paper, we study a point process, where the intensity is, potentially, unbounded over short time intervals. We coin this an intensity burst. We adopt a theoretical foundation, where the random intensity is decomposed into a base intensity, describing ``normal'' trading activity, and an exploding intensity, describing ``abnormal'' trading activity. In the normal state, the intensity is locally bounded and follows a doubly stochastic Poisson process. This implies that the probability of observing more than one trade in a short time interval is negligible, so none of the point clusters observed in a fixed window (interpreted as, e.g., a trading day) are substantially larger than the others. In the abnormal state, the intensity is unbounded in the vicinity of a stopping time, which permits the model to produce an extreme concentration of trades in a neighbourhood of that instant. However, the integrated intensity (i.e. the compensator) remains bounded, which ensures that the point process is non-explosive and well-defined.

We propose a framework for intensity burst detection by studying a pointwise test statistic, that is testing for an intensity burst near a single candidate time. We further refine this technique such that a burst in intensity can be separated from a jump in intensity. In contrast to standard inference for point processes, which normally proceeds in a ``long-span'' setting with time going to infinity, we attempt to detect abnormal surges in trading activity over short time intervals. Hence, we study the ``infill'' limit, where the time interval is fixed. To permit inference under the null hypothesis of no intensity burst, we appeal to a novel heavy traffic condition, where independent copies of the point process are stacked to generate an increasing numbers of events over a fixed time interval \citep*[e.g.][]{kingman:61a}. \citet*{clinet-potiron:18a} propose a related inference for the doubly stochastic self-exciting point process \citep*[see also][]{potiron-mykland:20a}, but they do not exploit it for intensity burst detection.

The heavy traffic assumption is a logical precursor for infill asymptotic theory widely employed in financial econometrics, such as nonparametric estimation of the volatility of arbitrage-free price processes \citep*[see, e.g.,][]{andersen-bollerslev:98a, barndorff-nielsen-shephard:02a, jacod-protter:12a}. In that setting, the number of observations over a fixed (or shrinking) time interval increases to infinity with the mesh going to zero. This fits directly into our framework. Indeed, a realization of our point process belongs to the class of stochastic sampling times analyzed in \citet*{hayashi-jacod-yoshida:11a}. We contribute to this field by adapting the observed asymptotic variance of \citet*{mykland-zhang:17a} to local estimation.

\citet*{christensen-oomen-reno:22a} suggest that the drift of an asset price (an It\^{o} semimartingale driven by a Brownian motion) can diverge without inducing a divergence in the price. We extend their concept of a burst to the intensity, which is a natural measure of drift for point processes. However, the construction of our test is much different due to some theoretical subtleties. Moreover, while drift burst detection relies on ultra high-frequency data of the price, our test statistic is based solely on the arrival times. It does not require a further mark in the form of a transaction price or midquote. This has the distinct advantage that our test statistic is unaffected by microstructure noise, which tends to impede high-frequency estimation of the drift and volatility \citep*[e.g.][]{hansen-lunde:06b}.\footnote{In fact, a branch of the high-frequency volatility estimation literature has advocated duration-based measures extracted from the point process generating the noisy transaction or quotation data to circumvent this problem \citep*[e.g.,][]{andersen-dobrev-schaumburg:08a, hong-nolte-taylor-zhao:21a}.}

\begin{figure}[ht!]
\begin{center}
\caption{Intensity burst in the EUR/USD with a drift burst.}
\label{figure:ib}
\begin{tabular}{cc}
\small{Panel A: Exchange rate.} & \small{Panel B: Transaction count.} \\
\includegraphics[height=7cm,width=0.48\textwidth]{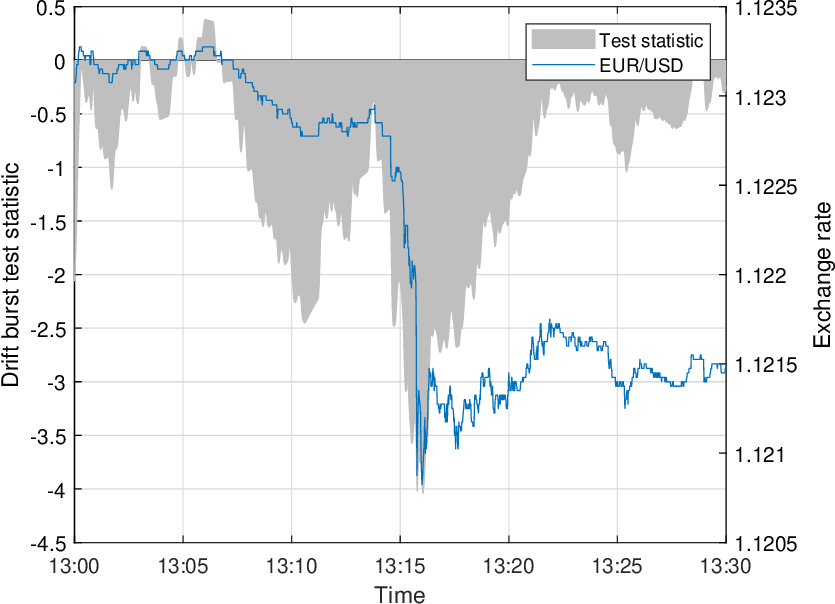} &
\includegraphics[height=7cm,width=0.48\textwidth]{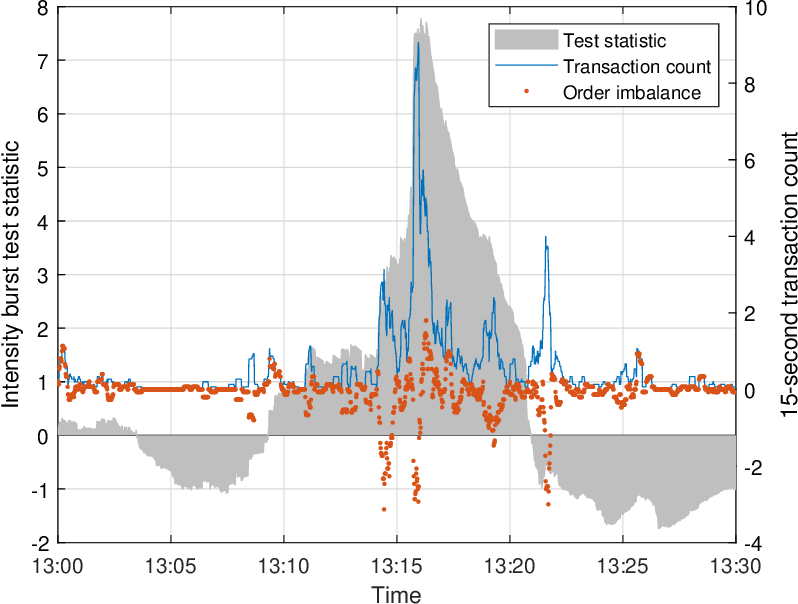}
\end{tabular}
\begin{scriptsize}
\parbox{\textwidth}{\emph{Note.} This figure shows the EUR/USD spot exchange rate on May 14, 2019, where we detect an intensity burst and a drift burst. In the left panel, we plot the exchange rate from 1:00am to 1:30am CET along with the drift burst test statistic proposed in \citet*{christensen-oomen-reno:22a}. In the right panel, we plot a nonparametric estimator of the time-varying trading intensity (as proxied by the number of transactions in a 15-second window) along with the intensity burst test statistic and a measure of order imbalance.}
\end{scriptsize}
\end{center}
\end{figure}

Our framework allows to independently screen financial high-frequency data in real-time for abnormal increases in trading activity that may---or may not---be associated with a drift burst or pockets of extreme return persistence, see \citet*{andersen-li-todorov-zhou:23a}. To illustrate, in Panel A of Figure \ref{figure:ib} we plot the spot EUR/USD exchange rate on May 14, 2019 from 1:00pm to 1:30pm Central European time (CET). The exchange rate drops sharply around 1:15pm. This is detected as a significant event by the drift burst test statistic. In Panel B of Figure \ref{figure:ib}, we report a nonparametric estimator of the time-varying trade intensity. It too accelerates, before reverting back toward a normal level. Interestingly, there are several short-lived spikes in sell-side activity around the time the exchange rate starts to depreciate, which is indicated by the order imbalance (buyer- minus seller-initiated trades). The figure also plots the intensity burst test statistic. As evident, our approach also identifies this as a significant event. In the empirical application, we shed further light on the dependence between drift and intensity, and we also show how the demand for and supply of liquidity affect this relationship.

\citet{rambaldi-pennesi-lillo:15a} and \citet{rambaldi-filimonov-lillo:18a} study a parametric self-exciting point process in order to explain clusters of events driven by an acceleration in trading activity. In their model, an ``intensity burst'' is an exogenous (potentially random) number of points added to the counting process at a single (potentially random) time. However, the intensity remains locally bounded, so none of the clusters can describe the enormous concentration of points observed during an intensity burst. Moreover, our approach is nonparametric and the number of burst points can be endogenous.

\citet{bollerslev-li-xue:18a} also study spot trading intensity using high-frequency data based on the nonparametric state-space model of \citet{li-xiu:16a}. They conduct an event study to examine the relationship between jumps in trading volume and return volatility around the release of macroeconomic information, as motivated by the differences of opinion literature \citep[e.g.,][]{kandel-pearson:95a}. As in our paper, a main ingredient for the analysis is the change in a local intensity estimator at different time points (in their case an ex-ante and ex-post measure before and after the announcement). In contrast to \citet{bollerslev-li-xue:18a}, we propose a test statistic for detecting a burst in the intensity process, as formalized in Assumption \ref{assumption:rate}. We show that our test statistic is robust to L\'{e}vy-style jump processes and roughness in the intensity. Furthermore, we develop a separate test procedure that builds on the ``change-of-frequency'' approach often employed in the volatility literature \citep*[e.g.,][]{ait-sahalia-jacod:09b, todorov-tauchen:10a}. The latter can be applied to separate a jump in intensity from a burst in intensity. In this framework, the detection of either an intensity burst or an intensity jump is essentially a two-sample mean problem. This is tightly related to the sharp regression discontinuity design (RDD) of \citet{thistlethwaite-campbell:60a}, which has a prominent history in econometric analysis. In our empirical application, there is overwhelming evidence toward our intensity burst test statistic not being driven by a jump component.

The roadmap of the paper is as follows. Section \ref{section:ib} introduces the theoretical foundation of point processes and describes the intensity burst model. Section \ref{section:identification} develops an inference strategy for the proposed test statistic. We also propose an estimator of the asymptotic variance of our test statistic. Section \ref{section:simulation} includes an extensive simulation study to demonstrate the finite sample properties of our procedure. An empirical application is conducted in Section \ref{section:empirical}, while we conclude in Section \ref{section:conclusion}. The proofs are deferred to the Appendix.

\section{Intensity burst of a point process} \label{section:ib}

We suppose a filtered probability space $( \Omega, \mathcal{F}, ( \mathcal{F}_{t})_{t \geq 0}, \mathbb{P})$ describes a random configuration of ordered points observed on the interval $[0,T]$, i.e. $0 < t_{1} < \dots < t_{m} < T$. The sequence of random variables $(t_{i})_{i = 1}^{m}$ is assumed to be a realization of a simple point process with associated counting process $N = (N_{t})_{t \geq 0}$, which is defined as $N_{t} = \sum_{i \leq t} \mathds{1}(t_{i} \leq t)$, where $\mathds{1}(\cdot)$ is the indicator function. In this paper, we assume $N$ to be a doubly stochastic Poisson---or \citet*{cox:55a}---process with associated random intensity---or rate---process $\lambda = ( \lambda_{t})_{t \geq 0}$, where $\lambda$ is an adapted and strictly positive real-valued stochastic process. That is, conditionally on $\lambda$, $N$ is an inhomogeneous Poisson process with rate $\lambda$, i.e. the conditional characteristic function of the increment $N_{t+ \Delta} - N_{t}$ is given by
\begin{equation} \label{equation:cox}
\varphi_{N_{t+ \Delta} - N_{t}}(u) = \mathbb{E} \Big( e^{iu(N_{t+ \Delta} - N_{t})} \mid \mathcal{F}_{t}^{ \lambda} \Big) = \exp \left( (e^{iu}-1) \int_{t}^{t+ \Delta} \lambda_{s} \mathrm{d}s \right) ,
\end{equation}
where $\mathcal{F}_{t}^{ \lambda} = \sigma( \{ \lambda_{s}; s \leq t \})$.

$\lambda_{t}$ can be thought of as the expected number of points arriving over the next short time interval $[t,t+ \Delta]$, based on available information about the rate process at time $t$, since
\begin{equation}
\lambda_{t} = \lim_{ \Delta \rightarrow 0} \frac{ \mathbb{E} \big[N_{t+ \Delta} - N_{t} \mid \mathcal{F}_{t}^{ \lambda} \big]}{ \Delta}.
\end{equation}
Hence, when $\lambda_{t}$ is locally bounded, the instantaneous expectation of the number of points is finite. In order to define intensity bursts, we therefore suppose that $\lambda_{t}$ is locally unbounded in the vicinity of a particular time instant, such that the expectation of the number of points in the vicinity of that time point is locally infinite.
\begin{assumption} \label{assumption:rate}
$\lambda_{t}$ can be decomposed as follows:
\begin{equation}
\lambda_{t} = \mu_{t} + \beta_{t},
\end{equation}
where $\mu = ( \mu_{t})_{t \geq 0}$ is a locally bounded and strictly positive real-valued stochastic process, and
\begin{equation} \label{equation:ib-theoretical}
\beta_{t} = \frac{ \sigma_{t}}{| \tau_{ \mathrm{ib}}-t|^{ \alpha}},
\end{equation}
for a stopping time $\tau_{ \mathrm{ib}}$, where $\sigma = ( \sigma_{t})_{t \geq 0}$ is a non-negative locally bounded and stochastically continuous random process and $0< \alpha <1$ is a constant.
\end{assumption}

In the context of an financial market, $N$ may describe the number of order submissions or trade executions in a security over the course of a trading session. With this interpretation, Assumption \ref{assumption:rate} therefore extends the notion of a drift burst in the asset log-price from \citet*{christensen-oomen-reno:22a} to the intensity process generating the discretely observed log-price record. $\mu$ is then the arrival rate of trades during ``normal'' market conditions, whereas $\beta$ represents the arrival rate of trades during ``distressed'' market conditions. The stopping time $\tau_{ \mathrm{ib}}$ is called the intensity burst time.

If $\sigma_{t} = 0$ almost surely for all $t$, the explosive term $\beta_{t}$ is absent and $N_{t}$ evolves as a simple Cox process. In particular, $N$ is orderly. That is, for any time $t$ it verifies the condition
\begin{equation} \label{equation:orderly}
\lim_{ \Delta \rightarrow 0} \frac{ \mathbb{P} \left( N_{t+ \Delta} - N_{t} > 1 \right)}{ \Delta} = 0,
\end{equation}
which implies that the probability of observing more than one event over a short time interval is negligible.

If $\sigma_{t} > 0$ almost surely, in a neighbourhood of $ \tau_{ \mathrm{ib}}$,
\begin{equation}
\int^{ \tau_{ \mathrm{ib}}+ \Delta}_{ \tau_{ \mathrm{ib}}- \Delta} \mu_{s} \mathrm{d}s = O_{p}( \Delta) \qquad \text{and} \qquad \int^{ \tau_{ \mathrm{ib}}+ \Delta}_{ \tau_{ \mathrm{ib}}- \Delta} \beta_{s} \mathrm{d}s = O_{p} \big( \Delta^{1- \alpha} \big),
\end{equation}
as $\Delta \rightarrow 0$. The last condition implies that the left-hand side of equation \eqref{equation:orderly} diverges. Hence, $N_{t}$ is not orderly. This implies that $N$ can generate a large cluster of points near each other in a small neighbourhood of $\tau_{ \mathrm{ib}}$. This property, which is crucial for generating extreme point clusters under the heavy traffic condition, cannot be reproduced by classical point processes (for example, a stationary Hawkes process), because any simple and stationary point process is orderly \citep[see, e.g., Section 3.3.V--IV in][]{daley-vere-jones:03a}. However, due to the integrability condition on $\beta_{t}$ in Assumption \ref{assumption:rate}, the compensator $\Lambda_{T} = \int_{0}^{T} \lambda_{s} \mathrm{d}s< \infty$ and $\mathbb{P}\left( N_{t+\Delta } - N_t > 1  \right) = o( \Delta^{1- \alpha})$ as $\Delta \rightarrow 0$, so $N$ itself is non-explosive and well-defined.

It is well-known that one cannot consistently estimate the intensity of a point process over a finite time interval, not even in the homogenous case. Hence, inference of point processes usually proceeds under long-span asymptotics ($T \rightarrow \infty$). However, as we are interested in identifying intensity bursts over small time intervals, we follow the convention in the high-frequency literature and assume $T$ is fixed. Instead, to do inference we impose an alternative ``heavy traffic'' assumption \citep*[e.g.,][]{kingman:61a}, in which the intensity of the underlying point process diverges under the null. To formalize this idea, we introduce an auxiliary parameter $n$ and consider a sequence of Cox processes $N^{n} = (N_{t}^{n})_{t \geq 0}$, for $n = 1, 2, \dots$. For a fixed $n$, the rate function of $N_{t}^{n}$ is equal to $n \lambda_{t}$, and the observed configuration of points is a realization of $N^{n}$. The asymptotic theory is then developed by supposing that $n \rightarrow \infty$.

To guarantee the existence of the required sequence of Cox processes, $N_{t}^{n}$ can conveniently be constructed as follows:
\begin{equation}\label{stacking}
N_{t}^{n} = \sum^{n}_{i=1} N_{t,i},
\end{equation}
where $(N_{t,i})_{i=1}^{n}$ are independent copies of $N_{t}$.

The heavy traffic condition is intimately connected with the literature on high-frequency estimation of volatility. Inference on realized variance typically proceeds under infill asymptotics \citep*[e.g.][]{andersen-bollerslev:98a, barndorff-nielsen-shephard:02a}. In that setting, we suppose a log-price process is observed at $m = m(n)$ discrete time points $t_{0}, t_{1}, \dots, t_{m}$ that partition the time interval $[0,T]$, where $m \overset{p}{ \longrightarrow} \infty$ such that $\sup_{i} | t_{i} - t_{i-1} | \overset{p}{ \longrightarrow} 0$ as $n \rightarrow \infty$. Thus, the point process $N^{n}$ together with the heavy traffic assumption is a standard model for generating stochastic sampling times satisfying the usual conditions alluded to in high-frequency econometrics. In fact, a realization of $N^{n}$ falls within the class of stochastic sampling schemes studied by \citet*{hayashi-jacod-yoshida:11a}. Hence, heavy traffic is a natural precursor for econometric analysis of financial high-frequency data.

Note that our theoretical exposition does not include self-exiting \citet*{hawkes:71a}-type processes, where the stochastic intensity process depends on the realization of $N$. In contrast to a Cox process, stacking independent self-exiting processes to generate an increasing number of events over a fixed time interval result in an poor model, because the correlation between increments of $N^{n}$ remain independent of $n$, which prevents inference of the local intensity. However, self-excitation can be embedded into our heavy traffic framework in several ways. \citet*{clinet-potiron:18a} study an exponential doubly-stochastic Hawkes process, where the base intensity and excitation function depend on $n$, so the correlation between increments of $N^{n}$ declines as $n$ increases. Alternatively, one may follow the stacking construction in \eqref{stacking} without the independence assumption. We postpone a theoretical treatment of this problem for future research. On a practical side, we explore the Hawkes processes as a model for the ``normal'' time event flow in the Monte Carlo section.

\section{Identification} \label{section:identification}

The detection of intensity bursts amounts to the following hypothesis:
\begin{equation} \label{equation:hypothesis}
\mathscr{H}_0: \omega \in \Omega_{0} \qquad \text{and} \qquad \mathscr{H}_{1}: \omega \in \Omega_{1},
\end{equation}
where $\Omega_{0}$ and $\Omega_{1}$ are complementary subsets of $\Omega$:
\begin{equation}
\Omega_{0} = \left\{ \omega \in \Omega : \int_{0}^{T} \beta_{t}^{2} \mathrm{d}t = 0 \right\} \qquad \text{and} \qquad \Omega_{1} = \left\{ \omega \in \Omega : \int_{0}^{T} \beta_{t}^{2} \mathrm{d}t > 0 \right\}.
\end{equation}
In this section, we propose a formal testing procedure to figure out which subset the realization of $N^{n}$ belongs to. In particular, we propose a test that conducts a statistical inference for the presence of an intensity burst near a single candidate time instance.

\subsection{Test statistic}

The identification of intensity bursts is based on a nonparametric, and backward-looking to allow for online detection, estimator of the local intensity, which is defined as:
\begin{equation} \label{equation:lambdba-hat}
\widehat{ \lambda}_{t} = \frac{N^{n}(t- \delta_{n},t)}{n \delta_{n}},
\end{equation}
where $N^{n}(a,b) = N_{b}^{n}-N_{a}^{n}$ is the number of points of the counting process on $(a, b]$, and $\delta_{n}$ is a bandwidth parameter.\footnote{The presence of $n$ in \eqref{equation:lambdba-hat} is inconvenient, since the parameter is not observed. However, it is only needed to show convergence of $\widehat{ \lambda}_{t}$. As we show below, $n$ drops out for the construction of the intensity burst test statistic.}

$\widehat{ \lambda}_{t}$ is nothing more than the realized intensity, i.e. the number of counts per time unit over an interval of length $\delta_{n}$, averaged across Cox processes. In a setup with light traffic (i.e. $n$ fixed), it follows that $\widehat{ \lambda}_{t} \overset{p}{ \longrightarrow} 0$. However, heavy traffic (i.e. $n \rightarrow \infty$) changes the stochastic properties of $\widehat{ \lambda}_{t}$.
\begin{lemma}\label{lemma:consistency}
Suppose that Assumption  \ref{assumption:rate} holds. Then, under $\mathscr{H}_{0}$, as $n \rightarrow \infty$ and $\delta_{n} \rightarrow 0$ such that $n \delta_{n} \rightarrow \infty$, it holds for all fixed $t \in [0,T]$ that
\begin{equation}
\widehat{ \lambda}_{t} \overset{p}{ \longrightarrow} \mu_{t-}.
\end{equation}
Moreover, under $\mathscr{H}_{1}$,
\begin{equation}
\widehat{ \lambda}_{ \tau_{ \mathrm{ib}}} = O_{p} \big( \delta_{n}^{- \alpha} \big).
\end{equation}
\end{lemma}
The lemma shows that, under general assumptions about the intensity process, the intensity estimator converges to the left-limit of the local intensity at time $t$ under the null hypothesis. As $n \rightarrow \infty$, the summation of independent copies of the Cox process in the heavy traffic limit ensures that, on average, the accumulation of points close to $t$ correspond approximately to the instantaneous arrival rate. To generate a sufficient number of points inside the estimation window $[t - \delta_{n}, t]$, we need $\delta_{n} n \rightarrow \infty$. A law of large numbers then shows that $\widehat{ \lambda}_{t}$ converges in probability and, as $\delta_{n} \rightarrow 0$, the limit is $\mu_{t-}$.

On the other hand, under the alternative, the arrival rate of the bursting process causes the estimator to explode around $ \tau_{ \mathrm{ib}}$.

Lemma \ref{lemma:consistency} highlights an important difference between the estimation of the local intensity of point processes and the estimation of the local drift of Brownian semimartingales. In the latter, the drift estimator is asymptotically unbiased but inconsistent \citep*{kristensen:10a}, because the increments of the Brownian motion, although mean zero, exhibit so much variation over short time intervals that any information about the drift is lost. This is exploited by \citet*{christensen-oomen-reno:22a} to construct a drift burst test statistic, which relies on the different rates of divergence of the drift estimator under the null and alternative. Their test statistic has the convenient side effect that it knocks out the true drift coefficient asymptotically. By contrast, under the null hypothesis $\widehat{ \lambda}_{t}$ is consistent for the local intensity of the point process in the heavy traffic limit. Hence, for intensity burst detection we propose a slightly different test statistic compared to \citet*{christensen-oomen-reno:22a}. In particular, to center the statistic we compare two local intensity estimators calculated from the nearest lagged block of observations:
\begin{equation}
\nabla \widehat{ \lambda}_{t} = \widehat{ \lambda}_{t} - \widehat{ \lambda}_{t - \delta_{n}}.
\end{equation}
As we prove below, $\nabla \widehat{ \lambda}_{t}$ converges in probability to zero under $\mathscr{H}_0$, but it is unbounded in probability under $\mathscr{H}_{1}$. Hence, one can derive an asymptotic confidence interval for $\nabla \widehat{ \lambda}_{t}$ and reject $\mathscr{H}_{0}$ if $\nabla \widehat{ \lambda}_{t}$  is outside of it. This suggests a standard test statistic of the form:
\begin{equation} \label{equation:test-statistic}
\phi_{t}^{ \text{ib}} = \frac{ \nabla \widehat{ \lambda}_{t}}{ \widehat{ \text{std}}( \nabla \widehat{ \lambda}_{t})},
\end{equation}
where $\widehat{ \text{std}}( \nabla \widehat{ \lambda}_{t})$ is an estimator of $\text{std}( \nabla \widehat{ \lambda}_{t})$ under $\mathscr{H}_{0}$.

It turns out, however, that the asymptotic variance of $\widehat{ \lambda}_{t}$ depends crucially on the interplay between $\delta_{n}$ and $n$, which follows from a central limit theorem (CLT) under $\mathscr{H}_{0}$. To derive this CLT, we require a bit of structure on the baseline intensity of $N$.

\begin{assumption} \label{assumption:mu}
$\mu$ is an It\^{o} semimartingale of the form:
\begin{equation}
\mu_{t} = \mu_{0}  + \int^{t}_{0} a_{s} \mathrm{d}s + \int^{t}_{0} \nu_{s} \mathrm{d}W_{s} +  \int_{0}^{t} \int_{ \abs{ \delta_{s}(x)} \leq 1} \delta_{s}(x) \left( \mathfrak{p}( \mathrm{d}s, \mathrm{d}x)- \mathfrak{q}( \mathrm{d}s, \mathrm{d}x) \right) + \int_{0}^{t} \int_{ \abs{ \delta_{s}(x)} > 1} \delta_{s}(x) \mathfrak{p}( \mathrm{d}s, \mathrm{d}x),
\end{equation}
where $\mu_{0}$ is $\mathcal{F}_{0}$-measurable, $a = (a_{t})_{t \geq 0}$ and $\nu = ( \nu_{t})_{t \geq 0}$ are adapted, c\`{a}dl\`{a}g stochastic processes,  $W$ is a standard Brownian motion, $\mathfrak{p}$ is a Poisson random measure on $\mathbb{R}_{+} \times \mathbb{R}$ with compensator $\mathfrak{q}( \mathrm{d}s, \mathrm{d}x) = \mathrm{d}s F( \mathrm{d}x)$, where $F$ is a $\sigma$-finite measure on $\mathbb{R}$, and $\delta_{s}(x)$ is a predictable function such that there exists a sequence of $\mathcal{F}_{t}$-stopping times $(\tau_{n})_{n=1}^{ \infty}$ with $\tau_{n} \rightarrow \infty$ and, for each $n$, a deterministic and non-negative function $\Gamma_{n}$ with $\min \left(| \delta_{t}(x)|, 1 \right) \leq \Gamma_{n}(x)$ for $t \leq \tau_{n}$, $\int_{ \mathbb{R}} \Gamma_{n}(x)^{2} F( \mathrm{d}x) < \infty$, and $\int_{ \left\{x: \Gamma_{n}(x) \leq \kappa \right\}} \Gamma_{n}(x)F( \mathrm{d}x) < \infty$, for all $(t,x)$ and $\kappa \in (0,1)$.
\end{assumption}

Assumption \ref{assumption:mu} is common in high-frequency analysis \citep*[e.g.][]{jacod-protter:12a}, since it allows to apply estimates for semimartingales to control the discretization error. It is a standard tool often used to describe arbitrage-free price processes, since these must be semimartingale \citep{delbaen-schachermayer:94a}. In that context, the assumption is---apart from the integrability condition on the driving processes---nonparametric, as it does not restrict the dynamic of the price and allows it to exhibit time-varying drift, stochastic volatility, and a more or less unrestricted jump component, which can be both infinitely active and of infinite variation. Moreover, the various terms can be arbitrarily dependent.

However, since we are looking at an intensity process and not the price of a tradable asset, we are not bound by the no free lunch principle. We therefore also inspect an alternative framework below, which replaces the above structural assumption by a general smoothness condition.

\begin{theorem} \label{theorem:clt}
Suppose that Assumptions \ref{assumption:rate} and \ref{assumption:mu} hold. Then, under $\mathscr{H}_{0}$, as $n \rightarrow \infty$ and $\delta_{n} \rightarrow 0$ such that $n \delta_{n} \rightarrow \infty$, it holds for all fixed $t \in [0,T]$ that
\begin{itemize}
\item[(i)] if  $\delta_{n} \sqrt{n} \rightarrow 0$,
\begin{equation}
\sqrt{n \delta_{n}} \left( \widehat{ \lambda}_{t} - \mu_{t-} \right) \overset{ \mathfrak{D}_{s}}{ \longrightarrow} \sqrt{ \mu_{t-}} \mathcal{Z},
\end{equation}
and
\begin{equation}
\sqrt{n \delta_{n}} \nabla \widehat{ \lambda}_{t} \overset{ \mathfrak{D}_{s}}{ \longrightarrow} \sqrt{2 \mu_{t-}} \mathcal{Z},
\end{equation}
\item[(ii)]  if  $\delta_{n} \sqrt{n} \rightarrow \theta>0$,
\begin{equation}
\frac{1}{ \sqrt{ \delta_{n}}} \left( \widehat{ \lambda}_{t} - \mu_{t-} \right) \overset{ \mathfrak{D}_{s}}{ \longrightarrow} \sqrt{ \frac{1}{ \theta^{2}} \mu_{t-} + \frac{1}{3} \nu_{t}^{2}} \mathcal{Z},
\end{equation}
and
\begin{equation}
\frac{1}{ \sqrt{ \delta_{n}}} \nabla \widehat{ \lambda}_{t} \overset{ \mathfrak{D}_{s}}{ \longrightarrow} \sqrt{ \frac{2}{ \theta^{2}} \mu_{t-} + \frac{8}{3} \nu_{t}^{2}} \mathcal{Z},
\end{equation}
\item[(iii)]  if  $\delta_{n} \sqrt{n} \rightarrow \infty$,
\begin{equation}
\frac{1}{ \sqrt{ \delta_{n}}} \left( \widehat{ \lambda}_{t} - \mu_{t-} \right) \overset{ \mathfrak{D}_{s}}{ \longrightarrow} \sqrt{ \frac{1}{3} \nu_{t}^{2}} \mathcal{Z},
\end{equation}
and
\begin{equation}
\frac{1}{ \sqrt{ \delta_{n}}} \nabla \widehat{ \lambda}_{t} \overset{ \mathfrak{D}_{s}}{ \longrightarrow} \sqrt{ \frac{8}{3} \nu_{t}^{2}} \mathcal{Z},
\end{equation}
\end{itemize}
where $\mathcal{Z}$ is a standard normal random variable, which is defined on an extension of $( \Omega, \mathcal{F}, ( \mathcal{F}_{t})_{t \geq 0}, \mathbb{P})$ and independent of $\mathcal{F}$, and $\overset{ \mathfrak{D}_{s}}{ \longrightarrow}$ is stable convergence in law \citep*[see, e.g.,][]{jacod-protter:12a}.
\end{theorem}

Theorem \ref{theorem:clt} shows that the asymptotic distribution of our estimator is mixed normal, but the convergence rate and limiting variance depends on the order of $\delta_{n}$. On the one hand, if $\delta_{n} \sqrt{n} \rightarrow 0$, the localization dominates and the variation of the intensity parameter along its sample path is immaterial. Hence, by the Poisson distribution, the variance is equal to the mean. On the other hand, if $\delta_{n} \sqrt{n} \rightarrow \infty$, the roles are reversed, and the variation of the intensity parameter controls the asymptotic variance. In both cases, the rate of convergence can be rather slow. The optimal convergence rate, $n^{-1/4}$, is achieved with $\delta_{n} \asymp n^{-1/2}$, where the opposing forces are balanced and both affect the limiting distribution.

The standard errors of $\nabla \widehat{ \lambda}_{t}$ appearing in the second part of (ii) and (iii) are larger than one would expect when comparing to (i). The explanation is that $n$ diverges so fast compared to the shrinking of $\delta_{n}$ that the lagged estimator $\widehat{ \lambda}_{t - \delta_{n}}$ is much more imprecise compared to $\widehat{ \lambda}_{t}$. This effect is not present in part (i) of the theorem.

Assumption \ref{assumption:mu} prevents the baseline intensity to be described by a non-semimartingale, such as a fractional Brownian motion. The latter is a popular model in the realm of asset return variation, since it can generate roughness and long memory \citep[e.g.][]{comte-renault:98a, gatheral-jaisson-rosenbaum:18a}. To the extent that trading activity and return volatility are connected, which holds empirically and is also a standard implication of the information flow interpretation of the mixture of distribution hypothesis \citep[e.g.,][]{andersen:96a, clark:73a, tauchen-pitts:83a}, we should feel uncomfortable with that restriction. It can, however, be replaced by an appropriate smoothness condition, such as assuming the sample paths of $\mu$ are H\"{o}lder continuous.

\begin{assumption} \label{assumption:holder}
There exists $H^{ \mu} > 0$, such that
\begin{equation}
\mathbb{E} \left[ \sup_{s \leq t \leq s + \Delta} | \mu_{s} - \mu_{t}| \right] \leq C \Delta^{H^{ \mu}},
\end{equation}
for some constant $C>0$ and all $0 \leq s \leq t \leq T$.
\end{assumption}

\begin{theorem} \label{theorem:clt-smoothness}
Suppose that Assumptions \ref{assumption:rate} and \ref{assumption:holder} hold. Then, under $\mathscr{H}_{0}$, as $n \rightarrow \infty$ and $\delta_{n} \rightarrow 0$ such that $n \delta_{n} \rightarrow \infty$ and $\delta_{n} n^{1- \frac{2H^{ \mu}}{3}} \rightarrow 0$, it holds for all fixed $t \in [0,T]$ that
\begin{equation}
\sqrt{n \delta_n } \left( \widehat{ \lambda}_{t} - \mu_{t-} \right) \overset{ \mathfrak{D}_{s}}{ \longrightarrow} \sqrt{ \mu_{t-}} \mathcal{Z},
\end{equation}
\end{theorem}
Theorem \ref{theorem:clt-smoothness} shows our spot intensity estimator is consistent and asymptotically normal under Assumption \ref{assumption:holder} with an unaltered convergence rate. However, it only covers the short bandwidth setting in part (i) of Theorem \ref{theorem:clt}, where the local variation of the intensity process does not affect the asymptotic conditional variance of the estimator. This is not true for part (ii) and (iii), and that is where the structure of Assumption \ref{assumption:mu} is helpful. Without such knowledge, it is difficult to extend Theorem \ref{theorem:clt-smoothness} in that direction.

The split points in Theorem \ref{theorem:clt} are dictated by the behavior of $\delta_{n} \sqrt{n}$, since for Brownian motion (the leading term is Assumption \ref{assumption:mu}), the probabilistic order of its increment is $O_{p}( \sqrt{ \delta_{n}})$. This condition is replaced in Theorem \ref{theorem:clt-smoothness} by $\delta_{n} n^{1- \frac{2H^{ \mu}}{3}}$. Thus, if $\delta_{n} n^{1- \frac{2H^{ \mu}}{3}} \rightarrow 0$, we are again in the short bandwidth setting vis-\`{a}-vis the H\"{o}lder regularity of the process. However, even for $H^{ \mu} = 1/2$ (the index of Brownian motion), the rate condition reduces to $\delta_{n} n^{2/3} \rightarrow 0$ and is more restrictive than before. This is because in the proof of Theorem \ref{theorem:clt}, we also exploit that Brownian motion is a martingale.

At last, we should stress that the dual rate conditions enforced by Theorem \ref{theorem:clt-smoothness} are in a direct conflict of interest. On the one hand, it should hold that $n \delta_{n} \rightarrow \infty$, so that there are a sufficient number of observations inside the estimation window. One the other hand, for $H^{ \mu}$ very small we more or less also need $n \delta_{n} \rightarrow 0$. This is because the smoothness condition imposed in kernel-based nonparametric analysis requires a shorter bandwidth for rough processes.\footnote{If the localization window is minuscule, there may be too few observations inside it to get the mechanics of the central limit theorem working, risking that the asymptotic theory is a poor approximation to the finite sampling distribution of the estimator. \citet{bugni-li-li:23a} also study non-price high-frequency data employing the nonparametric state-space model of \citet{li-xiu:16a}. They propose a permutation-based test to detect general distributional discontinuities, which includes the local two-sample mean problem. The advantage of their framework is its finite sample validity, which addresses potential concerns caused by roughness. However, developing such a permutation-based test is beyond the scope of the paper.}

\subsection{Observed asymptotic local variance}

In practice, picking the correct estimator of the asymptotic variance is difficult, because $n$ is not observed. To construct a feasible test statistic, we follow \citet{mykland-zhang:17a} by alluding to the notion of the observed asymptotic variance.\footnote{An alternative inference procedure is subsampling \citep[e.g.][]{politis-romano-wolf:99a}, which was adapted to high-frequency data in \citet{kalnina:11a} and \citet{christensen-podolskij-thamrongrat-veliyev:17a}.} We set $\delta_{n} = \ell_{n} \Delta_{n}$, where $\ell_{n}$ is a deterministic sequence of natural numbers and $\Delta_{n} > 0$ represents a small time interval. We further impose that $\Delta_{n} = n^{-1}$, but this is merely done for notational convenience. Nothing changes if $\Delta_{n} = O(n^{-1})$, except we introduce an additional tuning parameter.

$\widehat{ \lambda}_{t}$ can then be rewritten:
\begin{equation} \label{equation:lambda-hat-discrete}
\widehat{ \lambda}_{t} = \frac{1}{ \ell_{n}} \sum_{i \in \mathcal{D}_{t-}^{n}} \Delta_{i} N^{n},
\end{equation}
where $\Delta_{i} N^{n} = N^{n}(i \Delta_{n},(i-1)\Delta_{n})$ is the increment of $N_{t}^{n}$ over a short time interval and $\mathcal{D}_{t-}^{n} = \{ tn- \ell_{n}+1, tn- \ell_{n}+2, \dots, tn \}$.

The formulation in \eqref{equation:lambda-hat-discrete} follows the traditional high-frequency approach. It shows that the intensity estimator can be expressed as a local average of increments to the discretized point process $N_{t}^{n}$, which is a more convenient formulation for developing our asymptotic variance estimator.\footnote{In our framework, $N^{n}$ is observed in continuous-time. However, many existing datasets do not reveal the exact location of all events from a point process, but only report discrete observations. For example, the number of trades in every 1-second slot may be available. In such cases, the data is structured in the form of a discretized version of $N^{n}$ used in equation \eqref{equation:lambda-hat-discrete}. Thus, Corollary \ref{intensityCLT_elln} and subsequent theorems remains applicable to such data.} With this convention, Theorem \ref{theorem:clt} can be stated as follows.
\begin{corollary}\label{intensityCLT_elln}

Suppose that Assumptions \ref{assumption:rate} and \ref{assumption:mu} hold. Then, under $\mathscr{H}_{0}$, as $n \rightarrow \infty$ and $\ell_{n} \rightarrow \infty$ such that $\ell_{n} \Delta_{n} \rightarrow 0$, it holds for all fixed $t \in [0,T]$ that
\begin{equation}
\sqrt{ \rho_{n}} \nabla \widehat{ \lambda}_{t} \overset{ \mathfrak{D}_{s}}{ \longrightarrow} \sqrt{ \mathsf{avar}( \nabla \widehat{ \lambda}_{t})} \mathcal{Z},
\end{equation}
where
\begin{equation}
\left( \rho_{n}, \mathsf{avar} \big( \nabla \widehat{ \lambda}_{t} \big) \right) =
\begin{cases}
( \ell_{n},  2 \mu_{t-}), & \text{if } \ell_{n} \sqrt{ \Delta_{n}} \rightarrow 0, \\[0.10cm]
\big( ( \ell_{n} \Delta_{n})^{-1}, \frac{2}{ \theta^{2}} \mu_{t-} + \frac{8}{3} \nu_{t}^{2} \big), & \text{if } \ell_{n} \sqrt{ \Delta_{n}} \rightarrow \theta > 0, \\[0.10cm]
\big( ( \ell_{n} \Delta_{n})^{-1}, \frac{8}{3} \nu_{t}^{2} \big), & \text{if } \ell_{n} \sqrt{ \Delta_{n}} \rightarrow \infty.
\end{cases}
\end{equation}
\end{corollary}
\noindent Then, we propose to set
\begin{equation} \label{equation:observed-variance}
\widetilde{ \mathsf{avar}} \big( \nabla \widehat{ \lambda}_{t} \big) = \frac{ \rho_{n}}{K_{n}} \sum_{j=0}^{K_n-1} \Big( \widehat{ \lambda}_{t-2j \ell_{n} \Delta_{n}} - \widehat{ \lambda}_{t-( \ell_{n}+2j \ell_{n}) \Delta_{n}} \Big)^{2},
\end{equation}
where $K_{n}$ is another sequence of natural numbers.

The observed asymptotic local variance, $\widetilde{ \mathsf{avar}} \big( \nabla \widehat{ \lambda}_{t} \big)$, is the sample variance of $K_{n}$ estimates computed over non-overlapping blocks consisting of $2 \ell_{n}$ observations.  When $K_{n}$ is sufficiently large and the observation blocks are near $t$, $\widetilde{ \mathsf{avar}} \big( \nabla \widehat{ \lambda}_{t} \big)$ is consistent for $\mathsf{avar} \big( \nabla \widehat{ \lambda}_{t} \big)$.

\begin{theorem} \label{theorem:observed-asymptotic-variance}
Suppose that Assumptions \ref{assumption:rate} and \ref{assumption:mu} hold. Then, under $\mathscr{H}_{0}$, as $n \rightarrow \infty$, $\ell_{n} \rightarrow \infty$, $K_{n} \rightarrow \infty$ such that $\ell_{n} \Delta_{n} \rightarrow 0$ and $\ell_{n} K_{n} \Delta_{n} \rightarrow 0$, it holds for all fixed $t \in [0,T]$ that
\begin{equation}
\widetilde{ \mathsf{avar}} \big( \nabla \widehat{ \lambda}_{t} \big) \overset{p}{ \longrightarrow} \mathsf{avar} \big( \nabla \widehat{ \lambda}_{t} \big).
\end{equation}
\end{theorem}
That is, $\widetilde{ \mathsf{avar}} \big( \nabla \widehat{ \lambda}_{t} \big)$ converges in probability to $\mathsf{avar} \big( \nabla \widehat{ \lambda}_{t} \big)$ for any limiting behaviour of $\rho_{n}$ allowed in Corollary \ref{intensityCLT_elln}.

The condition $K_{n} \ell_{n} \Delta_{n} \rightarrow 0$ in Theorem \ref{theorem:observed-asymptotic-variance} is somewhat restrictive. However, it can be loosened with overlapping blocks in the definition of the observed asymptotic local variance. Hence, an alternative version of the estimator is the following:
\begin{equation} \label{equation:avar-efficient}
\widehat{ \mathsf{avar}} \big( \nabla \widehat{ \lambda}_{t} \big) = \frac{ \rho_{n}}{K_{n}} \sum_{j=0}^{K_{n}-1} \Big( \widehat{ \lambda}_{t- j \Delta_{n}} - \widehat{ \lambda}_{t-( \ell_{n}+j) \Delta_{n}} \Big)^{2}.
\end{equation}

\begin{theorem} \label{observed_variance_LLN2}
Suppose that Assumptions \ref{assumption:rate} and \ref{assumption:mu} hold. Then, under $\mathscr{H}_{0}$, as $n \rightarrow \infty$, $\ell_{n} \rightarrow \infty$, $K_{n} \rightarrow \infty$ such that $\ell_{n} \Delta_{n} \rightarrow 0$, $K_{n} \Delta_{n} \rightarrow 0$ and $\ell_{n} / K_{n} \rightarrow 0$, it holds for all fixed $t \in [0,T]$ that
\begin{equation}
\widehat{ \mathsf{avar}} \big( \nabla \widehat{ \lambda}_{t} \big) \overset{p}{ \longrightarrow} \mathsf{avar} \big( \nabla \widehat{ \lambda}_{t} \big).
\end{equation}
\end{theorem}
The condition $\ell_{n}/K_{n} \rightarrow 0$ in Theorem \ref{observed_variance_LLN2} indicates that the number of differences used to estimate the local asymptotic variance of $\nabla \widehat{ \lambda}_{t}$ should increase faster than the number of observations used to compute the local intensity estimate. The conditions $K_n\Delta_n \to 0$ and $\ell_{n} \Delta_{n} \rightarrow 0$ ensure that all observations remain close to $t$.

Thus, the test statistic in \eqref{equation:test-statistic} becomes:
\begin{equation}
\phi_{t}^{ \text{ib}} = \frac{ \sqrt{ \rho_{n}} \nabla \widehat{ \lambda}_{t}}{ \sqrt{ \widehat{ \mathsf{avar}} \big( \nabla \widehat{ \lambda}_{t} \big)}}.
\end{equation}

\begin{theorem} \label{theorem:test-statistic}
Suppose that the conditions of Theorems \ref{theorem:clt} and \ref{observed_variance_LLN2} hold. Then,
\begin{equation}
\begin{cases}
\phi_{t}^{ \mathrm{ib}} \overset{ \mathfrak{D}}{ \longrightarrow} N(0,1), & \text{conditional on } \mathscr{H}_{0}, \\[0.10cm]
\phi_{ \tau_{ \mathrm{ib}}}^{ \mathrm{ib}} \overset{p}{ \longrightarrow} \infty, & \text{conditional on } \mathscr{H}_{1}.
\end{cases}
\end{equation}
\end{theorem}
Hence, we reject $\mathscr{H}_{0}$ if $\phi_{t}^{ \text{ib}}$ exceeds a quantile in the right-hand tail of the standard normal distribution corresponding to a given significance level $\varsigma$. This ensures the test has size control under $\mathscr{H}_{0}$. On the other hand, under $\mathscr{H}_{1}$, $\phi_{t}^{ \text{ib}}$ is unbounded in probability as $t \rightarrow \tau_{ \mathrm{ib}}$, so the test is also consistent.\footnote{Theorem \ref{theorem:test-statistic} continues to hold if $\widehat{ \mathsf{avar}} \big( \nabla \widehat{ \lambda}_{t} \big)$ is replaced with $\widetilde{ \mathsf{avar}} \big( \nabla \widehat{ \lambda}_{t} \big)$ in the definition of $\phi_{t}^{ \text{ib}}$, provided the conditions of Theorem \ref{theorem:observed-asymptotic-variance} hold.}

It is important to emphasize that we do not need to choose $\rho_{n}$ to compute $\phi_{t}^{ \text{ib}}$. This follows by direct insertion, showing the test statistic can be expressed as:
\begin{equation}
\phi_{t}^{ \text{ib}} = \frac{N^{n}(t- \delta_{n}, t) - N^{n}(t-2 \delta_{n}, t-\delta_{n})}{ \sqrt{ \displaystyle \frac{1}{K_{n}} \sum_{j=0}^{K_{n}-1} \big(N^{n}(t-j \Delta_{n}- \delta_{n},t-j \Delta_{n}) - N^{n}(t-j \Delta_{n}-2 \delta_{n},t-j \Delta_{n}- \delta_{n}) \big)^2}}.
\end{equation}
In practice, we take $\Delta_{n}$ to be one second, although this is merely a convenient choice. The selection of $\ell_{n}$ and $K_{n}$ are studied in the Monte Carlo analysis presented in Section \ref{section:simulation}.

\begin{remark}
The local intensity estimator defined in \eqref{equation:lambda-hat-discrete} can be rewritten as a standard kernel-based estimator of the form
\begin{equation}
\widehat{ \lambda}_{t} =  \frac{1}{ \ell_{n}} \sum_{i=1}^{\lfloor T \Delta_{n} \rfloor} \mathcal{K} \bigg( \frac{i \Delta_{n} - t}{ \ell_{n}} \bigg) \Delta_{i} N^{n},
\end{equation}
where $\ell_{n}$ is the bandwidth and $\mathcal{K}$ is the indicator kernel $\mathcal{K}(x) = \mathbf{1}_{[-1,0]}(x)$. It is of course possible to select a different kernel. While the full development of a corresponding theory is outside of the scope of our paper, we conjecture that Theorem \ref{theorem:test-statistic} remains valid under suitable regularity conditions for the kernel and bandwidth \citep*[e.g.][]{kristensen:10a, christensen-oomen-reno:22a}. We investigate the exponential kernel in the simulation analysis and empirical application.
\end{remark}

\subsection{How to separate a burst from a jump?} \label{section:disentangle}

Theorem \ref{theorem:clt} and Theorem \ref{theorem:test-statistic} show that---in the limit---our test statistic is robust to jumps in the intensity process at the test time $t$. The intuition is that the main term in the numerator is $\widehat{ \lambda}_{t} - \widehat{ \lambda}_{t - \delta_{n}}$. With a fixed $t$ and a sufficiently small $\delta_{n}$, both $[t- \delta_{n}, t]$ and $[t-2 \delta_{n}, t- \delta_{n}]$ are close to $t$, so any jumps (of vanishing size as $\delta_{n} \rightarrow 0$) from a left neighbourhood of $t$ fall outside the estimation windows, so $\widehat{ \lambda}_{t}$ and $\widehat{ \lambda}_{t - \delta_{n}}$ converge to the left-limit $\mu_{t-}$, and the latter is not affected by the behaviour of the intensity process at time $t$.

In finite samples, $\delta_{n}$ is fixed. In that case, $\widehat{ \lambda}_{t} - \widehat{ \lambda}_{t - \delta_{n}}$ may numerically coincide with the difference between a forward- and backward-looking intensity estimator:
\begin{equation} \label{equation:backward-forward}
\widehat{ \lambda}^{(-)}_{t} = \frac{1}{ \ell_{n}} \sum_{i \in \mathcal{D}_{t-}^{n}} \Delta_{i} N^{n}, \qquad \mathrm{and} \qquad \widehat{ \lambda}^{(+)}_{t} = \frac{1}{ \ell_{n}} \sum_{i \in \mathcal{D}_{t+}^{n}} \Delta_{i} N^{n},
\end{equation}
where $\mathcal{D}_{t-}^{n} = \{ tn- \ell_{n}+1, tn- \ell_{n}+2, \dots, tn \}$ and  $\mathcal{D}_{t+}^{n} = \{ tn + 1, tn +2, \dots, tn + \ell_n \}$. $\widehat{ \lambda}^{(-)}_{t}$ is equivalent to $\widehat{ \lambda}_{t}$ and thus has probability limit $\mu_{t-}$, whereas $\widehat{ \lambda}^{(+)}_{t}$ converges to the right-limit $\mu_{t+} = \mu_{t}$. As such, one may suspect that the real cause of the rejection was in fact due to a jump in the intensity process, i.e. $\mu_{t} \neq \mu_{t-}$.

In this section, we dive into this issue and propose a framework that can separate a burst from a jump. We assume to know the existence of a time instant $\theta$, where there is a change-point in the intensity process--either a burst or a jump. $\theta$ can be a stopping time, a known point of pre-scheduled news announcements, or perhaps the center of an intensity burst that has already been detected with our testing procedure.

The crest of our approach is to study the limiting behaviour of a variant of our intensity estimator around the change point. We build on the ``change-of-frequency'' approach often employed in the volatility literature \citep*[e.g.,][]{ait-sahalia-jacod:09b, todorov-tauchen:10a} and propose a two-sided local intensity estimator:
\begin{equation} \label{equation:centered-estimator}
\widetilde{ \lambda}_{t}( \delta_{n}) = \frac{1}{2 \ell_{n}} \sum_{i = tn- \ell_{n}+1}^{tn+ \ell_{n}} \Delta_{i} N = \frac{N(t- \delta_{n},t+ \delta_{n})}{2n \delta_{n}}.
\end{equation}
We highlight the dependence of $\widetilde{ \lambda}_{t}$ on the bandwidth, because discrimination between a burst and a jump relies on sampling the estimator with different $\delta_{n}$.

To state a more concrete result, we let $\Omega_{0}^{H} \subseteq \Omega_{0}$ be the subset of the sample space, where the intensity process fulfills Assumption \ref{assumption:holder}:
\begin{equation}
\Omega_{0}^{H} = \left\{ \omega \in \Omega_{0} : \sup_{s \leq t \leq s + \Delta} | \mu_{s} - \mu_{t}| < L^{ \mu} \Delta^{H^{ \mu}} \right\},
\end{equation}
where $L^{ \mu}$ is an almost surely finite positive random variable. Note that there is no intensity burst on $\Omega_{0}^{H}$. Next, define $\Omega_{1}^{J} \subseteq \Omega$ such that:
\begin{equation}
\Omega_{1}^{J} = \left\{ \omega \in \Omega : \exists \theta \in (0,T) \, \exists C > 0 : | \Delta \mu_{ \theta}| > C \quad \mathrm{and} \quad \exists \epsilon > 0 : ( \mu_{t})_{t \in[ \theta- \epsilon, \theta+ \epsilon] \setminus \{ \theta \}} \in \Omega_{0}^{H} \right\},
\end{equation}
where $\Delta \mu_{t} = \mu_{t} - \mu_{t-}$. The interpretation is that on $\Omega_{1}^{J}$, there is a jump in $\mu$ of size $\Delta \mu_{ \theta}$ at time $\theta$, but otherwise the intensity process is smooth (and, in particular, there is no intensity burst) in a closed ball around this point. At last, we consider the set $\Omega_{1}^{B} \subseteq \Omega_{1}$ such that there is an intensity burst with center at $\tau_{ \mathrm{ib}} = \theta$.

As we show below, it turns out that the pathwise properties of $\widetilde{ \lambda}_{t}( \delta_{n})$ are different on $\Omega_{1}^{J}$ and $\Omega_{1}^{B}$. This can be exploited to discriminate an intensity burst from an intensity jump.  However, to develop an asymptotic theory for $\widetilde{ \lambda}_{t}( \delta_{n})$ in the presence of an intensity burst, we need a smoothness condition on $\sigma$ (the volatility of the burst process) vis-\'{a}-vis the above.
\begin{assumption} \label{assumption:sigma}
There exists $H^{ \sigma} > \alpha/2$, such that
\begin{equation}
\sup_{s \leq t \leq s + \Delta} | \sigma_{s} - \sigma_{t}| < L^{ \sigma} \Delta^{H^{ \sigma}},
\end{equation}
where $L^{ \sigma}$ is an almost surely finite positive random variable, and $\alpha$ is the explosion rate of the intensity burst from Assumption \ref{assumption:rate}.
\end{assumption}

Now, we are able to establish the following lemma.
\begin{lemma} \label{lemma:cof}
Suppose that Assumptions \ref{assumption:rate} and \ref{assumption:sigma} hold and let $k > 0$ be a positive number. As $n \rightarrow \infty$ and $\delta_{n} \rightarrow 0$ such that $n \delta_{n} \rightarrow \infty$ and $\delta_{n} n^{1- \frac{2H^{ \mu}}{3}} \rightarrow 0$, it holds that
\begin{itemize}
\item[(i)] Conditional on $\Omega_{1}^{B}$,
\begin{equation}
\delta_{n}^{ \alpha/2} \left( \widetilde{ \lambda}_{ \theta}(k \delta_{n}) - \sigma_{ \theta} \left(2k \delta_{n} \right)^{- \alpha} \right) \overset{p}{ \longrightarrow} 0 \qquad \mathrm{and} \qquad \frac{ \widetilde{ \lambda}_{ \theta}(k \delta_{n})}{ \widetilde{ \lambda}_{ \theta}( \delta_{n})} \overset{p}{ \longrightarrow} k^{- \alpha}.
\end{equation}
\item[(ii)] Conditional on $\Omega_{1}^{J}$,
\begin{equation}
\widetilde{ \lambda}_{ \theta}(k \delta_{n}) \overset{p}{ \longrightarrow} \mu_{ \theta-} + \frac{ \Delta \mu_{ \theta}}{2} \qquad \mathrm{and} \qquad \frac{ \widetilde{ \lambda}_{ \theta}(k \delta_{n})}{ \widetilde{ \lambda}_{ \theta}( \delta_{n})} \overset{p}{ \longrightarrow} 1.
\end{equation}
\end{itemize}
\end{lemma}

The first part of Lemma \ref{lemma:cof} suggests that on $\Omega_{1}^{B}$ we can construct a direct estimator of the explosion rate of an intensity burst, $\alpha$, via a logarithmic transformation:
\begin{equation} \label{equation:alpha-estimator}
\widehat{ \alpha} = - \frac{ \log \left( \widetilde{ \lambda}_{ \theta}(k \delta_{n}) \right) - \log \left( \widetilde{ \lambda}_{ \theta}( \delta_{n}) \right)}{ \log k} \overset{p}{ \longrightarrow} \alpha.
\end{equation}
On $\Omega_{1}^{J}$, this estimator converges to zero in probability as it should be in the absence of an intensity burst, i.e. $\hat{ \alpha} \overset{p}{ \longrightarrow} 0$.

A more efficient estimator of $\alpha$ can be retrieved by calculating $\log \left( \widetilde{ \lambda}_{ \theta}(k \delta_{n}) \right)$ for a range of $k$ and regressing these against $\log(k)$. The negative of the slope estimate is then $\hat{ \alpha}$. We explore this possibility in the empirical application.

The testing procedure to separate the competing hypothesis of a jump or a burst in the intensity process can now be conducted with either $\Omega_{1}^{B}$ or $\Omega_{1}^{J}$ as the null, but it requires a formal derivation of the asymptotic distribution of the ratio statistic in that state. If the prior is that intensity has jumped, then a statistical test to reject this conviction can be deduced from the following result.

\begin{theorem} \label{theorem:cof-jump}
Suppose that Assumption \ref{assumption:rate} holds. Conditional on $\Omega_{1}^{J}$, as $n \rightarrow \infty$ and $\delta_{n} \rightarrow 0$ such that $n \delta_{n} \rightarrow \infty$ and $\delta_{n} n^{1- \frac{2H^{ \mu}}{3}} \rightarrow 0$, it follows that
\begin{equation} \label{equation:test-statistic-jump}
\sqrt{n \delta_n } \left( \frac{ \widetilde{ \lambda}_{ \theta}(k \delta_{n})}{ \widetilde{ \lambda}_{ \theta}( \delta_{n})} - 1 \right) \overset{ \mathfrak{D}_{s}}{ \longrightarrow} \sqrt{ \mathrm{avar_{k}}} \mathcal{Z},
\end{equation}
where $\mathcal{Z}$ is as in Theorem \ref{theorem:clt}, and
\begin{equation}
\mathrm{avar_{k}} = \left( \mu_{ \theta-} + \dfrac{ \Delta \mu_{ \theta}}{2} \right)^{-1} \left( \frac{1}{2} + \frac{1}{2k} - \left[ k \left( \mu_{ \theta-} + \dfrac{ \Delta \mu_{ \theta}}{2} \right) \right]^{-1} \right).
\end{equation}
\end{theorem}
Conversely, $\frac{ \widetilde{ \lambda}_{ \theta}(k \delta_{n})}{ \widetilde{ \lambda}_{ \theta}( \delta_{n})} \overset{p}{ \longrightarrow} k^{- \alpha}$ conditional on $\Omega_{1}^{B}$ such that $\sqrt{n \delta_{n}} \left( \frac{ \widetilde{ \lambda}_{ \theta}(k \delta_{n})}{ \widetilde{ \lambda}_{ \theta}( \delta_{n})} - 1 \right) \rightarrow - \infty$.

\section{Simulation study} \label{section:simulation}

We investigate the small sample properties of our intensity burst test statistic by a Monte Carlo approach. We set $T = 1$ and interpret [0,1] as a standard 6.5 hour trading day on a U.S. stock exchange, which we partition with a discretization step of $\mathrm{d}t = 1 / (23{,}400 \times 100)$, corresponding to one-hundredth of a second.

We simulate the aggregate intensity as a multiplicative process of the form $\mu_{t} = \mu_{s,t} \mu_{d,t}$, where the first term is a stationary process and the second term is a recurrent diurnal component. We assume the stationary part is a continuous-time realization of independent point processes representing the ``normal'' and ``burst'' intensity, $N_{s,t} = N_{s,t}^{ \text{normal}} + N_{s,t}^{ \text{burst}}$, where the latter is only active under the alternative.

To gauge the robustness of our approach, we inspect several choices of $N_{s,t}^{ \text{normal}}$. We start with a standard homogenous Poisson process that has constant intensity $\mu_{s,t} = 1$ (measured on a second-by-second basis).

The second extends the above to an inhomogenous Poisson process with stochastic intensity of the \citet*{cox-ingersoll-ross:85a} form:
\begin{equation} \label{equation:xir}
\mu_{s,t} = \lambda(1-e^{- \kappa t}) + \mu_{s,0}e^{- \kappa t} + \gamma \int_{0}^{t} e^{- \kappa(u-s)} \sqrt{ \mu_{s,u}} \mathrm{d}W_{u}.
\end{equation}
Here, the intensity is time-varying and mean-reverting. If the current value of the process, $\mu_{s,t}$, is above its steady state level, $\lambda$, the intensity has a tendency to go down, and vice versa. $\kappa$ controls the speed of mean reversion. $\gamma$ is the volatility. The scaling with $\sqrt{ \mu_{s,t}}$ inside the stochastic integral prevents the process from being negative. This model has been applied to describe the default intensity in a credit risk framework by \citet*{duffie:05a} and time-varying volatility in \citet*{heston:93a}. We select the parameter vector as $(\lambda, \kappa, \gamma) = (1.00, 0.03, 0.20)$ following \citet*{li-todorov-tauchen:13a}. To ensure the process is stationary with $\mathbb{E}( \mu_{s,t}) = 1$, we draw $\mu_{s,0}$ at random from $\text{Gamma}(2 \lambda \kappa \gamma^{-2},2 \kappa \gamma^{-2})$.

To generate a more challenging data-generating process under the null, as our last choice we assume the intensity of $N_{s,t}^{ \text{normal}}$ is an exponential Hawkes process:\footnote{Although the intensity of the Hawkes process fulfills Assumption \ref{assumption:mu}, it falls outside of the Cox framework in \eqref{equation:cox}. This part of the simulations is therefore also meant to illustrate the robustness of our testing procedure.}
\begin{equation} \label{equation:hawkes}
\mu_{s,t} = \lambda_{0} + \int^{t}_{0} \theta e^{-\kappa(u-s)} \mathrm{d}N_{s,u}^{ \text{normal}},
\end{equation}
where $\lambda_0$, $\theta$ and $\kappa$ are parameters. $\lambda_{0}$ is called the background intensity, which is a lower bound on $\mu_{s,t}$, while $f( \tau) = \theta e^{- \kappa \tau}$ is the excitation function. Our choice of kernel follows the original article by \citet*{hawkes:71a}.

A Hawkes process is self-exciting. That is, after the arrival of an event the intensity of $N_{s,t}^{ \text{normal}}$ inclines by $\theta$. Hence, the probability of an increment in the next time interval $t + \mathrm{d}t$ increases. Tempering by the excitation function pulls the intensity back toward its baseline level $\lambda_{0}$ until further events occur. $\kappa$ controls the rate of decay. If $\theta / \kappa < 1$ the self-excitation is held in check by the mean reversion, and the Hawkes process is non-explosive and stationary with
\begin{equation}
\mathbb{E}( \mu_{t}) = \frac{ \lambda_{0}}{1-\theta / \kappa}.
\end{equation}
The Hawkes process is also capable of generating event clusters. However, none of them are as large as one expects during an intensity burst. In the former, the intensity remains bounded, almost surely, whereas it diverges in the latter. Nevertheless, this may be confused by our test statistic as an intensity burst. The purpose is to see if this distorts our test statistic. As in the diffusion model, the mean reversion is $\kappa = 0.030$. We further select $\lambda_{0} = 0.500$ and $\theta = (1-\lambda_{0}) \kappa = 0.015$, which again ensures that $\mathbb{E}( \mu_{s,t}) = 1$.

In practice, trading intensity (and return volatility) follows a deterministic U- or inverted J-shape at the intraday horizon. There is heightened activity in financial markets near the opening and closing of trading.\footnote{In global markets, where trading is spread out across the world (such as foreign exchange rates), there can be multiple intraday curves aligned with the start and end of trading in the various geographical regions (such as Asia, Europe, and the US).} At a theoretical level, this can be explained in models such as \citet{hong-wang:00a}. To capture this effect, we borrow from the volatility literature \citep*[e.g.,][]{andersen-dobrev-schaumburg:12a, hasbrouck:99a} and assume that
\begin{equation}
\mu_{d,t} = C + Ae^{-a_{1}t} + Be^{-a_{2}(1-t)},
\end{equation}
where $A = 0.75$, $B = 0.25$, $C = 0.89998744$ and $a_{1} = a_{2} = 10$. The choice of $C$ ensures that $\int_{0}^{1} \mu_{d,u} \mathrm{d}u = 1$.

Prior to the calculation of the test statistic, we deflate the increment of the counting process using a nonparametric estimator of the intraday intensity curve with a direct adaption of \citet*{taylor-xu:97a} from the volatility setting. We base the latter on 518 randomly selected Monte Carlo replica out of the 1,000 in total. This corresponds to the sample size in our empirical part and is therefore intended to ensure that the size of the measurement error corresponds to what we face in the application.

The above configurations imply that there is on average one event per second in the normal state, which is slightly higher compared to the daily transaction count observed in our empirical high-frequency data.

The intensity of $N_{s,t}^{ \text{burst}}$ is defined as:
\begin{equation} \label{equation:ib}
\beta_{t} = \frac{ \sigma}{| \tau_{ \mathrm{ib}}-t|^{ \alpha}}, \quad \mathrm{for} \ t \in [ \tau_{ \mathrm{ib}}-0.05, \tau_{ \mathrm{ib}}+0.05],
\end{equation}
and $\beta_{t} = 0$ otherwise. $\sigma$ and $\alpha$ are constant. We position $\tau_{ \mathrm{ib}}$ at random in the interval $[0.05,0.95]$ and make the duration of the intensity burst cover a 10-minute window. To get events of varying magnitude, we take $\alpha \in \{0.25,0.50,0.75 \}$ and calibrate $\sigma$ such that $N_{s,t}^{ \text{burst}}$ generates on average a fraction $c$ of the points produced by $N_{s,t}^{ \text{normal}}$, where $c \in \{0.000, 0.025, 0.050, 0.100 \}$.\footnote{Solving for $\sigma$ in $c \int_{0}^{1} \mu_{t} \mathrm{d}t = \int_{ \tau_{ \text{ib}}- \Delta}^{ \tau_{ \text{ib}}+ \Delta} \beta_{t} \mathrm{d}t$, where $\Delta$ is the duration of the intensity burst and $\beta_{t} = \frac{ \sigma}{| \tau_{ \mathrm{ib}}-t|^{ \alpha}}$, we get $\sigma = c \frac{1- \alpha}{2 \Delta^{1- \alpha}} \int_{0}^{1} \mu_{t} \mathrm{d}t$, such that $c = 0$ corresponds to no intensity burst.} This is called no, small, medium, or large intensity burst.

\begin{figure}[t!]
\begin{center}
\caption{Example of a simulated intensity burst.}
\label{figure:example_ib}
\begin{tabular}{cc}
\small{Panel A: Intensity process.} & \small{Panel B: Intensity estimate}\\
\includegraphics[height=7cm,width=0.48\textwidth]{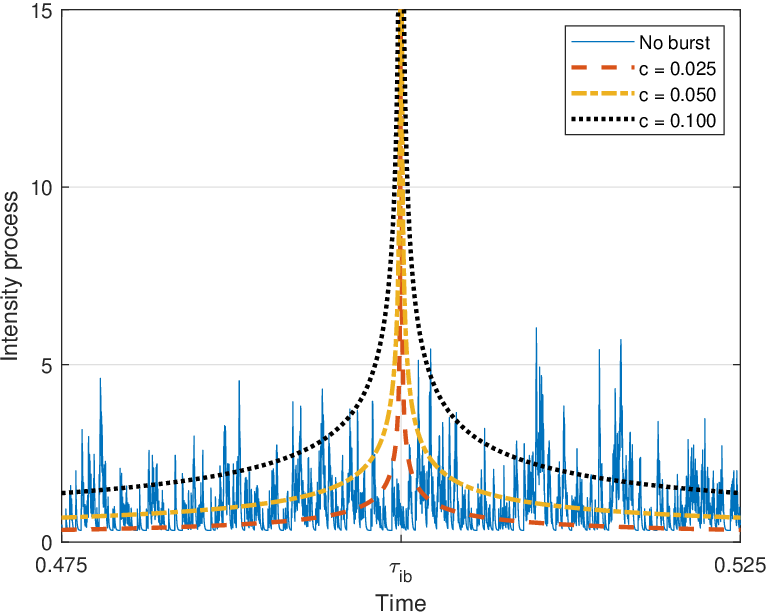} &
\includegraphics[height=7cm,width=0.48\textwidth]{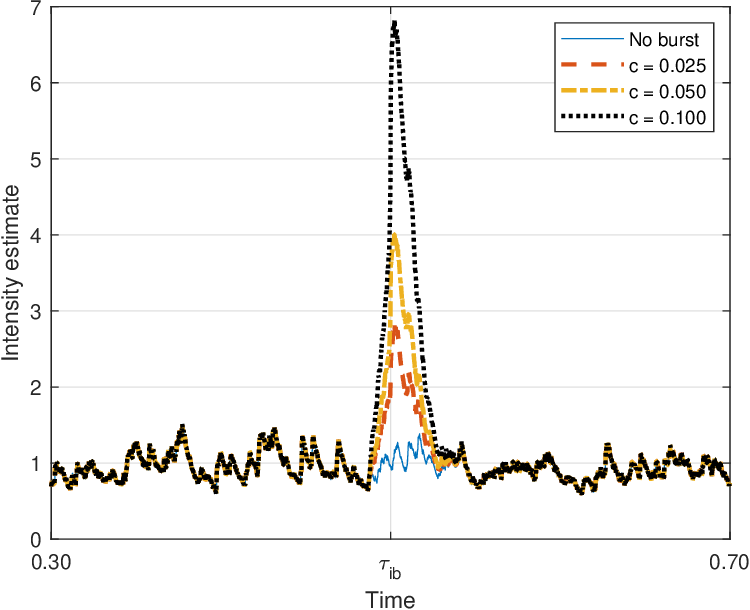} \\
\end{tabular}
\begin{scriptsize}
\parbox{\textwidth}{\emph{Note.} This figure shows the true (in Panel A) and estimated (in Panel B) intensity of the simulated Hawkes process with no burst, and a small, medium and large intensity burst.}
\end{scriptsize}
\end{center}
\end{figure}

We compute $\phi_{t}^{ \text{ib}}$ at $t = \tau_{ \mathrm{ib}}$, which is also the intensity burst time under the alternative. The bandwidth is $\ell_{n} \in \{60,300,600 \}$ seconds for the calculation of $\widehat{ \lambda}_{t}$ (corresponding to a 1, 5, and 10-minute block) and $K_{n} = 10 \ell_{n}$ (corresponding to a 5, 25, and 50-minute block) for $\widehat{ \mathsf{avar}} \big( \nabla \widehat{ \lambda}_{t} \big)$. We experiment both with the indicator kernel $\mathcal{K}(x) = \mathbf{1}_{[-1,0]}(x)$, as implicit in \eqref{equation:lambda-hat-discrete}, and the exponential kernel $\mathcal{K}(x) = \exp(x)$, for $x \leq 0$.

In Panel A of Figure \ref{figure:example_ib}, we illustrate the simulated intensity of the Hawkes process for $N_{s,t}^{ \text{normal}}$ (no burst, or $c = 0.000$) together with a small ($c = 0.025$), medium ($c = 0.050$) and large ($c = 0.100$) intensity burst of $N_{s,t}^{ \text{burst}}$ with $\alpha = 0.50$. In Panel B, we show the associated intensity estimate for $\ell_{n} = 300$. Across burst regimes, the maximal local intensity estimate is roughly $3$, $4$, and $7$ times larger than the average rate of the Hawkes process. In comparison, the intensity burst displayed in Figure \ref{figure:ib} is much larger. Hence, our simulated intensity bursts are conservative relative to many of those observed in empirical data.

Tables \ref{table:sim-ib-poisson.tex} -- \ref{table:sim-ib-hawkes.tex} collects the results of the Monte Carlo analysis by reporting rejection rates for the test statistic across 1,000 independent trials. The left-hand side is for the indicator kernel, while the right-hand side is for the exponential kernel. Throughout, the test statistic is evaluated against critical values from the standard normal distribution, i.e. the $p$th percentile $z_{p} = \Phi^{-1}(p)$ for $p \in \{ 0.950, 0.975, 0.990 \}$.

Looking at Panel A in Tables \ref{table:sim-ib-poisson.tex} -- \ref{table:sim-ib-hawkes.tex}, we observe that in the absence of an intensity burst the rejection rates are close to the nominal level across bandwidth choices. Stochastic intensity creates a minuscule size distortion, in particular for the shortest bandwidth and exponential kernel. However, the effect vanishes once we calculate the intensity estimator over a slightly longer window. What is also interesting to note is that the self-excitation of the Hawkes intensity is not a separate source of additional size distortion.

Turning to the alternative in Panels B -- D, the rejection rates of the test are nearly perfect for a large intensity burst, irrespective of the dynamic of $N_{t}^{ \text{normal}}$. This is as expected, since the test statistic is calculated at the center of the burst process. In contrast, detection of small bursts remains challenging if they account for an insignificant fraction of the total transaction count---as measured by a smaller value of $c$---or are slow to materialize---as measured by a smaller value of $\alpha$. Moreover, there is an intuitive trade-off embedded in the selection of the bandwidth parameter. We observe that a longer bandwidth is preferable if the explosion rate of the intensity burst is small, and vice versa.

Overall, the test statistic catches the vast majority of medium and large intensity bursts, whereas smaller ones with slow rates of divergence may go unnoticed. This is not a cause of too much concern, since we are primarily interested in the behavior of large and rapid surges in trading activity.

\begin{sidewaystable}[ht!]
\setlength{ \tabcolsep}{0.115cm}
\begin{center}
\caption{Rejection rate for the Poisson process.
\label{table:sim-ib-poisson.tex}}
\begin{tabular}{lrrrrrrrrrrrrrrrrrrrrrrr}
\hline \hline
& \multicolumn{11}{c}{Indicator kernel} && \multicolumn{11}{c}{Exponential kernel}\\
\cline{2-12} \cline{14-24}
& \multicolumn{3}{c}{$\ell_{n} = 60$} & & \multicolumn{3}{c}{$\ell_{n} = 300$} & & \multicolumn{3}{c}{$\ell_{n} = 600$} & & \multicolumn{3}{c}{$\ell_{n} = 60$} & & \multicolumn{3}{c}{$\ell_{n} = 300$} & & \multicolumn{3}{c}{$\ell_{n} = 600$}\\
& $z_{0.950}$& $z_{0.990}$& $z_{0.995}$ & & $z_{0.950}$& $z_{0.990}$& $z_{0.995}$ & & $z_{0.950}$& $z_{0.990}$& $z_{0.995}$ & & $z_{0.950}$& $z_{0.990}$& $z_{0.995}$ & & $z_{0.950}$& $z_{0.990}$& $z_{0.995}$ & & $z_{0.950}$& $z_{0.990}$& $z_{0.995}$\\
\cline{2-4} \cline{6-8} \cline{10-12} \cline{14-16} \cline{18-20} \cline{22-24}
\multicolumn{18}{l}{ \textit{Panel A: size ($c = 0.000$, no burst)}}\\
& 6.0 & 1.2 & 0.6  & & 4.8 & 1.1 & 0.7  & & 5.1 & 0.9 & 0.4  & & 5.5 & 1.5 & 0.9  & & 5.1 & 0.8 & 0.8  & & 5.8 & 1.1 & 0.4 \\
\\
\multicolumn{18}{l}{ \textit{Panel B: power ($c = 0.025$, small burst)}}\\
$\alpha = 0.25$& 59.2 & 32.0 & 24.3  & & 100.0 & 100.0 & 100.0  & & 100.0 & 100.0 & 100.0  & & 83.2 & 58.6 & 48.0  & & 100.0 & 100.0 & 100.0  & & 100.0 & 100.0 & 100.0 \\
$\alpha = 0.50$& 100.0 & 100.0 & 99.9  & & 100.0 & 100.0 & 100.0  & & 100.0 & 100.0 & 100.0  & & 100.0 & 100.0 & 100.0  & & 100.0 & 100.0 & 100.0  & & 100.0 & 100.0 & 100.0 \\
$\alpha = 0.75$& 100.0 & 100.0 & 100.0  & & 100.0 & 100.0 & 100.0  & & 100.0 & 100.0 & 100.0  & & 100.0 & 100.0 & 100.0  & & 100.0 & 100.0 & 100.0  & & 100.0 & 100.0 & 100.0 \\
\\
\multicolumn{18}{l}{ \textit{Panel C: power ($c = 0.050$, medium burst)}}\\
$\alpha = 0.25$& 83.1 & 54.9 & 41.9  & & 100.0 & 100.0 & 100.0  & & 100.0 & 100.0 & 100.0  & & 97.9 & 83.9 & 75.3  & & 100.0 & 100.0 & 100.0  & & 100.0 & 100.0 & 100.0 \\
$\alpha = 0.50$& 100.0 & 100.0 & 100.0  & & 100.0 & 100.0 & 100.0  & & 100.0 & 100.0 & 100.0  & & 100.0 & 100.0 & 100.0  & & 100.0 & 100.0 & 100.0  & & 100.0 & 100.0 & 100.0 \\
$\alpha = 0.75$& 100.0 & 100.0 & 100.0  & & 100.0 & 100.0 & 100.0  & & 100.0 & 100.0 & 100.0  & & 100.0 & 100.0 & 100.0  & & 100.0 & 100.0 & 100.0  & & 100.0 & 100.0 & 100.0 \\
\\
\multicolumn{18}{l}{ \textit{Panel D: power ($c = 0.100$, large burst)}}\\
$\alpha = 0.25$& 96.1 & 72.6 & 57.9  & & 100.0 & 100.0 & 100.0  & & 100.0 & 100.0 & 100.0  & & 100.0 & 97.1 & 91.7  & & 100.0 & 100.0 & 100.0  & & 100.0 & 100.0 & 100.0 \\
$\alpha = 0.50$& 100.0 & 100.0 & 100.0  & & 100.0 & 100.0 & 100.0  & & 100.0 & 100.0 & 100.0  & & 100.0 & 100.0 & 100.0  & & 100.0 & 100.0 & 100.0  & & 100.0 & 100.0 & 100.0 \\
$\alpha = 0.75$& 100.0 & 100.0 & 100.0  & & 100.0 & 100.0 & 100.0  & & 100.0 & 100.0 & 100.0  & & 100.0 & 100.0 & 100.0  & & 100.0 & 100.0 & 100.0  & & 100.0 & 100.0 & 100.0 \\
\hline \hline
\end{tabular}
\smallskip
\begin{scriptsize}
\parbox{\textwidth}{\emph{Note.}
We simulate a point process as the sum of a Poisson process (``normal'') and an intensity burst (``burst''), $N_{t} = N_{t}^{ \text{normal}} + N_{t}^{ \text{burst}}$.
$\alpha$ measures the explosion rate of the latter.
We test the hypothesis of no intensity burst.
$c$ is the proportion of the total transaction count induced by burst process, so $c = 0.000$ corresponds to $\mathcal{H}_{0}$.
The test is calculated at $t = \tau_{ \text{ib}}$, which is the intensity burst time under $\mathcal{H}_{a}$.
$\ell_{n}$ is the bandwidth (in seconds) for the intensity estimator.
The bandwidth for the observed local asymptotic variance is $K_{n} = 10 \ell_{n}$.
The table reports rejection rates of the intensity burst test statistic evaluated against the $(1- \varsigma)$th percentile of the standard normal distribution, $z_{(1- \varsigma)}$, for $\varsigma = (0.050,0.010,0.005)$.
}
\end{scriptsize}
\end{center}
\end{sidewaystable}

\begin{sidewaystable}[ht!]
\setlength{ \tabcolsep}{0.115cm}
\begin{center}
\caption{Rejection rate for the Heston process.
\label{table:sim-ib-heston.tex}}
\begin{tabular}{lrrrrrrrrrrrrrrrrrrrrrrr}
\hline \hline
& \multicolumn{11}{c}{Indicator kernel} && \multicolumn{11}{c}{Exponential kernel}\\
\cline{2-12} \cline{14-24}
& \multicolumn{3}{c}{$\ell_{n} = 60$} & & \multicolumn{3}{c}{$\ell_{n} = 300$} & & \multicolumn{3}{c}{$\ell_{n} = 600$} & & \multicolumn{3}{c}{$\ell_{n} = 60$} & & \multicolumn{3}{c}{$\ell_{n} = 300$} & & \multicolumn{3}{c}{$\ell_{n} = 600$}\\
& $z_{0.950}$& $z_{0.990}$& $z_{0.995}$ & & $z_{0.950}$& $z_{0.990}$& $z_{0.995}$ & & $z_{0.950}$& $z_{0.990}$& $z_{0.995}$ & & $z_{0.950}$& $z_{0.990}$& $z_{0.995}$ & & $z_{0.950}$& $z_{0.990}$& $z_{0.995}$ & & $z_{0.950}$& $z_{0.990}$& $z_{0.995}$\\
\cline{2-4} \cline{6-8} \cline{10-12} \cline{14-16} \cline{18-20} \cline{22-24}
\multicolumn{18}{l}{ \textit{Panel A: size ($c = 0.000$, no burst)}}\\
& 7.2 & 3.4 & 2.4  & & 5.7 & 2.6 & 1.1  & & 4.9 & 1.5 & 1.0  & & 8.2 & 3.8 & 3.0  & & 7.1 & 2.5 & 1.7  & & 6.0 & 1.9 & 1.2 \\
\\
\multicolumn{18}{l}{ \textit{Panel B: power ($c = 0.025$, small burst)}}\\
$\alpha = 0.25$& 18.1 & 8.5 & 6.4  & & 61.7 & 39.0 & 30.2  & & 37.9 & 15.0 & 11.1  & & 25.4 & 13.6 & 10.6  & & 73.3 & 47.5 & 40.3  & & 60.6 & 34.1 & 27.2 \\
$\alpha = 0.50$& 69.0 & 43.9 & 36.7  & & 66.4 & 43.8 & 34.9  & & 40.2 & 17.8 & 12.2  & & 89.1 & 75.3 & 68.6  & & 86.0 & 62.9 & 55.0  & & 67.9 & 42.3 & 33.8 \\
$\alpha = 0.75$& 99.3 & 98.1 & 97.0  & & 78.7 & 55.7 & 48.4  & & 51.0 & 26.3 & 18.9  & & 100.0 & 100.0 & 100.0  & & 97.4 & 86.9 & 83.7  & & 83.3 & 64.2 & 55.3 \\
\\
\multicolumn{18}{l}{ \textit{Panel C: power ($c = 0.050$, medium burst)}}\\
$\alpha = 0.25$& 36.4 & 16.5 & 11.5  & & 97.2 & 88.7 & 83.5  & & 84.0 & 62.5 & 52.3  & & 55.2 & 30.0 & 23.5  & & 99.6 & 96.8 & 93.0  & & 97.5 & 90.7 & 84.8 \\
$\alpha = 0.50$& 97.5 & 90.8 & 87.2  & & 98.3 & 92.7 & 88.2  & & 85.8 & 67.2 & 58.7  & & 99.8 & 99.4 & 99.1  & & 99.9 & 99.6 & 98.5  & & 98.9 & 94.9 & 92.1 \\
$\alpha = 0.75$& 100.0 & 100.0 & 100.0  & & 99.3 & 98.0 & 96.0  & & 93.6 & 81.3 & 75.0  & & 100.0 & 100.0 & 100.0  & & 100.0 & 100.0 & 100.0  & & 100.0 & 99.6 & 99.2 \\
\\
\multicolumn{18}{l}{ \textit{Panel D: power ($c = 0.100$, large burst)}}\\
$\alpha = 0.25$& 69.0 & 35.1 & 25.8  & & 100.0 & 100.0 & 100.0  & & 99.7 & 99.0 & 98.1  & & 89.0 & 65.3 & 53.2  & & 100.0 & 100.0 & 100.0  & & 100.0 & 100.0 & 100.0 \\
$\alpha = 0.50$& 100.0 & 99.9 & 99.8  & & 100.0 & 100.0 & 100.0  & & 99.7 & 99.3 & 98.5  & & 100.0 & 100.0 & 100.0  & & 100.0 & 100.0 & 100.0  & & 100.0 & 100.0 & 100.0 \\
$\alpha = 0.75$& 100.0 & 100.0 & 100.0  & & 100.0 & 100.0 & 100.0  & & 100.0 & 99.7 & 99.6  & & 100.0 & 100.0 & 100.0  & & 100.0 & 100.0 & 100.0  & & 100.0 & 100.0 & 100.0 \\
\hline \hline
\end{tabular}
\smallskip
\begin{scriptsize}
\parbox{\textwidth}{\emph{Note.}
We simulate a point process as the sum of a Heston process (``normal'') and an intensity burst (``burst''), $N_{t} = N_{t}^{ \text{normal}} + N_{t}^{ \text{burst}}$.
$\alpha$ measures the explosion rate of the latter.
We test the hypothesis of no intensity burst.
$c$ is the proportion of the total transaction count induced by burst process, so $c = 0.000$ corresponds to $\mathcal{H}_{0}$.
The test is calculated at $t = \tau_{ \text{ib}}$, which is the intensity burst time under $\mathcal{H}_{a}$.
$\ell_{n}$ is the bandwidth (in seconds) for the intensity estimator.
The bandwidth for the observed local asymptotic variance is $K_{n} = 10 \ell_{n}$.
The table reports rejection rates of the intensity burst test statistic evaluated against the $(1- \varsigma)$th percentile of the standard normal distribution, $z_{(1- \varsigma)}$, for $\varsigma = (0.050,0.010,0.005)$.
}
\end{scriptsize}
\end{center}
\end{sidewaystable}

\begin{sidewaystable}[ht!]
\setlength{ \tabcolsep}{0.115cm}
\begin{center}
\caption{Rejection rate for the Hawkes process.
\label{table:sim-ib-hawkes.tex}}
\begin{tabular}{lrrrrrrrrrrrrrrrrrrrrrrr}
\hline \hline
& \multicolumn{11}{c}{Indicator kernel} && \multicolumn{11}{c}{Exponential kernel}\\
\cline{2-12} \cline{14-24}
& \multicolumn{3}{c}{$\ell_{n} = 60$} & & \multicolumn{3}{c}{$\ell_{n} = 300$} & & \multicolumn{3}{c}{$\ell_{n} = 600$} & & \multicolumn{3}{c}{$\ell_{n} = 60$} & & \multicolumn{3}{c}{$\ell_{n} = 300$} & & \multicolumn{3}{c}{$\ell_{n} = 600$}\\
& $z_{0.950}$& $z_{0.990}$& $z_{0.995}$ & & $z_{0.950}$& $z_{0.990}$& $z_{0.995}$ & & $z_{0.950}$& $z_{0.990}$& $z_{0.995}$ & & $z_{0.950}$& $z_{0.990}$& $z_{0.995}$ & & $z_{0.950}$& $z_{0.990}$& $z_{0.995}$ & & $z_{0.950}$& $z_{0.990}$& $z_{0.995}$\\
\cline{2-4} \cline{6-8} \cline{10-12} \cline{14-16} \cline{18-20} \cline{22-24}
\multicolumn{18}{l}{ \textit{Panel A: size ($c = 0.000$, no burst)}}\\
& 4.6 & 1.2 & 1.0  & & 5.2 & 1.6 & 1.0  & & 5.6 & 1.1 & 0.8  & & 5.0 & 1.5 & 0.9  & & 5.5 & 1.6 & 0.7  & & 5.5 & 1.2 & 0.8 \\
\\
\multicolumn{18}{l}{ \textit{Panel B: power ($c = 0.025$, small burst)}}\\
$\alpha = 0.25$& 49.1 & 25.2 & 18.2  & & 100.0 & 100.0 & 99.7  & & 99.3 & 94.5 & 93.2  & & 74.5 & 47.7 & 38.0  & & 100.0 & 100.0 & 100.0  & & 100.0 & 99.7 & 99.6 \\
$\alpha = 0.50$& 99.8 & 99.4 & 98.3  & & 100.0 & 100.0 & 100.0  & & 99.5 & 96.9 & 94.0  & & 100.0 & 100.0 & 100.0  & & 100.0 & 100.0 & 100.0  & & 100.0 & 99.9 & 99.7 \\
$\alpha = 0.75$& 100.0 & 100.0 & 100.0  & & 100.0 & 100.0 & 100.0  & & 99.7 & 99.2 & 98.1  & & 100.0 & 100.0 & 100.0  & & 100.0 & 100.0 & 100.0  & & 100.0 & 100.0 & 100.0 \\
\\
\multicolumn{18}{l}{ \textit{Panel C: power ($c = 0.050$, medium burst)}}\\
$\alpha = 0.25$& 77.1 & 47.7 & 36.8  & & 100.0 & 100.0 & 100.0  & & 100.0 & 100.0 & 100.0  & & 95.5 & 78.1 & 65.3  & & 100.0 & 100.0 & 100.0  & & 100.0 & 100.0 & 100.0 \\
$\alpha = 0.50$& 100.0 & 100.0 & 100.0  & & 100.0 & 100.0 & 100.0  & & 100.0 & 100.0 & 100.0  & & 100.0 & 100.0 & 100.0  & & 100.0 & 100.0 & 100.0  & & 100.0 & 100.0 & 100.0 \\
$\alpha = 0.75$& 100.0 & 100.0 & 100.0  & & 100.0 & 100.0 & 100.0  & & 100.0 & 100.0 & 100.0  & & 100.0 & 100.0 & 100.0  & & 100.0 & 100.0 & 100.0  & & 100.0 & 100.0 & 100.0 \\
\\
\multicolumn{18}{l}{ \textit{Panel D: power ($c = 0.100$, large burst)}}\\
$\alpha = 0.25$& 94.7 & 67.3 & 54.4  & & 100.0 & 100.0 & 100.0  & & 100.0 & 100.0 & 100.0  & & 100.0 & 95.9 & 89.5  & & 100.0 & 100.0 & 100.0  & & 100.0 & 100.0 & 100.0 \\
$\alpha = 0.50$& 100.0 & 100.0 & 100.0  & & 100.0 & 100.0 & 100.0  & & 100.0 & 100.0 & 100.0  & & 100.0 & 100.0 & 100.0  & & 100.0 & 100.0 & 100.0  & & 100.0 & 100.0 & 100.0 \\
$\alpha = 0.75$& 100.0 & 100.0 & 100.0  & & 100.0 & 100.0 & 100.0  & & 100.0 & 100.0 & 100.0  & & 100.0 & 100.0 & 100.0  & & 100.0 & 100.0 & 100.0  & & 100.0 & 100.0 & 100.0 \\
\hline \hline
\end{tabular}
\smallskip
\begin{scriptsize}
\parbox{\textwidth}{\emph{Note.}
We simulate a point process as the sum of a Hawkes process (``normal'') and an intensity burst (``burst''), $N_{t} = N_{t}^{ \text{normal}} + N_{t}^{ \text{burst}}$.
$\alpha$ measures the explosion rate of the latter.
We test the hypothesis of no intensity burst.
$c$ is the proportion of the total transaction count induced by burst process, so $c = 0.000$ corresponds to $\mathcal{H}_{0}$.
The test is calculated at $t = \tau_{ \text{ib}}$, which is the intensity burst time under $\mathcal{H}_{a}$.
$\ell_{n}$ is the bandwidth (in seconds) for the intensity estimator.
The bandwidth for the observed local asymptotic variance is $K_{n} = 10 \ell_{n}$.
The table reports rejection rates of the intensity burst test statistic evaluated against the $(1- \varsigma)$th percentile of the standard normal distribution, $z_{(1- \varsigma)}$, for $\varsigma = (0.050,0.010,0.005)$.
}
\end{scriptsize}
\end{center}
\end{sidewaystable}

\clearpage

\section{Empirical application} \label{section:empirical}

\begin{figure}[t!]
\caption{Daily price and log-return in EUR/USD.} \label{figure:ebs-eurusd}
\begin{center}
\begin{tabular}{cc}
\small{Panel A: Exchange rate.} & \small{Panel B: Daily log-return.} \\
\includegraphics[height=7cm,width=0.48\textwidth]{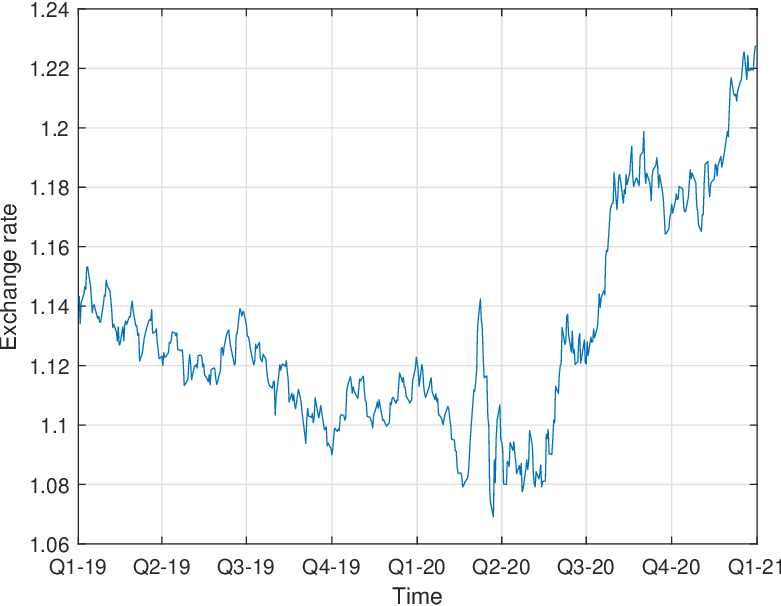} &
\includegraphics[height=7cm,width=0.48\textwidth]{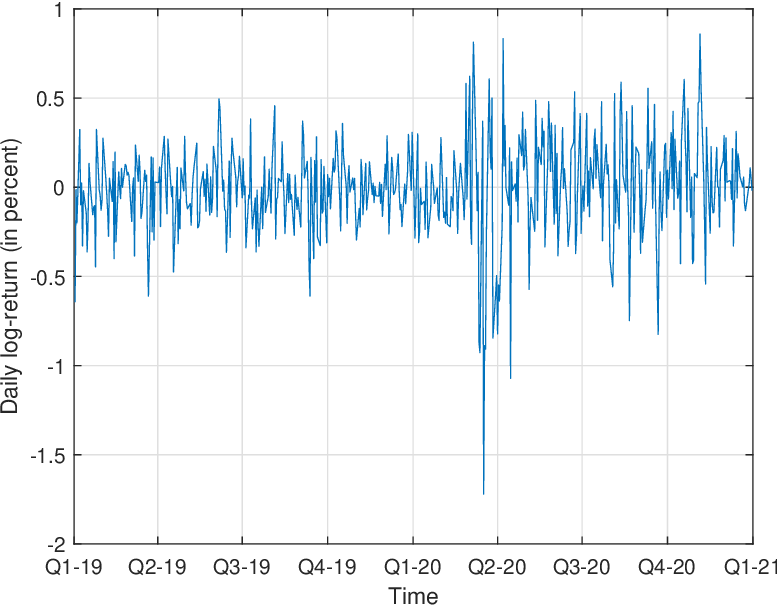} \\
\end{tabular}
\begin{scriptsize}
\parbox{\textwidth}{\emph{Note.} In Panel A, we plot the EUR/USD spot exchange rate at noon CET for the sample period January 1, 2019 to December 31, 2020. In Panel B, we show the associated daily log-return in percent.}
\end{scriptsize}
\end{center}
\end{figure}

We examine high-frequency data obtained via Electronic Broking Services (EBS) on the Chicago Mercantile Exchange (CME). EBS is an electronic platform operating at the wholesale level and is the main interdealer venue in the foreign exchange market. Our dataset consists of all spot transactions and every tick-by-tick update to the first ten levels in the limit order book of the EUR/USD. The sample period January 1, 2019 to December 31, 2020. We exclude days with limited trading (e.g. weekends and public holidays) and retain $T = 518$ days. We also restrict attention to the most active trading hours from 8:00am to 4:00pm Central European Time (CET), covering the European and American session. In Panel A of Figure \ref{figure:ebs-eurusd}, we plot the daily midquote of the spot exchange rate at noon, while Panel B shows the associated log-return.

\begin{figure}[t!]
\caption{Trading activity in EUR/USD.} \label{figure:ebs-eurusd-trading-activity}
\begin{center}
\begin{tabular}{cc}
\small{Panel A: Interday.} & \small{Panel B: Intraday.} \\
\includegraphics[height=7cm,width=0.48\textwidth]{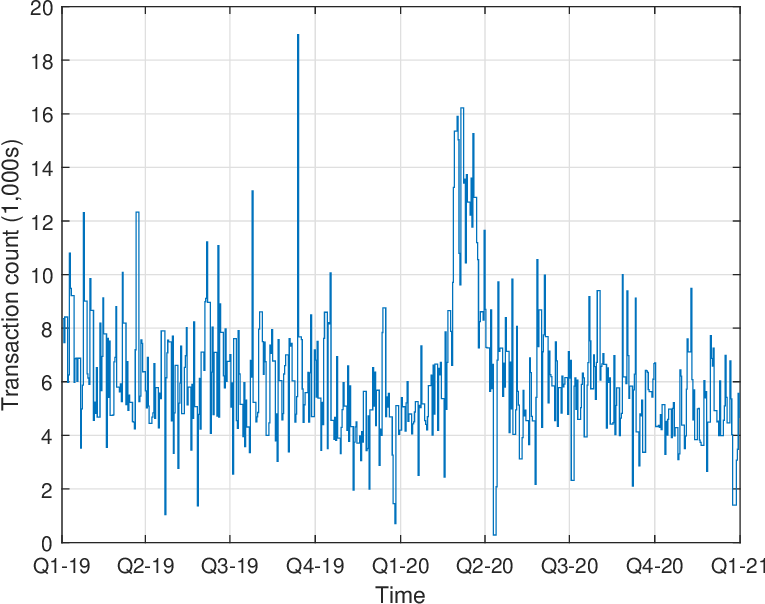} &
\includegraphics[height=7cm,width=0.48\textwidth]{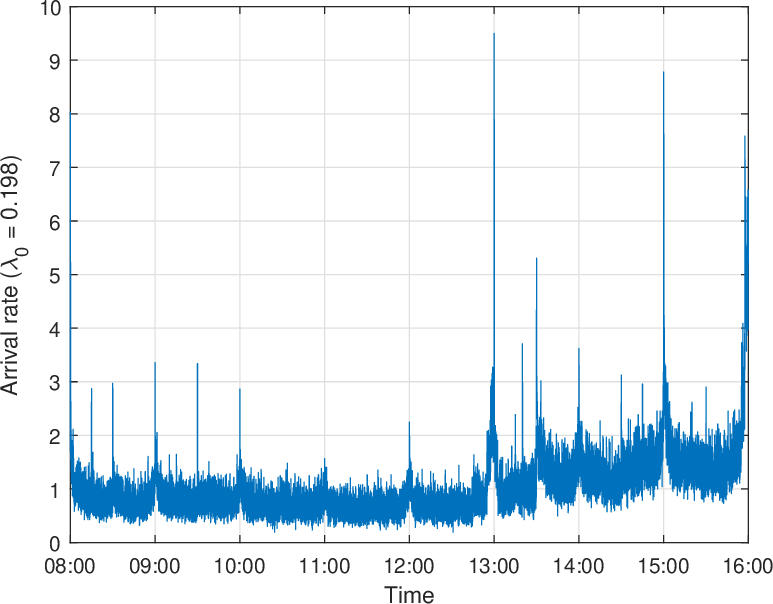} \\
\end{tabular}
\begin{scriptsize}
\parbox{\textwidth}{\emph{Note.} In Panel A, we report the daily number of transactions (in 1,000s) on the EBS platform in the EUR/USD spot exchange rate. In Panel B, we show a nonparametric estimate of the time-of-the-day mean transaction intensity between 8:00am and 4:00pm CET, which is normalized to sum to one. $\lambda_{0}$ is an estimate of the unconditional arrival rate over the whole sample.}
\end{scriptsize}
\end{center}
\end{figure}

In Panel A of Figure \ref{figure:ebs-eurusd-trading-activity}, we show the total number of transactions per day in the retained sample. There are typically thousands of trades. To account for the expected increases in trading activity that are associated with the normal periodicity of this market or regular releases of pre-scheduled news announcements, we start by extracting a nonparametric estimate of the time-of-the-day mean transaction count. We calculate the latter as the average number of transactions in each one-second bucket, where the average is taken across days in the sample. The sequence of these estimates is then normalized to sum to one. The seasonality in the intraday transaction count is reported in Panel B of Figure \ref{figure:ebs-eurusd-trading-activity}.

\begin{figure}[t!]
\caption{Candidate times for an intensity burst.} \label{figure:ebs-ib-time}
\begin{center}
\begin{tabular}{cc}
\small{Panel A: July 2, 2019.} & \small{Panel B: Whole sample.} \\
\includegraphics[height=7cm,width=0.48\textwidth]{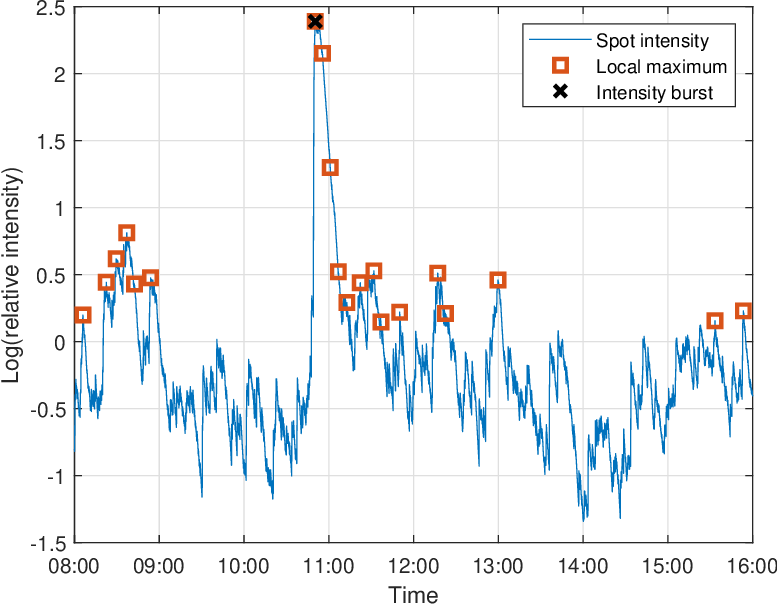} &
\includegraphics[height=7cm,width=0.48\textwidth]{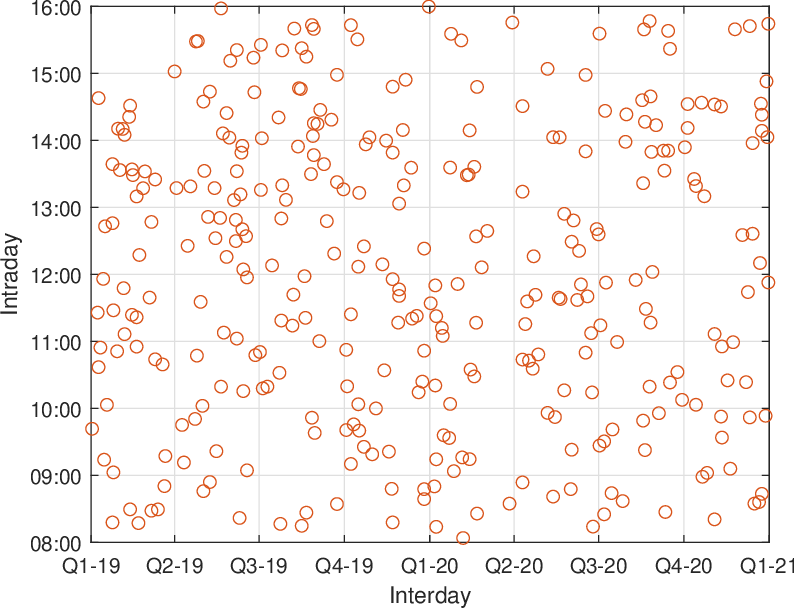} \\
\end{tabular}
\begin{scriptsize}
\parbox{\textwidth}{\emph{Note.} In Panel A of this figure, we show the candidate times for an intensity burst on July 2, 2019 highlighted with a square. These are defined as the 20 largest local maxima of the estimated spot intensity within a 5-minute window (identical to the bandwidth of the estimator). The crossed square is identified as an intensity burst. In Panel B, we show how the intensity burst test statistics with a value greater than 5 are scattered at the intraday level over the whole sample.}
\end{scriptsize}
\end{center}
\end{figure}

We compute the spot intensity estimator in \eqref{equation:lambdba-hat} each second from 8:00am to 4:00pm CET with a bandwidth of $\ell_{n} = 300$ and the exponential kernel. Each estimate is then divided by the appropriate seasonality factor to remove the inherent diurnal variation. Next, to avoid the computational load of running the test statistic at such a refined grid, every day we select the 20 largest local maxima of the diurnal-corrected spot intensity estimator, allowing for at most one maximum in a 5-minute neighbourhood---equal to the bandwidth of the intensity estimator---to avoid double-counting. In Panel A of Figure \ref{figure:ebs-ib-time}, we show how this rule operates on July 2, 2019. The spot intensity is shown from 8:00am to 4:00pm. It is scaled by the average intensity on that day and log-transformed, as further motivated below.

In the subsequent analysis, we limit our attention to the points indicated with a square, which are the most probable intensity burst times.\footnote{By Lemma \ref{lemma:consistency}, the unobserved intensity diverges at an intensity burst time. Thus, if an intensity burst occurs near a given point, the local intensity estimator should be at its highest here.} We implement the test here with $K_{n} = 10 \ell_{n}$ for the observed asymptotic local variance estimator, as in the simulation section. In doing so, only the crossed square is identified as an intensity burst. Overall, we monitor a total of 10,360 test statistics across the 518 days. The median value of the sequence is 1.24, which is more than a standard deviation away from the unconditional mean of zero of the asymptotic standard normal distribution. There are 913, 331, 119, and 36 observations with a value higher than 4, 5, 6, and 7, amounting to 8.81\%, 3.20\%, 1.15\%, and 0.35\%. Hence, intensity bursts---even more extreme ones---happen on a frequent basis.

In the remainder of the empirical investigation, we confine the analysis to a subsample of intensity bursts, where the test statistic is greater than 5. This value is large enough to be highly significant and small enough to retain a sufficient number of observations.\footnote{We face a theoretical concern in looking at the test statistic at many local maxima of the intensity estimator, since this can cause over-rejection under the null hypothesis of no intensity burst. We therefore do not evaluate it against a customary quantile from the standard normal distribution, but we select a much higher value following the multiple comparison literature to control the family-wise error rate at the $\varsigma = 0.01$ level of significance. On the one hand, we conduct 10,380 tests in total, so a Bonferroni correction recommends a critical value of $\Phi^{-1}(1- 0.01/10{,}380) = 4.7610$. On the other hand, according to extreme value theory, the normalized maximum test statistic converges in law to a Gumbel distribution, i.e. $\frac{\max_{1 \leq i \leq m} \phi_{t_{i}}^{ \text{ib}} - a_{m}}{b_{m}} \overset{ \mathfrak{D}}{ \longrightarrow} \xi$ as $m \rightarrow \infty$, where $F_{ \xi}(x) = \mathbb{P}( \xi \leq x) = \exp(- \exp(-x))$, $a_{m} = \sqrt{2 \log(m)} - \frac{ \log( \pi)+ \log( \log(m))}{2 \sqrt{2 \log(m)}}$, and $b_{m} = (2 \log(m))^{-1/2}$. Hence, if $\phi_{t_{i}}^{ \text{ib}}$ is computed at a one-second grid, as consistent with our implementation, and the daily maximum test statistic is extracted from this sequence, the critical value for the nonnormalized maximum test statistic is $5.1846$. We base our inference on the average of these approaches by adopting a critical value of $5$.} The reduced sample of 331 events are highlighted in Panel B of Figure \ref{figure:ebs-ib-time}. There is no systematic placement in the points, which are more or less uniformly distributed at the intraday level.\footnote{We calculated the fraction of the total daily transaction count that occurs over a 10-minute interval around each such event. The sample average is 3.60\% (median of 2.91\%) with a 1st and 3rd quartile of 1.93\% and 8.95\%. This crudely aligns with the our choices of $c$ made in the simulation section.}

\subsection{A burst or jump?} \label{section:burstorjump}

We first explore whether the above events correspond to genuine intensity bursts or are, in fact, more likely to be associated with false alarms caused by discrete changes in the intensity process, as discussed in Section \ref{section:disentangle}.

\begin{figure}[t!]
\caption{Histogram of ratio statistic and explosion rate.}
\label{figure:burst-or-jump}
\begin{center}
\begin{tabular}{cc}
\small{Panel A: Ratio statistic.} & \small{Panel B: Explosion rate.} \\
\includegraphics[height=7cm,width=0.48\textwidth]{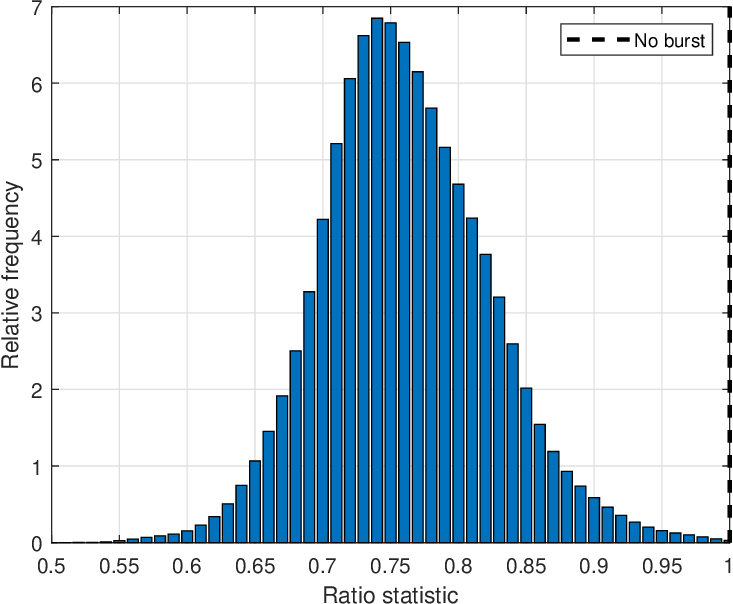} &
\includegraphics[height=7cm,width=0.48\textwidth]{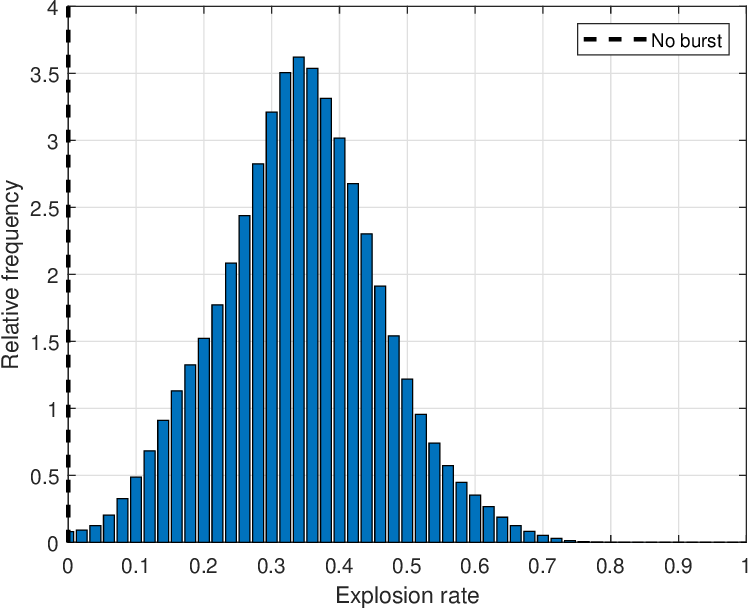} \\
\end{tabular}
\begin{scriptsize}
\parbox{\textwidth}{\emph{Note.} In Panel A, we plot a histogram of the ratio statistic from Lemma \ref{lemma:cof}, $\widetilde{ \lambda}_{ \theta}(k \delta_{n}) / \widetilde{ \lambda}_{ \theta}( \delta_{n})$, with $k = 2$. It is centered around one in the absence of a burst, and strictly smaller than one otherwise. In Panel B, we report the distribution of the estimated explosion rate, $\hat{ \alpha}$, based on the negative slope coefficient of an OLS regression through the origin of $\log \big( \widetilde{ \lambda}_{ \theta}(k \delta_{n}) \big)$ on $\log(k)$ for $k = 1, 2, 3, 5$, and $10$.}
\end{scriptsize}
\end{center}
\end{figure}

We calculate the two-sided intensity estimator from \eqref{equation:centered-estimator} with $k = 2$ and construct the ratio statistic from Lemma \ref{lemma:cof}, where the baseline in the denominator corresponds to an implicit $k = 1$. It should be centered around one if the event coincided with a jump in intensity, whereas it should be strictly smaller than one---or $k^{- \alpha}$---for an intensity burst. In Panel A of Figure \ref{figure:burst-or-jump}, we plot a histogram of the quotient. As evident, the empirical density function is concentrated around 0.75 with little probability mass near one. In total, only three of the ratio statistics are greater than or equal to one, where it is 1.05 on average. This strongly indicates that our test statistic is not driven by a jump component.

Conditional on an intensity burst, we can convert the ratio statistic to an estimate of the explosion rate, $\alpha$, as described in \eqref{equation:alpha-estimator} and the subsequent comment. We regress $\log \big( \widetilde{ \lambda}_{ \theta}(k \delta_{n}) \big)$ through the origin against $\log(k)$ for $k = 1, 2, 3, 5$, and $10$. The distribution of the negative slope estimates are reported in Panel B of Figure \ref{figure:burst-or-jump}. As consistent with the previous analysis, they are strictly above the no burst lower bound of zero. The overall impression is that an intensity burst is often modest with an explosion rate in the middle of the permissible range.

\subsection{State of the market near an intensity burst} \label{section:market}

\begin{figure}[t!]
\begin{center}
\caption{Trading activity and return volatility near intensity burst.}
\label{figure:before-and-after}
\begin{tabular}{cc}
\small{Panel A: Dynamic of spot intensity.} & \small{Panel B: Association with spot volatility.}\\
\includegraphics[height=7cm,width=0.48\textwidth]{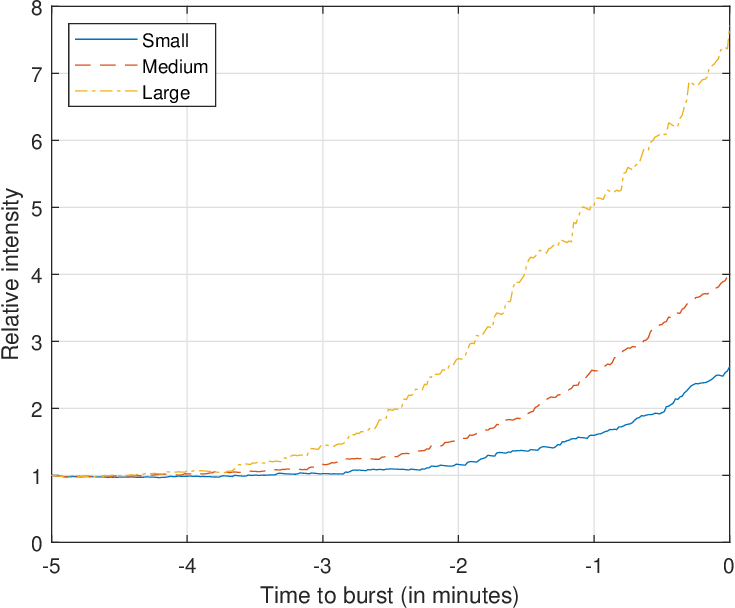} &
\includegraphics[height=7cm,width=0.48\textwidth]{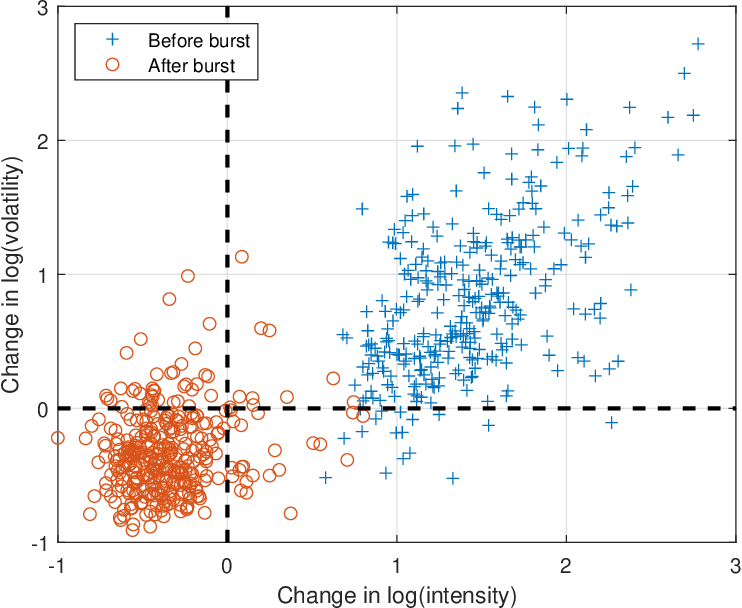} \\
\end{tabular}
\begin{scriptsize}
\parbox{\textwidth}{\emph{Note.} In Panel A, we plot the evolution of the relative intensity estimator each second in a 5-minute window prior an intensity burst time. The sample has been split in four based on the magnitude of this measure and we then report the median intensity in the first (small), second--third (medium), and fourth (large) quartile. In Panel B, we plot the change in log-spot intensity against the change in log-spot volatility of the EUR/USD exchange rate process before and after the burst.}
\end{scriptsize}
\end{center}
\end{figure}

We next take a closer look at how various state variables evolve in the vicinity of an intensity burst. In Panel A of Figure \ref{figure:before-and-after}, we show the evolution of trading activity. We compute a proxy for the burst size as the relative change in the intensity estimator at each second in a 5-minute interval prior to the intensity burst time. We split the sample in four based on the magnitude of this measure and report the median intensity in the first (small), second--third (medium), and fourth (large) quartile. As expected, there is a steady rise in trading activity prior to a detected event. The curvature indicates a good fit with an intensity burst of the form \eqref{equation:ib-theoretical}, while the level is consistent with the simulation of this model illuminated in Figure \ref{figure:example_ib}.

In Panel B of the figure, we plot the change in log-spot intensity against the change in log-spot volatility of the EUR/USD exchange rate process. The latter is calculated with the price jump- and microstructure noise-robust estimator from \citet*{christensen-oomen-reno:22a}. To construct the increment, we measure the spot processes at the 5-minute mark before and after the intensity burst time, together with an estimate at the peak. As readily seen, there is a strong positive association between the change in intensity and change in volatility. This is consistent with the mixture of distribution hypothesis. The correlation is stronger before an intensity burst, around 0.55, which is on par with \citet*{stoltenberg-mykland-zhang:22a}. However, there is a significant drop in the level of the correlation, around 0.20, in the aftermath of an intensity burst.

\begin{figure}[t!]
\begin{center}
\caption{Realized illiquidity and transaction volume near intensity burst.}
\label{figure:illiquidity}
\begin{tabular}{cc}
\small{Panel A: Realized illiquidity.} & \small{Panel B: Transaction volume.}\\
\includegraphics[height=7cm,width=0.48\textwidth]{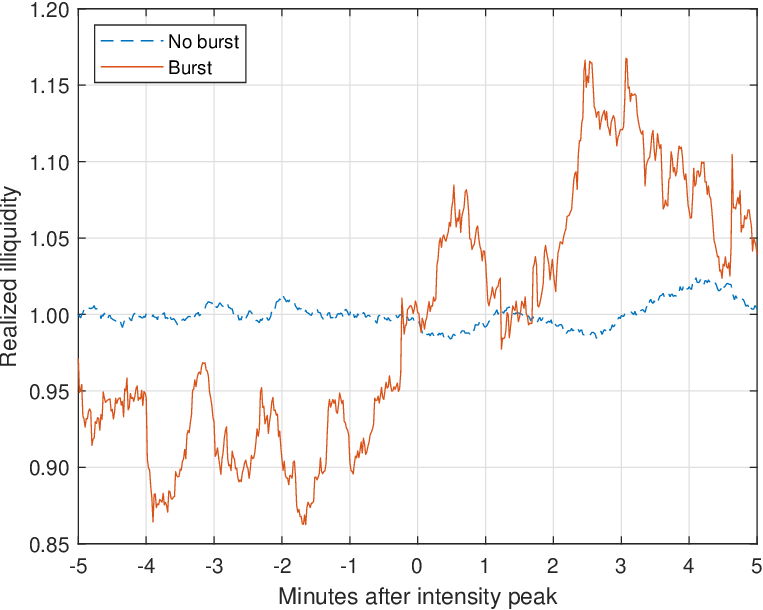} &
\includegraphics[height=7cm,width=0.48\textwidth]{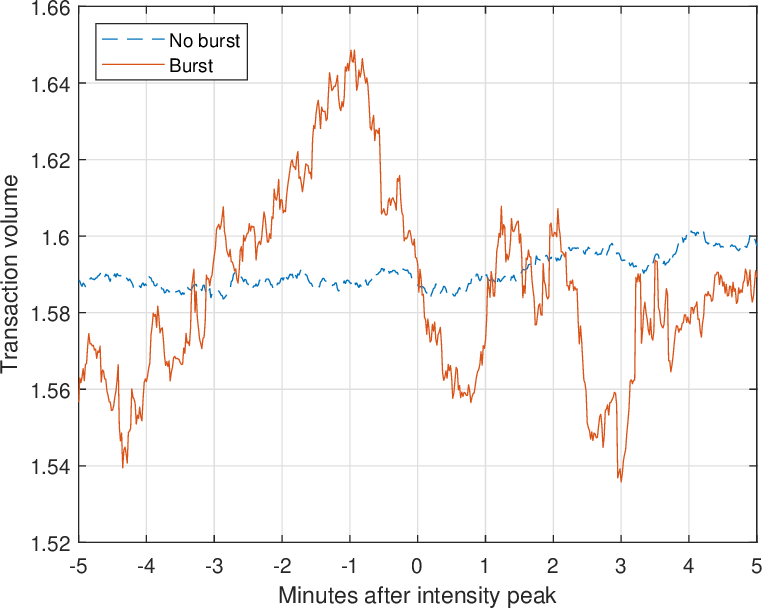} \\
\end{tabular}
\begin{scriptsize}
\parbox{\textwidth}{\emph{Note.} In Panel A, we show the average realized illiquidity of \citet{lacave-ranaldo-magistris:23a} over a 10-minute interval centered at the candidate times associated with the local maximum intensity estimates from Panel A in Figure \ref{figure:ebs-ib-time}. The sample has been divided based on whether the intensity burst statistic at these points is greater than 5 [burst] or not [no burst]. In Panel B, the average transaction volume (in million \euro) is reported. The latter is proportional to the denominator of the realized illiquidity measure.}
\end{scriptsize}
\end{center}
\end{figure}

How does liquidity change around an intensity burst? To understand this, we construct the realized illiquidity of \citet{lacave-ranaldo-magistris:23a}, which is a high-frequency version of the \citet{amihud:02a} measure. In Panel A of Figure \ref{figure:illiquidity}, we chart of the average evolution of this measure on a second-by-second basis. At each time point, we extract one-minute log-returns over the preceding 5-minute interval and calculate the statistic as the realized power variation (i.e. the sum of absolute log-returns) divided by the transaction volume over that window. We normalize it to one on average for ease of interpretation. We restrict attention to the candidate times from Panel A of Figure \ref{figure:ebs-ib-time} and split the sample dependent on whether the intensity burst statistic is greater than 5 (burst) or not (no burst). While there is no material change in liquidity for an insignificant peak in the intensity estimator, during an intensity burst we observe a substantial rise in the illiquidity index both before and after the intensity burst time. In Panel B of the figure, we plot the cumulative transaction volume in million \euro \ (i.e. the denominator of realized illiquidity). There is no discernible shift in transaction volume, which hovers about in a very narrow interval confined to a few percent around its average level. This suggests that the upsurge in the realized illiquidity is mainly induced by the increase in volatility (i.e. the numerator of the statistic).

\subsection{What is the relationship with a drift burst?} \label{section:drift-burst}

\begin{figure}[t!]
\begin{center}
\caption{Intensity burst in the EUR/USD without a drift burst.}
\label{figure:ib-nodb}
\begin{tabular}{cc}
\small{Panel A: Exchange rate.} & \small{Panel B: Transaction count.} \\
\includegraphics[height=7cm,width=0.48\textwidth]{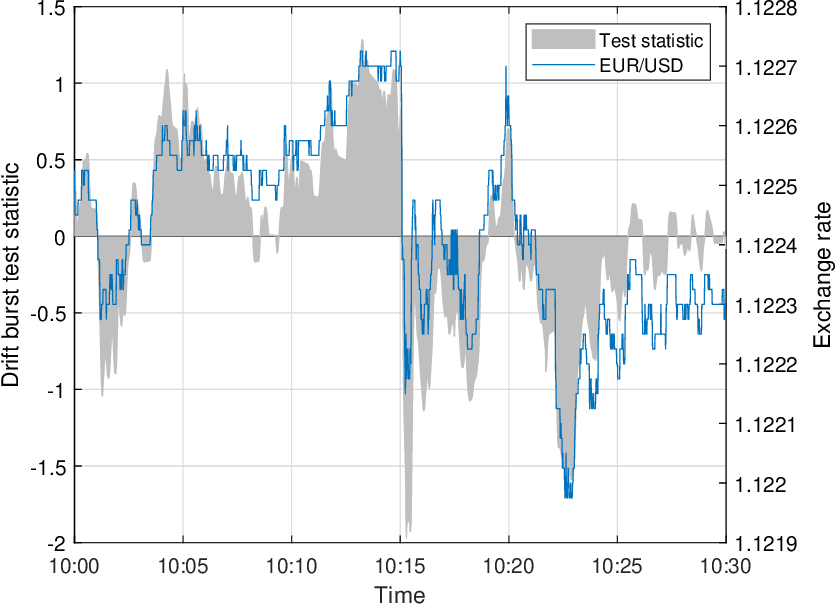} &
\includegraphics[height=7cm,width=0.48\textwidth]{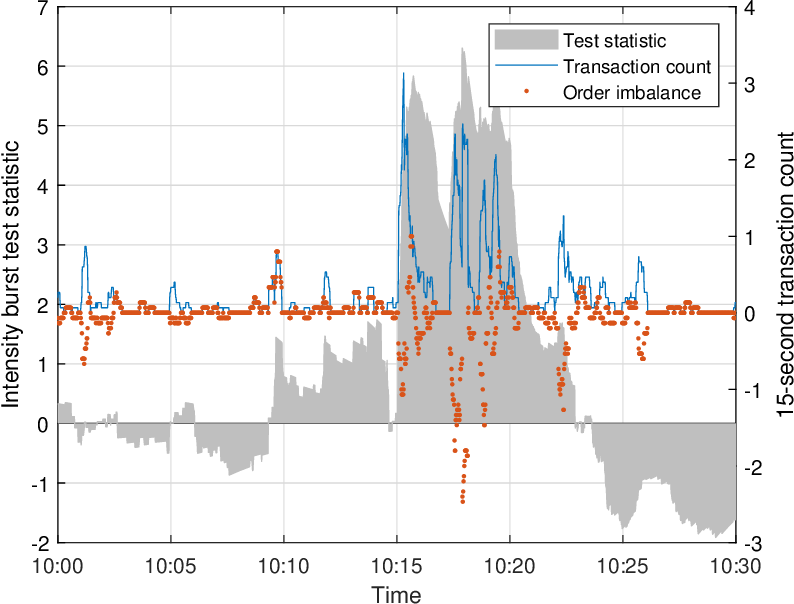}
\end{tabular}
\begin{scriptsize}
\parbox{\textwidth}{\emph{Note.} This figure shows the EUR/USD spot exchange rate on July 10, 2019, where we detect an intensity burst but no a drift burst. In the left panel, we plot the exchange rate from 10:00am to 10:30am CET along with the drift burst test statistic proposed in \citet*{christensen-oomen-reno:22a}. In the right panel, we plot a nonparametric estimator of the time-varying trading intensity (as proxied by the number of transactions in a 15-second window) along with the intensity burst test statistic and a measure of order imbalance.}
\end{scriptsize}
\end{center}
\end{figure}

In Figure \ref{figure:ib-nodb}, we follow the recipe of May 14, 2019 displayed in Figure \ref{figure:ib} and report the price and transaction data in the EUR/USD from 10:00am to 10:30am on July 10, 2019. As in the previous example, this day contains an intensity burst, which is started by sell-side activity around 10:15am. However, there is only a minuscule downtick in the exchange rate, which does not get caught as a drift burst. So here, the intensity burst is an isolated event.

Why is the market sometimes able to withstand the abnormal trading activity observed during an intensity burst with no discernible movement in the price, while at other times it goes through a severe correction? It is possible that a piece of news (or public signal) does not alter the average opinion of investors, even if their interpretation of the information differ. This is related to the differences-of-opinion literature \citep[e.g.,][]{bollerslev-li-xue:18a, kandel-pearson:95a}. Roughly speaking, in these models price change is related to the average opinion (or 1st moment), while trading volume is related to the dispersion of opinion (or 2nd moment). So there can be a lot of trading without a big impact on price, because mostly disagreeing investors trade with each other. However, absolute price change and trading volume should still be positively correlated \citep{harris-raviv:93a}.\footnote{A small trade can cause a relatively big change in price if the market impact is large, such as in the presence of asymmetric information \citep[e.g.][]{glosten-milgrom:85a, kyle:85a}.}

To explore this question in more detail, we extract the drift burst test statistic at a one-second frequency in a 10-minute window centered at the intensity burst time.\footnote{We do not pretend that there is a cause and effect from an intensity burst to a drift burst. In practice, these events are the result of a highly complicated and interrelated process. The aim here is merely to determine correlation with suspected drivers of the drift burst test statistic during an intensity burst.} The attractiveness of the drift burst test statistic is that it is detects abnormal price changes on a volatility-adjusted basis, since the test statistic is calculated as the average log-price increment divided by spot volatility over an interval.

\begin{figure}[t!]
\begin{center}
\caption{Properties of relative trading intensity.} \label{figure:trading-intensity}
\begin{tabular}{cc}
\small{Panel A: Histogram.} & \small{Panel B: Scatter with drift burst statistic.} \\
\includegraphics[height=7cm,width=0.48\textwidth]{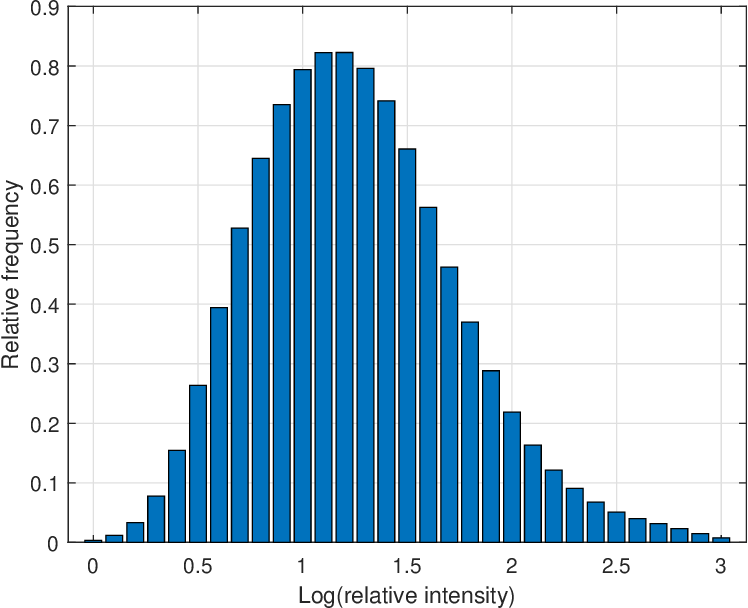} &
\includegraphics[height=7cm,width=0.48\textwidth]{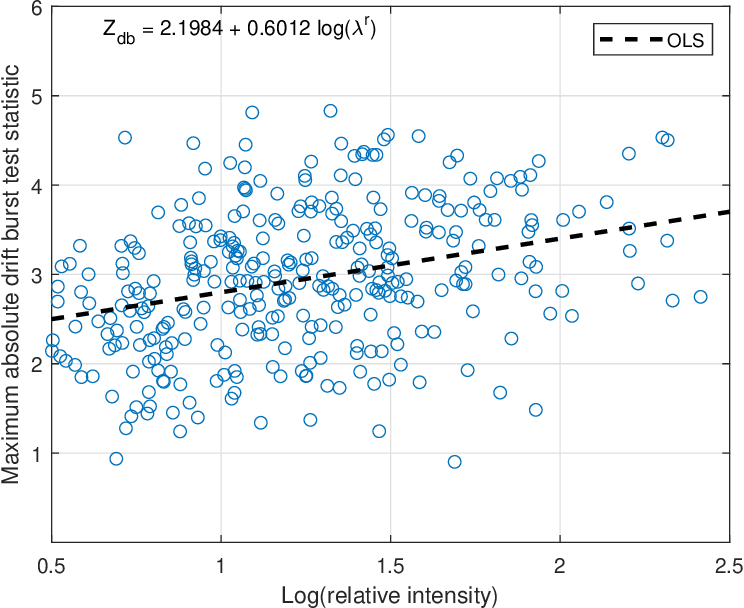}
\end{tabular}
\begin{scriptsize}
\parbox{\textwidth}{\emph{Note.} In Panel A, we plot a histogram of the logarithm of relative trading intensity at the peak of an intensity burst. In Panel B, we show how the log-relative intensity is related to the maximum absolute value of the drift burst statistic in a 10-minute interval around the intensity burst time. A least squares fit is superimposed as a reference. The slope coefficient is significant with a $P$-value less than a percent. The sample is conditional on an intensity burst statistic being greater than 5. It contains a total of 331 events.}
\end{scriptsize}
\end{center}
\end{figure}

We employ a separate linear regression with the minimum (most negative) and maximum (most positive) value of the drift burst test statistic as dependent variable. As covariates, we include the level of trading activity that should, everything else equal, inflate the magnitude of the drift burst test statistic. The rationale is, as stated above, that it requires trades (or quote updates) to move the price. We therefore add the intensity estimated at the peak of the intensity burst as the first predictor. However, because there is a great variety in the absolute level of trading over time, we scale the diurnal-corrected spot estimator by the average level of intensity for the entire day to construct a relative trading activity, which is more homogenous across events. As shown in Panel A of Figure \ref{figure:trading-intensity}, a log-transformation renders this variable close to normally distributed, apart from a slightly elevated right-tail. In Panel B of the figure, we plot the logarithmic measure against the maximum absolute value of the drift burst statistic over the calculation window, which indeed shows a positive relationship.

Our transaction data also contains an indicator variable of whether a deal was buyer- or seller-initiated. One-sided order flow is a good proxy for price pressure---that can be caused by either a change in the average opinion of investors or asymmetric information---so we include the order imbalance in the 5-minute interval before the intensity burst time as our second predictor. We calculate this as the number of buyer trades minus the number of seller trades. A positive value is therefore indicative of buy-side activity, and vice versa. We scale the difference by the spot intensity to again get a relative measure.

As our third predictor, we construct an intuitive concept of market depth derived from the information in the limit order book. We compute the elasticity of the price with respect to quantity following the procedure described in \cite*{naes-skjeltorp:06a}. This is referred to as the slope of the order book.\footnote{\cite*{naes-skjeltorp:06a} look at the elasticity of quantity with respect to price. We employ the reciprocal of their formula for the slope of the order book by reversing the role of price and volume.} It accounts for the resilience of the market by indicating whether it can absorb relatively large amounts of trading volume with minimal price impact. In particular, a high price elasticity suggests that the market is in a very fragile state, such that a sudden burst in trading activity can induce pronounced price volatility. The calculation is as follows:
\begin{equation}
\mathcal{S}_{ \text{ob}} = \frac{1}{L} \left( \frac{|P_{1}/M-1|}{v_{1}} + \sum_{l=1}^{L} \frac{|P_{l+1}/P_{l}-1|}{v_{l+1}/v_{l}-1} \right),
\end{equation}
where $L$ is the number of levels in the limit order book (10 here), $M$ is the midquote of the best bid and offer, $P_{l}$ and $v_{l}$ are the price and natural logarithm of accumulated volume at tick level $l$, for $l = 1, \dots, L$. We calculate the slope separatively for the bid and ask side at a one-second frequency and include the moving average of the sequence over the 5-minute interval preceding the intensity burst in the regression.\footnote{The one-sided slope calculation is a bit sensitive to the flip-flopping of large orders up and down in tick level, when the market bounces around at the inside spread. The slight averaging alleviates this problem.}

We estimate the model:
\begin{equation} \label{equation:regression}
Z_{\text{db}}^{ \mp} = a + b_{1} \log( \lambda^{r}) + b_{2} \mathcal{O} + b_{3} \mathcal{S}_{ \text{ob}}^{ \mp} + b_{4} \sqrt{RV} + b_{5} \mathcal{P}_{0} + \epsilon,
\end{equation}
where $Z_{\text{db}}^{-}$ ($Z_{\text{db}}^{+}$) is the minimum (maximum) drift burst test statistic, $\lambda^{r}$ is the relative intensity estimate, $\mathcal{O}$ is the relative order imbalance, and $\mathcal{S}_{ \text{ob}}^{-}$ ($\mathcal{S}_{ \text{ob}}^{+}$) is the slope of the order book on the bid (ask) side.

There are a few controls in \eqref{equation:regression}. $\sqrt{RV}$ is the square-root of 5-minute realized variance computed between three- to one-hour before the intensity burst, while $\mathcal{P}_{0}$ is the percent of daily 5-minute log-returns that are zero. The latter is a proxy of staleness in the aggregate trading environment, see \citet*{bandi-kolokolov-pirino-reno:20a}. We first run the regression without these two additional variables and then add them as a robustness check.\footnote{We also estimated a regression, where the realized illiquidity was included as a covariate. However, it was highly insignificant, so we removed it again. This is rather intuitive, since the measure was found to be driven mostly by volatility and the drift burst test statistic controls for this effect.}

\begin{table}[t!]
\setlength{\tabcolsep}{1.00cm}
\begin{center}
\caption{Regression with drift burst test statistic.} \label{table:drift-burst-regression}
\begin{tabular}{lrrrr}
\hline
\hline
Constant & $\underset{(-4.591)}{-1.145}$ & $\underset{(-3.936)}{-1.318}$ & $\underset{(1.618)}{0.525}$ & $\underset{(1.885)}{0.722}$ \\
$\log( \lambda_{r})$ & $\underset{(-2.512)}{-0.430}$ & $\underset{(-2.657)}{-0.456}$ & $\underset{(1.942)}{0.318}$ & $\underset{(2.013)}{0.330}$ \\
$\mathcal{O}$ & $\underset{(3.121)}{0.594}$ & $\underset{(3.217)}{0.610}$ & $\underset{(2.587)}{0.476}$ & $\underset{(2.289)}{0.423}$ \\
$\mathcal{S}_{ \text{ob}}$ & $\underset{(-2.495)}{-0.713}$ & $\underset{(-3.273)}{-1.062}$ & $\underset{(4.036)}{2.281}$ & $\underset{(4.459)}{2.601}$ \\
$\sqrt{RV}$ & & $\underset{(0.431)}{0.689}$ & & $\underset{(-0.701)}{-1.069}$ \\
$\mathcal{P}_{0}$ & & $\underset{(2.214)}{6.467}$ & & $\underset{(-2.086)}{-5.260}$ \\
\\
$R^{2}$ & 0.082 & 0.098 & 0.109 & 0.123\\
\hline
\hline
\end{tabular}
\begin{scriptsize}
\parbox{\textwidth}{\emph{Note.} We estimate the linear regression: $Z_{ \text{db}}^{ \mp} = a + b_{1} \log( \lambda^{r}) + b_{2} \mathcal{O} + b_{3} \mathcal{S}_{ \text{ob}}^{ \mp} + b_{4} \sqrt{RV} + b_{5} \mathcal{P}_{0} + \varepsilon$, where $Z_{ \text{db}}^{-}$ ($Z_{ \text{db}}^{+}$) is the minimum (maximum) drift burst test statistic in a 10-minute window centered at the intensity burst time, $\lambda^{r}$ is the relative intensity, $ \mathcal{O}$ is the relative order imbalance, $\mathcal{S}_{ \text{ob}}^{-}$ ($\mathcal{S}_{ \text{ob}}^{+}$) is the bid slope (ask slope) of the order book, $\sqrt{RV}$ is the 5-minute realized variance between three- to one-hour before the intensity burst, while $\mathcal{P}_{0}$ is the percent of daily 5-minute log-returns that are zero. The table reports parameter estimates with the $t$-statistic shown in parenthesis underneath. The sample size is 331. $R^{2}$ is the coefficient of determination.}
\end{scriptsize}
\end{center}
\end{table}

Table \ref{table:drift-burst-regression} presents the outcome of the regression estimated for the minimum (maximum) drift burst statistic in the left-hand (right-hand) side using the bid (ask) slope as price elasticity. We only comment on the results for the minimum drift burst test statistic, while summarizing the main differences to the other regression below.

The results are compelling. In particular, the sign of the slope coefficient estimates are in line with our economic intuition. The logarithm of relative intensity makes the minimum drift burst test statistic more negative, and the effect is significant. This is perhaps a bit surprising, since we are conditioning on an intensity burst. What also ought to matter is the ability of the market to handle the surge in trading activity, as proxied by order imbalance and price elasticity. The coefficient on order flow is positive and highly significant. This implies that sell-side activity (a negative value of this variable) moves the minimum drift burst statistic farther away from zero. In addition, a large price elasticity, as measured by the bid slope of the order book, reinforces this effect and pushes the drift burst statistic even further into negative territory.\footnote{We also incorporated an interaction effect between order imbalance and price elasticity. However, it was insignificant both in the restricted and full version of the model. We therefore omitted it again.}

The maximum regression shows identical results, apart from a few expected sign changes in appropriate places, i.e. the intercept, and the coefficient in front of log-relative intensity, the order book slope, square-root realized variance, and the fraction of zeros in the log-return series. The main highlight is that the $t$-statistic on the log-relative intensity and order imbalance are a bit more compressed, although both are still at the margin of being significant. This can possibly be reconciled by a costly market presence, as developed in \citet*{huang-wang:09a}. In their framework, if market participation has a cost, the need of financial intermediaries to rebalance their equilibrium portfolios in response to news creates an endogenous order imbalance, which always leads to excessive selling. This effect exacerbates the demand for cash during a sell-off, while rallies remove a smaller portion of volume resting on the ask side. Hence, trading activity and order imbalances put less strain on the system during an upsurge. On the other hand, since EUR/USD is an exchange rate, it is not evident how to define a sell-off.

The additional controls do not alter the coefficients estimates of the above predictors much. The realized volatility reduces the magnitude of the drift burst test statistic, even if the effect is insignificant. Moreover, zero returns further attenuates it. The explanation is probably that in a slow market with limited trading, a large burst in relative trading activity may still be economically small and incapable of shifting the price much.

In summary, our analysis shows it is asymmetric trading combined with a vulnerable order book that causes the absolute drift burst test statistic to inflate during an intensity burst, thus increasing the likelihood of observing a drift burst.

\section{Conclusion} \label{section:conclusion}

In this paper, we propose a model with a locally unbounded intensity for point processes. We allow the spot intensity to explode such that the integrated intensity is finite, ensuring the point process is non-explosive. This is the definition of an intensity burst. We show the model is capable of generating rather extreme clusters of observations over small time intervals that are far more concentrated compared to what a standard doubly stochastic Poisson or self-exciting Hawkes processes with locally bounded intensity can produce.

We develop an inference strategy for detecting an intensity burst. The theory relies on a heavy traffic assumption, which permits to consistently estimate and draw inference about the properties of a point process over a finite time interval. It resembles the standard in-fill condition for asymptotic theory of realized variance. In absence of an intensity burst (null hypothesis), the asymptotic distribution of our test statistic is standard normal, but the rate of convergence depends on a nuisance parameter, the speed with which the process accumulates points under the heavy traffic condition. Hence, the feasible implementation of our test statistic is based on an automatic inference procedure, where we adapt the observed asymptotic variance of \citet*{mykland-zhang:17a} to spot estimation. Conditional on an intensity burst (alternative hypothesis) our test statistic diverges, so the power of the testing procedure converges to one.

A simulation study demonstrates that the test statistic has good finite sample properties for a variety of point processes. It controls size and has high power under the alternative. With our numerical experiments, we also show that with self-exciting exponential \citet*{hawkes:71a} process, a more general model than the theoretical framework allows, the test statistic does not produce false positives. Furthermore, we demonstrate robustness to diurnal variation in the intensity process based on a deflated intensity estimator.

At last, we implement the test statistic on an extensive set of EUR/USD foreign exchange rate high-frequency data extracted from the electronic broking services (EBS) platform, where it captures abnormal surges in trading activity. We detect a nontrivial amount of intensity bursts in these data and describe their basic properties. We examine how intensity is related to other fundamental indicators, such as volatility and liquidity. We also show that the more excessive the trading activity observed during an intensity burst is, the higher is the chance to observe a drift burst. This effect is stronger with an unbalanced order flow and when the price elasticity of the limit order book is high. We leave a further investigation of these findings to future research.

\clearpage

\appendix \label{section:appendix}

\section{Proofs}

We here derive the theoretical results listed in the main text. We note that under Assumption \ref{assumption:rate} -- \ref{assumption:mu}, we can appeal to the localization procedure described in \citet*[][Section 4.4.1]{jacod-protter:12a} to bound the processes $a_{t}$, $\nu_{t}$, and $\delta_{t}(x)$ as $(t,x)$ vary within $[0,T] \times \mathbb{R}$. Moreover, we can assume that $| \delta_{t}(x)| \leq \bar{ \Gamma}(x)$, where $\bar{ \Gamma}$ is a bounding function such that $\int_{ \mathbb{R}} \bar{ \Gamma}(x)^{2} F( \mathrm{d}x) < \infty$ and, for all $\kappa \in (0,1)$, $\int_{ \left\{x: \bar{ \Gamma}(x) \leq \kappa \right\}} \bar{ \Gamma}(x) F( \mathrm{d}x) < \infty$.

We make extensive use of a universal positive constant, $C$, whose value changes from equation-to-equation and line-to-line without notification. As a shorthand notation, we write the intensity burst time as $\tau_{ \mathrm{ib}} \equiv \tau$.

To establish stable convergence of the local intensity estimator in \eqref{equation:lambdba-hat}, we need an auxiliary result, which is a reproduction of \citet[][Lemma 8]{alvarez-panloup-pontier-savy:12a}.

\begin{lemma} \label{lemma:stable-convergernce}
Let $(\Omega, \mathcal{F}, \mathbb{P})$ denote a probability space. For each $n \geq 1$, suppose $\zeta_{2}^{n}, \zeta_{3}^{n}, \dots, \zeta_{k_{n}}^{n}$ are martingale increments with respect to the sub-$\sigma$-fields of $\mathcal{F}$: $\mathcal{F}_{n,1} \subseteq \mathcal{F}_{n,2} \subseteq \dots \subseteq \mathcal{F}_{n,k_{n}}$. Set $S_{n} = \sum_{i=2}^{k_{n}} \zeta_{i}^{n}$ and $\mathcal{G} = \cap_{n \geq 1} \mathcal{F}_{n,1}$.  Assume that $n \mapsto \mathcal{F}_{n,k_{n}}$ is a non-increasing sequence of $\sigma$-fields such that $\cap_{n \geq 1} \mathcal{F}_{n,k_{n}} = \mathcal{G}$. Then, if the following conditions hold:
\begin{itemize}
\item[(i)] There exists a $\mathcal{G}$-measurable random variable $\eta$ such that, as $n \rightarrow \infty$,
\begin{equation*}
\sum_{i=2}^{k_{n}} \mathbb{E} \Big[( \zeta_{i}^{n})^{2} \mid \mathcal{F}_{n,i-1} \Big] \overset{p}{ \longrightarrow} \eta,
\end{equation*}
\item[(ii)] For every $\epsilon > 0$,
\begin{equation*}
\sum_{i=2}^{k_{n}} \mathbb{E} \Big[ (\zeta_{i}^{n})^{2} \mathbf{1}_{ \{( \zeta_{i}^{n})^{2} \geq \epsilon \}} \mid \mathcal{F}_{n,i-1} \Big] \overset{p}{ \longrightarrow} 0,
\end{equation*}
then $S_{n} \overset{ \mathfrak{D}_{s}}{ \longrightarrow} S$, defined on an extension of $(\Omega, \mathcal{F}, \mathbb{P})$, such that conditionally on $\mathcal{F}$, the distribution of $S$ is a centered Gaussian with variance $\eta$.
\end{itemize}
\end{lemma}

\subsection{Proof of Lemma \ref{lemma:consistency}}

Theorem \ref{theorem:clt} implies that, under $\mathscr{H}_{0}$,
\begin{equation} \label{equation:op1}
\widehat{ \lambda}_{t} - \lambda_{t-} = o_{p}(1),
\end{equation}
which shows the first statement.

Next, under $\mathscr{H}_{1}$, we notice that the observed process can be decomposed as
\begin{equation*}
N_{t}^{n} = N_{t}^{n, \mu} + N_{t}^{n, \beta},
\end{equation*}
where $N_{t}^{n, \mu}$ and $N_{t}^{n, \beta}$ are inhomogeneous Poisson processes with rates $n \mu_{t}$ and $n \beta_{t}$.

Therefore, at time $\tau$,
\begin{equation*}
\widehat{ \lambda}_{ \tau} = \frac{N^{n, \mu}( \tau- \delta_{n}, \tau)}{n \delta_{n}} + \frac{N^{n, \beta}( \tau - \delta_{n}, \tau)}{n \delta_{n}} = \widehat{ \mu}_{ \tau} + \widehat{ \beta}_{ \tau}.
\end{equation*}
It follows from \eqref{equation:op1} that
\begin{equation*}
\widehat{ \mu}_{ \tau} = \mu_{ \tau-} + o_{p}(1),
\end{equation*}
so it is enough to look at $\widehat{ \beta}_{ \tau}$. To prove that $\widehat{ \beta}_{ \tau} = O_{p}( \delta_{n}^{- \alpha})$, by the definition of stochastic orders in probability, it suffices to show that for every $\epsilon > 0$ there exists a $0 < \Delta_{ \epsilon} < \infty$ and $N_{ \epsilon} \in \mathbb{N}$ such that for all $n \geq N_{ \epsilon} : \mathbb{P} \big( \delta_{n}^{ \alpha} \widehat{ \beta}_{ \tau} > \Delta_{ \epsilon} \big) < \epsilon$.

By Markov's inequality:
\begin{equation*}
\mathbb{P} \Big( \delta_{n}^{\alpha} \widehat{ \beta}_{ \tau} > \Delta_{ \epsilon} \Big) \leq \frac{ \delta_{n}^{ \alpha} \mathbb{E}( \widehat{ \beta}_{ \tau})}{ \Delta_{ \epsilon}}.
\end{equation*}
Next, conditioning on $\sigma$ and employing the law of iterated expectations:
\begin{align*}
\mathbb{E} \big( \widehat{ \beta}_{ \tau} \big) &= \mathbb{E} \bigg[ \frac{N^{n, \beta}( \tau-\delta_{n}, \tau)}{n \delta_{n}} \bigg] = \mathbb{E} \bigg[ \mathbb{E} \bigg( \frac{N^{n, \beta}( \tau- \delta_{n}, \tau)}{n \delta_{n}} \mid \mathcal{F}^{ \sigma} \bigg) \bigg] = \mathbb{E} \bigg[ \frac{1}{ \delta_{n}} \int_{ \tau- \delta_{n}}^{ \tau} \sigma_{u} | \tau-u|^{- \alpha} \mathrm{d}u \bigg] \\[0.10cm] &\leq \frac{C}{ \delta_{n}} \int_{ \tau- \delta_{n}}^{ \tau} | \tau-u|^{- \alpha} \mathrm{d}u = \frac{C}{1- \alpha} \delta_{n}^{- \alpha},
\end{align*}
where $\mathcal{F}^{\sigma}$ is the $\sigma$-algebra generated by $\sigma$ and $C>0$ is a constant that bounds the process from above in light of the localization procedure. Thus,
\begin{equation*}
\mathbb{P} \Big( \delta_{n}^{\alpha} \widehat{ \beta}_{ \tau} > \Delta_{ \epsilon} \Big) \leq \frac{C}{ \Delta_{ \epsilon} (1- \alpha)}.
\end{equation*}
Thus, for every $\epsilon > 0$, we can choose $\displaystyle \Delta_{ \varepsilon} > \frac{C}{ \epsilon(1- \alpha)}$ to make $\mathbb{P} \big( \delta_{n}^{ \alpha} \widehat{ \beta}_{ \tau} > \Delta_{ \epsilon} \big) < \epsilon$. \qed

\subsection{Proof of Theorem \ref{theorem:clt}}

We show the univariate statement in (i), (ii), and (iii). The second half follows from the calculations in the proof of Theorem \ref{theorem:observed-asymptotic-variance}.

To this end, we define the sequence $\rho_n$ as follows:
\begin{equation*}
\rho_{n} =
\begin{cases}
\delta_{n} n, & \text{if } \delta_{n} \sqrt{n} \rightarrow 0, \\[0.10cm]
\delta_{n}^{-1}, & \text{if } \delta_{n} \sqrt{n} \rightarrow \theta \text{ or } \delta_{n} \sqrt{n} \rightarrow \infty,
\end{cases}
\end{equation*}
and study the difference $\sqrt{ \rho_{n}}( \widehat{ \lambda}_{t} - \lambda_{t-})$.

We assume $\delta_{n}$ can be expressed as $\delta_{n} = \ell_{n} \Delta_{n}$, where $\Delta_{n} = n^{-1}$ and $\ell_{n}$ is a deterministic sequence of positive integers. Then, $\widehat{ \lambda}_{t}$ can be expressed as
\begin{equation*}
\widehat{\lambda}_{t} = \frac{1}{ \ell_{n}} \sum_{i \in \mathcal{D}_{t-}^{n}} \Delta_{i} N^{n},
\end{equation*}
where $\Delta_{i} N^{n} = N^{n}(0,i \Delta_{n}) - N^{n}(0,(i-1) \Delta_{n})$ are the increments of the process $N^{n}$ over the short time intervals of length $\Delta_{n}$, and
\begin{equation*}
\mathcal{D}^{n}_{t-} = \{ tn- \ell_{n}+1,tn- \ell_{n}+2, \dots,tn \}.
\end{equation*}
We approximate $\Delta_{i} N^{n}$ by the increments of an inhomogeneous Poisson process $\widetilde{N}^{n}$ with piecewise constant intensity, $\Delta_{i} \widetilde{N}^{n}$. This is done as follows. By the random time-change theorem for point processes \citep*[see, e.g.,][Theorem 7.4.I]{daley-vere-jones:03a}, there exists a homogeneous unit intensity Poisson process $N^{( \star),(i)}$, such that
\begin{equation*}
\Delta_{i} N^{n} = N^{(\star),(i)} \Bigg(n \int_{(i-1) \Delta_{n}}^{i \Delta_{n}} \mu_{s} \mathrm{d}s \Bigg).
\end{equation*}
We set $\Delta_{i} \widetilde{N}^{n} = N^{( \star),(i)} \Big( n \int_{(i-1) \Delta_{n}}^{i \Delta_{n}} \mu_{(i-1) \Delta_{n}} \mathrm{d}s \Big)$.
Hence, $\widetilde{N}^{n}(k \Delta_{n}) = \sum_{i \leq k} \Delta_{i} \widetilde{N}^{n}$ is an inhomogeneous Poisson process with piecewise constant intensity, such that
\begin{equation*}
\mathbb{E} \Big[ \Delta_{i} \widetilde{N}^{n} \mid \mathcal{F}_{(i-1) \Delta_{n}} \Big] = n \int_{(i-1) \Delta_{n}}^{i \Delta_{n}} \mu_{(i-1) \Delta_{n}} \mathrm{d}s = \mu_{(i-1) \Delta_{n}}.
\end{equation*}
Moreover, the absolute value of the approximation error $| \Delta_{i} N^{n} - \Delta_{i} \widetilde{N}^{n}|$ can be expressed as an increment of the process $N^{( \star),(i)}$:
\begin{equation*}
| \Delta_{i} N^{n} - \Delta_{i} \widetilde{N}^{n}| = N^{( \star),(i)}( \underline{t}, \overline{t}),
\end{equation*}
where
\begin{equation*}
\underline{t} = n \Bigg( \int_{(i-1) \Delta_{n}}^{i \Delta_{n}} \mu_{s} \mathrm{d}s \wedge \int_{(i-1) \Delta_{n}}^{i \Delta_{n}} \mu_{(i-1) \Delta_{n}} \mathrm{d}s \Bigg) \quad \text{and} \quad \overline{t} = n \Bigg( \int_{(i-1) \Delta_{n}}^{i \Delta_{n}} \mu_{s} \mathrm{d}s \vee \int_{(i-1) \Delta_{n}}^{i \Delta_{n}} \mu_{(i-1) \Delta_{n}} \mathrm{d}s \Bigg).
\end{equation*}
Then, we can write
\begin{equation*}
\widehat{ \lambda}_{t} - \mu_{t-} = \underbrace{ \frac{1}{ \ell_{n}} \sum_{i \in \mathcal{D}_{t-}^{n}} \big( \Delta_{i} N^{n} - \Delta_{i} \widetilde{N}^{n} \big)}_{ \mathrm{(I)}} + \underbrace{ \frac{1}{ \ell_{n}} \sum_{i \in \mathcal{D}_{t-}^{n}} \big( \Delta_{i} \widetilde{N}^{n} - \mu_{(i-1) \Delta_{n}} \big)}_{ \mathrm{(II)}} + \underbrace{ \frac{1}{ \ell_{n}} \sum_{i \in \mathcal{D}_{t-}^{n}} \big( \mu_{(i-1) \Delta_{n}} - \mu_{t-} \big)}_{ \mathrm{(III)}}.
\end{equation*}
(I) arises by approximating the observed process with a point process that has a locally constant intensity, (II) is the deviation of the approximating process from its conditional expectation, and (III) denotes the error due to the variation in the rate process.

(I) can be further decomposed as follows:
\begin{align*}
\mathrm{(I)} &= \frac{1}{ \ell_{n}} \sum_{i \in \mathcal{D}_{t-}^{n}} \Big( \Delta_{i} N^{n} - \Delta_{i} \widetilde{N}^{n} - \mathbb{E} \big[ \Delta_{i} N^{n} - \Delta_{i} \widetilde{N}^{n} \mid \mathcal{F}_{(i-1) \Delta_{n}} \big] \Big) \\[0.10cm] &+ \frac{1}{ \ell_{n}} \sum_{i \in \mathcal{D}_{t-}^{n}} \mathbb{E} \big[ \Delta_{i} N^{n} - \Delta_{i} \widetilde{N}^{n} \mid \mathcal{F}_{(i-1) \Delta_{n}} \big].
\end{align*}
Moreover, due to Assumption \ref{assumption:mu}, (III) can be split into a drift and volatility part:
\begin{equation*}
\mathrm{(III)} = \frac{1}{ \ell_{n}} \sum_{i \in \mathcal{D}_{t-}^{n}} (tn- \ell_{n}-i)(A_{i \Delta_{n}} - A_{(i-1) \Delta_{n}}) + \frac{1}{ \ell_{n}} \sum_{i \in \mathcal{D}_{t-}^{n}} (tn- \ell_{n}-i)(M_{i \Delta_{n}} - M_{(i-1) \Delta_{n}}),
\end{equation*}
where $A_{t} = \int_{0}^{t} a_{s}^{*} \mathrm{d}s + \int_{0}^{t} \int_{\mathbb{R}} \delta_{s}(x) \left( \mathfrak{p}( \mathrm{d}s, \mathrm{d}x) - \mathfrak{q}( \mathrm{d}s, \mathrm{d}x) \right)$, where $a_{s}^{s} = a_{s} + \int_{| \delta_{s}(x)| > 1} \delta_{s}(x) F( \mathrm{d}x)$, and $M_{t} = \int_{0}^{t} \nu_{s} \mathrm{d}W_{s}$.

Consequently,
\begin{equation*}
\sqrt{ \rho_{n}} ( \widehat{ \lambda}_{t} - \mu_{t-}) = \Lambda_{1}^{n}(t)  + \Lambda_{2}^{n}(t)  + \Lambda_{3}^{n}(t) + \Lambda_{4}^{n}(t) + \Lambda_{5}^{n}(t),
\end{equation*}
where
\begin{align*}
\Lambda_{1}^{n}(t) &= \frac{ \sqrt{ \rho_{n}}}{ \ell_{n}} \sum_{i \in \mathcal{D}_{t-}^{n}} \big( \Delta_{i} N^{n} - \Delta_{i} \widetilde{N}^{n} - \mathbb{E} \big[ \Delta_{i} N^{n} - \Delta_{i} \widetilde{N}^{n} \mid \mathcal{F}_{(i-1) \Delta_{n}} \big] \big), \\[0.10cm]
\Lambda_{2}^{n}(t) &= \frac{ \sqrt{ \rho_{n}}}{ \ell_{n}} \sum_{i \in \mathcal{D}_{t-}^{n}} \mathbb{E} \big[ \Delta_{i} N^{n} - \Delta_{i} \widetilde{N}^{n} \mid \mathcal{F}_{(i-1) \Delta_{n}} \big], \\[0.10cm]
\Lambda_{3}^{n}(t) &= \frac{ \sqrt{ \rho_{n}}}{ \ell_{n}} \sum_{i \in \mathcal{D}_{t-}^{n}} \big( \Delta_{i} \widetilde{N}^{n} - \mu_{(i-1) \Delta_{n}} \big), \\[0.10cm]
\Lambda_{4}^{n}(t) &= \frac{ \sqrt{ \rho_{n}}}{ \ell_{n}} \sum_{i \in \mathcal{D}_{t-}^{n}} (tn- \ell_{n}-i)(A_{i \Delta_{n}} - A_{(i-1) \Delta_{n}}).
\\[0.10cm]
\Lambda_{5}^{n}(t) &= \frac{ \sqrt{ \rho_{n}}}{ \ell_{n}} \sum_{i \in \mathcal{D}_{t-}^{n}} (tn- \ell_{n}-i)(M_{i \Delta_{n}} - M_{(i-1) \Delta_{n}}), \end{align*}
Now, we show that the sum $\Lambda_{3}^{n}(t) + \Lambda_{5}^{n}(t)$ converges stably in law, while the other terms ($\Lambda_{1}^{n}(t)$, $\Lambda_{2}^{n}(t)$, and $\Lambda_{4}^{n}(t)$) are asymptotically negligible.

Set
\begin{equation*}
\Lambda_{3}^{n}(t) + \Lambda_{5}^{n}(t) = \sum_{ i \in \mathcal{D}_{t-}^{n}} \zeta_{i}^{n}(3)+ \zeta_{i}^{n}(5),
\end{equation*}
where
\begin{equation*}
\zeta_{i}^{n}(3) = \frac{ \sqrt{ \rho_{n}}}{ \ell_{n}} \big( \Delta_{i} \widetilde{N}^{n} - \mu_{(i-1) \Delta_{n}} \big) \quad \text{and} \quad \zeta^n_i(5) =   \frac{ \sqrt{ \rho_{n}}}{ \ell_{n}} (tn- \ell_{n}-i)(M_{i \Delta_{n}} - M_{(i-1) \Delta_{n}}).
\end{equation*}
By construction, $\zeta_{i}^{n}(3)$ and $\zeta_{i}^{n}(5)$ are uncorrelated $\mathcal{F}_{(i-1) \Delta_{n}}$-martingale differences:
\begin{equation*}
\mathbb{E} \big[ \zeta_{i}^{n}(3) \mid \mathcal{F}_{(i-1) \Delta_{n}} \big] = \mathbb{E} \big[ \zeta_{i}^{n}(5) \mid \mathcal{F}_{(i-1) \Delta_{n}} \big] = \mathbb{E} \big[ \zeta_{i}^{n}(3) \zeta_{i}^{n}(5) \mid \mathcal{F}_{(i-1) \Delta_{n}} \big] = 0.
\end{equation*}
As a result,
\begin{equation*}
\sum_{i \in \mathcal{D}_{t-}^{n}} \mathbb{E} \Big[ \big( \zeta_{i}^{n}(3) + \zeta_{i}^{n}(5) \big)^{2} \mid \mathcal{F}_{(i-1) \Delta_{n}} \Big] = \sum_{i \in \mathcal{D}_{t-}^{n}} \mathbb{E} \Big[ \big( \zeta_{i}^{n}(3) \big)^{2} \mid \mathcal{F}_{(i-1) \Delta_{n}} \Big] + \sum_{i \in \mathcal{D}_{t-}^{n}} \mathbb{E} \Big[ \big( \zeta_{i}^{n}(5) \big)^{2} \mid \mathcal{F}_{(i-1) \Delta_{n}} \Big].
\end{equation*}
Since $\Delta_{i} \widetilde{N}^{n}$ is Poisson distributed, we deduce that:
\begin{equation*}
\mathbb{E} \Big[ \big( \zeta_{i}^{n}(3) \big)^{2} \mid \mathcal{F}_{(i-1) \Delta_{n}} \Big] = \frac{ \rho_{n}}{ \ell_{n}^{2}} \mu_{(i-1) \Delta_{n}},
\end{equation*}
and, as $\delta_{n} \rightarrow 0$,
\begin{equation*}
\frac{1}{ \ell_{n}} \sum_{i \in \mathcal{D}_{t-}^{n}} \mu_{(i-1) \Delta_{n}} \overset{a.s}{ \longrightarrow} \mu_{t-}.
\end{equation*}
From the definition of $\rho_{n}$:
\begin{equation*}
\frac{ \rho_{n}}{ \ell_{n}} \rightarrow
\begin{cases}
1, & \text{if } \delta_{n} \sqrt{n} \rightarrow 0, \\[0.10cm]
\displaystyle \frac{1}{ \theta^{2}}, & \text{if } \delta_{n} \sqrt{n} \rightarrow \theta, \\[0.10cm]
0, & \text{if } \delta_{n} \sqrt{n} \rightarrow \infty.
\end{cases}
\end{equation*}
Therefore,
\begin{equation*}
\sum_{i \in \mathcal{D}_{t-}^{n}} \mathbb{E} \Big[ \big( \zeta_{i}^{n}(3) \big)^{2} \mid \mathcal{F}_{(i-1) \Delta_{n}} \Big] \overset{p}{ \longrightarrow}
\begin{cases}
\mu_{t-}, & \text{if } \delta_{n} \sqrt{n} \rightarrow 0, \\[0.10cm]
\displaystyle \frac{1}{ \theta^{2}} \mu_{t-}, & \text{if } \delta_{n} \sqrt{n} \rightarrow \theta, \\[0.10cm]
0, & \text{if } \delta_{n} \sqrt{n} \rightarrow \infty.
\end{cases}
\end{equation*}
Next,
\begin{equation*}
\mathbb{E} \Big[ \big( \zeta_{i}^{n}(5) \big)^{2} \mid \mathcal{F}_{(i-1) \Delta_{n}} \Big] = \frac{ \rho_{n}}{ \ell_{n}^{2}}(tn- \ell_{n}-i)^{2} \int_{(i-1) \Delta_{n}}^{i \Delta_{n}} \mathbb{E} \big[ \nu_{s}^{2} \mid \mathcal{F}_{(i-1) \Delta_{n}} \big] \mathrm{d}s.
\end{equation*}
Assumption \ref{assumption:mu} implies that, as $n \rightarrow \infty$,
\begin{equation*}
\mathbb{E} \bigg[ \sup_{s \in [t- \delta_{n},t]} | \nu_{s}^{2} - \nu_{t}^{2} | \bigg] \rightarrow 0,
\end{equation*}
which further means that
\begin{equation*}
\sum_{i \in \mathcal{D}_{t-}^{n}} \bigg( \mathbb{E} \Big[ \big( \zeta_{i}^{n}(5) \big)^{2} \mid \mathcal{F}_{(i-1) \Delta_{n}} \Big] - \frac{ \rho_{n}}{ \ell_{n}^{2}}(tn- \ell_{n}-i)^{2} \Delta_{n} \nu_{t}^{2} \bigg) \overset{p}{ \longrightarrow} 0.
\end{equation*}
On the other hand, multiplying and dividing by $\delta_{n}$,
\begin{equation*}
\sum_{i \in \mathcal{D}_{t-}^{n}} \frac{ \rho_{n}}{ \ell_{n}^{2}}(tn- \ell_{n}-i)^{2} \Delta_{n} \nu_{t}^{2} = \rho_{n} \delta_{n} \nu_{t}^{2} \frac{1}{ \ell_{n}^{3}} \sum_{i \in \mathcal{D}_{t-}^{n}} (tn- \ell_{n}-i)^{2}.
\end{equation*}
The last part is convergent with limit
\begin{equation*}
\frac{1}{ \ell_{n}^{3}} \sum_{i \in \mathcal{D}_{t-}^{n}} (tn- \ell_{n}-i)^{2} = \frac{1}{ \ell_{n}^{3}} \sum_{j=1}^{ \ell_{n}} j^{2} = \frac{1}{ \ell_{n}^{3}} \bigg[ \frac{ \ell_{n}( \ell_{n}+1)(2 \ell_{n}+1)}{6} \bigg] \rightarrow \frac{1}{3}.
\end{equation*}
Moreover,
\begin{equation*}
\rho_{n} \delta_{n} =
\begin{cases}
\delta_{n}^{2} n \rightarrow 0, & \text{if } \delta_{n} \sqrt{n} \rightarrow 0, \\[0.10cm]
 1, & \text{if } \delta_{n} \sqrt{n} \rightarrow \theta \text{ or } \delta_{n} \sqrt{n} \rightarrow \infty.
\end{cases}
\end{equation*}
Putting it together,
\begin{equation*}
\sum_{i \in \mathcal{D}_{t-}^{n}} \mathbb{E} \Big[ \big( \zeta_{i}^{n}(5) \big)^{2} \mid \mathcal{F}_{(i-1) \Delta_{n}} \Big] \overset{p}{ \longrightarrow}
\begin{cases}
0, & \text{if } \delta_{n} \sqrt{n} \rightarrow 0, \\[0.10cm]
\displaystyle \frac{1}{3} \nu_{t}^{2} & \text{if } \delta_{n} \sqrt{n} \rightarrow \theta \text{ or } \delta_{n} \sqrt{n} \rightarrow \infty,
\end{cases}
\end{equation*}
which implies that
\begin{equation*}
\sum_{i \in \mathcal{D}_{t-}^{n}} \mathbb{E} \Big[ \big( \zeta_{i}^{n}(3) \big)^{2} + \big( \zeta_{i}^{n}(5) \big)^{2} \mid \mathcal{F}_{(i-1) \Delta_{n}} \Big]   \overset{p}{ \longrightarrow}
\begin{cases}
\mu_{t-}, & \text{if } \delta_{n} \sqrt{n} \rightarrow 0, \\[0.25cm]
\displaystyle \frac{1}{ \theta^{2}} \mu_{t-} + \frac{1}{3} \nu_{t}^{2} & \text{if } \delta_{n} \sqrt{n} \rightarrow \theta, \\[0.25cm]
\displaystyle \frac{1}{3} \nu_{t}^{2} & \text{if } \delta_{n} \sqrt{n} \rightarrow \infty.
\end{cases}
\end{equation*}
To establish the asymptotic distribution, we prove a Lindeberg condition of the form:
\begin{equation*}
\sum_{i \in \mathcal{D}_{t-}^{n}} \mathbb{E} \Big[ \big( \zeta_{i}^{n}(3) + \zeta_{i}^{n}(5) \big)^{2} \mathbf{1}_{ \big\{( \zeta_{i}{n}(3) + \zeta_{i}^{n}(5))^{2} \geq \epsilon \big\}} \mid \mathcal{F}_{(i-1) \Delta_{n}} \Big] \overset{a.s.}{ \longrightarrow} 0, \quad \forall \epsilon > 0.
\end{equation*}
By the Cauchy-Schwarz and Chebyshev's inequalities,
\begin{align*}
\mathbb{E} \Big[ \big( \zeta_{i}^{n}(3) + \zeta_{i}^{n}(5) \big)^{2} & \mathbf{1}_{ \big\{( \zeta_{i}^{n}(3) + \zeta_{i}^{n}(5))^{2} \geq \epsilon \big\}} \mid \mathcal{F}_{(i-1) \Delta_{n}} \Big] \\[0.10cm] &\leq \sqrt{ \mathbb{E} \Big[ \big( \zeta_{i}^{n}(3) + \zeta_{i}^{n}(5) \big)^{4} \mid \mathcal{F}_{(i-1) \Delta_{n}} \Big] \mathbb{E} \Big[ \mathbf{1}_{ \big\{( \zeta_{i}^{n}(3) + \zeta_{i}^{n}(5))^{2} \geq \epsilon \big\}} \mid \mathcal{F}_{(i-1) \Delta_{n}} \Big]} \\[0.10cm]
&= \sqrt{ \mathbb{E} \Big[ \big( \zeta_{i}^{n}(3) + \zeta_{i}^{n}(5) \big)^{4} \mid \mathcal{F}_{(i-1) \Delta_{n}} \Big] \mathbb{P} \Big( \big( \zeta_{i}^{n}(3) + \zeta_{i}^{n}(5) \big)^{2} \geq \epsilon \mid \mathcal{F}_{(i-1) \Delta_{n}} \Big)} \\[0.10cm]
&\leq \frac{1}{ \epsilon} \mathbb{E} \Big[ \big( \zeta_{i}^{n}(3) + \zeta_{i}^{n}(5) \big)^{4} \mid \mathcal{F}_{(i-1) \Delta_{n}} \Big] \\[0.10cm]
&\leq \frac{8}{ \epsilon} \left( \mathbb{E} \Big[ \big( \zeta_{i}^{n}(3) \big)^{4} \mid \mathcal{F}_{(i-1) \Delta_{n}} \Big] + \mathbb{E} \Big[ \big( \zeta_{i}^{n}(5) \big)^{4} \mid \mathcal{F}_{(i-1) \Delta_{n}} \Big] \right),
\end{align*}
where the last line follows from the inequality $|x+y|^{p} \leq 2^{p-1}(|x|^{p} + |y|^{p})$, for all real $x$ and $y$ and any $p \geq 1$.

Since $\Delta_i \widetilde{N}^n$ follows a Poisson distribution with bounded intensity:
\begin{align*}
\sum_{i \in \mathcal{D}_{t-}^{n}} \mathbb{E} \Big[ \big( \zeta_{i}^{n}(3) \big)^{4} \mid \mathcal{F}_{(i-1) \Delta_{n}} \Big] &= \sum_{i \in \mathcal{D}_{t-}^{n}} \frac{ \rho_{n}^{2}}{ \ell_{n}^{4}} \mathbb{E} \Big[ \Delta_{i} \widetilde{N}^{n} - \mu_{(i-1) \Delta_{n}}^{4} \mid \mathcal{F}_{(i-1) \Delta_{n}} \Big] \\[0.10cm]
&= \sum_{i \in \mathcal{D}_{t-}^{n}} \frac{ \rho_{n}^{2}}{ \ell_{n}^{4}} \mu_{(i-1) \Delta_{n}}(1 + 3 \mu_{(i-1) \Delta_{n}}) \\[0.10cm]
& \leq C \frac{ \rho_{n}^{2}}{ \ell_{n}^{3}} \rightarrow 0,
\end{align*}
for both choices of $\rho_{n}$.

On the other hand, using the boundedness of $\nu$,
\begin{align*}
\sum_{i \in \mathcal{D}_{t-}^{n}} \mathbb{E} \Big[ \big( \zeta_{i}^{n}(4) \big)^{4} \mid \mathcal{F}_{(i-1) \Delta_{n}} \Big] &= \sum_{i \in \mathcal{D}_{t-}^{n}} \frac{ \rho_{n}^{2}}{ \ell_{n}^{4}}(tn- \ell_{n}-i)^{4} \ \mathbb{E} \Big[ \big( M_{i \Delta_{n}} - M_{(i-1) \Delta_{n}} \big)^{4} \mid \mathcal{F}_{(i-1) \Delta_{n}} \Big] \\[0.10cm]
&= 3 \frac{ \rho_{n}^{2}}{ \ell_{n}^{4}} \sum_{i \in \mathcal{D}_{t-}^{n}} (tn- \ell_{n}-i)^{4} \ \mathbb{E} \Bigg[ \bigg( \int_{(i-1) \Delta_{n}}^{i \Delta_{n}} \nu_{s}^{2} \mathrm{d}s \bigg)^{2} \mid \mathcal{F}_{(i-1) \Delta_{n}} \Bigg] \\[0.10cm]
&\leq C \frac{ \rho_{n}^{2} \Delta_{n}^{2}}{ \ell_{n}^{4}} \sum_{i \in \mathcal{D}_{t-}^{n}}(tn- \ell_{n}-i)^{4} = C \frac{ \rho_{n}^{2} \Delta_{n}^{2}}{ \ell_{n}^{4}} \frac{6 \ell_{n}^{5}+15 \ell_{n}^{4}+10 \ell_{n}^{3}- \ell_{n}}{30} \rightarrow 0,
\end{align*}
again for both choices of $\rho_{n}$. Hence, Lindeberg's condition holds.

By Lemma \ref{lemma:stable-convergernce}, we therefore conclude that
\begin{equation*}
\Lambda_{3}^{n}(t) + \Lambda_{5}^{n}(t)  \overset{ \mathcal{D}_{s}}{ \longrightarrow}
\begin{cases}
\sqrt{ \mu_{t-}} \mathcal{Z}, & \text{if } \delta_{n} \sqrt{n} \rightarrow 0, \\[0.10cm]
\displaystyle \sqrt{ \frac{1}{ \theta^{2}} \mu_{t-} + \frac{1}{3} \nu_{t}^{2}} \mathcal{Z}, & \text{if } \delta_{n} \sqrt{n} \rightarrow \theta, \\[0.10cm]
\displaystyle \sqrt{ \frac{1}{3} \nu_{t}^{2}} \mathcal{Z}, & \text{if } \delta_{n} \sqrt{n} \rightarrow \infty,
\end{cases}
\end{equation*}
where $\mathcal{Z} \sim N(0,1)$ independent of $\mathcal{F}$.

To end the proof, we next demonstrate asymptotic negligibility of the remaining terms. We start with $\Lambda_{1}^{n}(t)$, which we express as follows:
\begin{equation*}
\Lambda_{1}^{n}(t) = \sum_{i \in \mathcal{D}_{t-}^{n}} \zeta_{i}^{n}(1),
\end{equation*}
where
\begin{equation*}
\zeta_{i}^{n}(1) = \frac{ \sqrt{ \rho_{n}}}{ \ell_{n}} \Big( \Delta_{i} N^{n} - \Delta_{i} \widetilde{N}^{n} - \mathbb{E} \Big[ \Delta_{i} N^{n} - \Delta_{i} \widetilde{N}^{n} \mid \mathcal{F}_{(i-1) \Delta_{n}} \Big] \Big)
\end{equation*}
is an $\mathcal{F}_{(i-1) \Delta_{n}}$-martingale difference sequence by design. Hence,
\begin{align*}
\mathbb{E} \big[| \Lambda_{1}^{n}(t)|^{2} \big] &= \frac{ \rho_{n}}{ \ell_{n}^{2}} \sum_{i \in \mathcal{D}_{t-}^{n}} \mathbb{E} \bigg[ \Big( \Delta_{i} N^{n} - \Delta_{i} \widetilde{N}^{n} - \mathbb{E} \big[ \Delta_{i} N^{n} - \Delta_{i} \widetilde{N}^{n} \mid \mathcal{F}_{(i-1) \Delta_{n}} \big] \Big)^{2} \bigg] \\[0.10cm]
&= \frac{ \rho_{n}}{ \ell_{n}^{2}} \sum_{i \in \mathcal{D}_{t-}^{n}} \mathbb{E} \bigg[n \int_{(i-1) \Delta_{n}}^{i \Delta_{n}} \mu_{s} - \mu_{(i-1) \Delta_{n}} \mathrm{d}s \bigg] \\[0.10cm]
&\leq \frac{ \rho_{n} n}{ \ell_{n}^{2}} \sum_{i \in \mathcal{D}_{t-}^{n}} \int_{(i-1) \Delta_{n}}^{i \Delta_{n}} \mathbb{E} \big[| \mu_{s} - \mu_{(i-1) \Delta_{n}}| \big] \mathrm{d}s,
\end{align*}
where the tower property of conditional expectation was used. Now, Assumption \ref{assumption:mu} and standard estimates for semimartingales imply the existence of a constant $C>0$, such that
\begin{equation} \label{equation:semimartingale}
\mathbb{E} \big[| \mu_{s} - \mu_{(i-1) \Delta_{n}}| \big] \leq C \sqrt{ \Delta_{n}}.
\end{equation}
Thus, for any definition of $\rho_{n}$,
\begin{equation*}
\mathbb{E} \big[| \Lambda_{1}^{n}(t)|^{2} \big] \leq \frac{ \rho_{n} C \sqrt{ \Delta_{n}}}{ \ell_{n}} \rightarrow 0,
\end{equation*}
which implies the asymptotic negligibility of $\Lambda_{1}^{n}(t)$.

Next,
\begin{equation*}
\Lambda_{2}^{n}(t) = \sum_{i \in \mathcal{D}_{t-}^{n}} \zeta_{i}^{n}(2),
\end{equation*}
where
\begin{equation*}
\zeta_{i}^{n}(2) = \frac{ \sqrt{ \rho_{n}}}{ \ell_{n}} \mathbb{E} \Big[ \Delta_{i} N^{n} - \Delta_{i} \widetilde{N}^{n} \mid \mathcal{F}_{(i-1) \Delta_{n}} \Big].
\end{equation*}
Employing the estimate in \eqref{equation:semimartingale},
\begin{equation*}
\mathbb{E} \big[| \Lambda_{2}^{n}(t)| \big] \leq \frac{ \sqrt{ \rho_{n}}}{ \ell_{n}} \sum_{i \in \mathcal{D}_{t-}^{n}} \mathbb{E} \bigg[| n \int_{(i-1) \Delta_{n}}^{i \Delta_{n}} \mu_{s} - \mu_{(i-1) \Delta_{n}} \mathrm{d}s| \mid \mathcal{F}_{(i-1) \Delta_{n}} \bigg] \leq C \sqrt{ \rho_{n} \Delta_{n}} \rightarrow 0,
\end{equation*}
so $\Lambda_{2}^{n}(t)$ also vanishes.

Now, $\Lambda_{4}^{n}(t)$ can be expressed as follows:
\begin{equation*}
\Lambda_{4}^{n}(t) = \sum_{i \in \mathcal{D}_{t-}^{n}} (tn- \ell_{n}-i) \big( \zeta_{i}^{n}(4,1) + \zeta_{i}^{n}(4,2) \big),
\end{equation*}
where
\begin{equation*}
\zeta_{i}^{n}(4,1) = \frac{ \sqrt{ \rho_{n}}}{ \ell_{n}}  \int_{(i-1) \Delta_{n}}^{i \Delta_{n}} a_{s}^{*} \mathrm{d}s,
\end{equation*}
and
\begin{equation*}
\zeta_{i}^{n}(4, 2) = \frac{ \sqrt{ \rho_{n}}}{ \ell_{n}} \int_{(i-1) \Delta_{n}}^{i \Delta_{n}} \int_{ \mathbb{R}} \delta_{s}(x) \left( \mathfrak{p}( \mathrm{d}s, \mathrm{d}x) - \mathfrak{q}( \mathrm{d}s, \mathrm{d}x) \right).
\end{equation*}
Yet again, employing standard estimates for semimartingales from \citet*{jacod-protter:12a} means that there exists a constant $C > 0$ such that
\begin{equation*}
\mathbb{E} \left[ \Big| \int_{(i-1) \Delta_{n}}^{i \Delta_{n}} a_{s}^{*} \mathrm{d}s \Big| \right] \leq C \Delta_{n},
\end{equation*}
so that
\begin{align*}
\mathbb{E} \left[ \bigg| \sum_{i \in \mathcal{D}_{t-}^{n}} (tn- \ell_{n}-i) \zeta_{i}^{n}(4,1) \bigg| \right] & \leq C \frac{ \sqrt{ \rho_{n}}}{ \ell_{n}} \sum_{i \in \mathcal{D}_{t-}^{n}} (tn- \ell_{n}-i) \Delta_{n} \\
&= C \frac{ \Delta_{n} \sqrt{ \rho_{n}}}{ \ell_{n}} \sum_{j=1}^{ \ell_{n}} j \\
&= C \Delta_{n} \sqrt{ \rho_{n}} ( \ell_{n}+1) \\[0.10cm]
&\sim C \,
\begin{cases}
\delta_{n}^{3/2} \sqrt{n} & \text{if } \delta_{n} \sqrt{n} \rightarrow 0, \\[0.10cm]
\delta_{n}^{1/2} & \text{if } \delta_{n} \sqrt{n} \rightarrow \theta \text{ or } \delta_{n} \sqrt{n} \rightarrow \infty.
\end{cases}
\end{align*}
Thus,
\begin{equation*}
\mathbb{E} \left[ \bigg| \sum_{i \in \mathcal{D}_{t-}^{n}} (tn- \ell_{n}-i) \zeta_{i}^{n}(4,1) \bigg| \right] \rightarrow 0.
\end{equation*}
This implies that the first sum in $\Lambda_{4}^{n}(t)$ is asymptotically negligible.

The second sum in $\Lambda_{4}^{n}(t)$ is denoted by $R^{n}(t) = \sum_{i \in \mathcal{D}_{t-}^{n}} (tn- \ell_{n}-i) \zeta_{i}^{n}(4,2)$ and observe that $\zeta_{i}^{n}(4,2)$ is a pure-jump martingale with $\mathbb{E} \left[ \zeta_{i}^{n}(4,2) \right] = 0$. We take a $\kappa \in(0,1)$ and decompose $\zeta_{i}^{n}(4,2)$ into a ``small jump'' and ``big jump'' component defined through the bounding function $\bar{ \Gamma}$, $\zeta_{i}^{n}(4,2) = \chi_{i}^{n, \prime}( \kappa) + \chi_{i}^{n, \prime \prime}( \kappa)$, where
\begin{equation*}
\chi_{i}^{n, \prime}( \kappa) = \frac{ \sqrt{ \rho_{n}}}{ \ell_{n}} \int_{(i-1) \Delta_{n}}^{i \Delta_{n}} \int_{ \left\{x: \bar{ \Gamma}(x) \leq \kappa \right\}} \delta_{s}(x) \big( \mathfrak{p}( \mathrm{d}s, \mathrm{d}x) - \mathfrak{q}( \mathrm{d}s, \mathrm{d}x) \big),
\end{equation*}
and
\begin{equation*}
\chi_{i}^{n, \prime \prime}( \kappa) = \frac{ \sqrt{ \rho_{n}}}{ \ell_{n}} \int_{(i-1) \Delta_{n}}^{i \Delta_{n}} \int_{ \left\{x: \bar{ \Gamma}(x) > \kappa \right\}} \delta_{s}(x) \big( \mathfrak{p}( \mathrm{d}s, \mathrm{d}x) - \mathfrak{q}( \mathrm{d}s, \mathrm{d}x) \big).
\end{equation*}
Next, define the subset $\Omega_{n}( \psi, \kappa) \subseteq \Omega$ such that the Poisson process $\mathfrak{p} \big([0,t] \times \{x: \overline{ \Gamma}(x) > \kappa \} \big)$ has no jumps in the interval $[t- \ell_{n}^{ \psi} \Delta_{n},t]$ for all $0 < \psi < 1$ and $0 < \kappa < 1$. As $n \rightarrow \infty$, we note that $\Omega_{n}( \psi, \kappa) \rightarrow \Omega$, because $\ell_{n}^{ \psi} \Delta_{n} \rightarrow 0$.

Now, for all $c > 0$, the law of total probability and Markov's inequality imply that
\begin{align*}
\mathbb{P} \big(|R^{n}(t)| > c \big) &= \mathbb{P} \Big( |R^{n}(t)| > c \mid \Omega_{n}^{ \complement}( \psi, \kappa) \Big) + \mathbb{P} \big(|R^{n}(t)| > c \mid \Omega_{n}( \psi, \kappa) \big) \\
&\leq \mathbb{P} \Big(|R^{n}(t)| > c \mid \Omega_{n}^{ \complement}( \psi, \kappa) \Big) + \frac{ \mathbb{E} \left[ \big( R^{n}(t) \big)^{2} \mid \Omega_{n}( \psi, \kappa)\right]}{c^{2}},
\end{align*}
where $\Omega_{n}^{ \complement}( \psi, \kappa)$ is the complement of $\Omega_{n}( \psi, \kappa)$.

As there are no common jumps in $\chi_{i}^{n, \prime}( \kappa)$ and $\chi_{i}^{n, \prime \prime}( \kappa)$:
\begin{equation*}
\mathbb{E} \left[ \big( R^{n}(t) \big)^{2} \mid \Omega_{n}( \psi, \kappa) \right] = \sum_{i \in \mathcal{D}_{t-}^{n}} (tn- \ell_{n}-i)^{2} \left( \mathbb{E} \left[ \left( \chi_{i}^{n, \prime}( \kappa) \right)^{2} \mid \Omega_{n}( \psi, \kappa) \right] + \mathbb{E} \left[ \left( \chi_{i}^{n, \prime \prime}( \kappa) \right)^{2} \mid \Omega_{n}( \psi, \kappa) \right] \right).
\end{equation*}
The small jump term can be controlled as
\begin{equation*}
\mathbb{E} \left[ \left( \chi_{i}^{n, \prime}( \kappa) \right)^{2} \right] \leq C \frac{ \rho_{n}}{ \ell_{n}^{2}} \Delta_{n} \int_{ \left\{x: \bar{ \Gamma}(x) \leq \kappa \right\}} \bar{ \Gamma}(x)^{2} F( \mathrm{d}x).
\end{equation*}
Thus,
\begin{align*}
\sum_{i \in \mathcal{D}_{t-}^{n}} (tn- \ell_{n}-i)^{2} \mathbb{E} \left[ \left( \chi_{i}^{n, \prime}( \kappa) \right)^{2} \right] &\leq   C \frac{ \rho_{n}}{ \ell_{n}^{2}} \Delta_{n} \int_{ \left\{x: \bar{ \Gamma}(x) \leq \kappa \right\}} \bar{ \Gamma}(x)^{2} F( \mathrm{d}x) \sum_{i \in \mathcal{D}_{t-}^{n}} (tn- \ell_{n}-i)^{2} \\
&= C \bigg[ \frac{ \rho_{n} \Delta_{n} \ell_{n}( \ell_{n}+1)(2 \ell_{n}+1)}{6 \ell_{n}^{2}} \bigg] \int_{ \left\{x: \bar{ \Gamma}(x) \leq \kappa \right\}} \bar{ \Gamma}(x)^{2}F( \mathrm{d}x),
\end{align*}
where
\begin{equation*}
\bigg[ \frac{ \rho_{n} \Delta_{n} \ell_{n}( \ell_{n}+1)(2 \ell_{n}+1)}{6 \ell_{n}^{2}} \bigg] \sim \rho_{n} \delta_{n} =
\begin{cases}
\delta_{n}^{2} n, & \text{if } \delta_{n} \sqrt{n} \rightarrow 0, \\
1, &  \text{if } \delta_{n} \sqrt{n} \rightarrow \theta \text{ or } \delta_{n} \sqrt{n} \rightarrow \infty.
\end{cases}
\end{equation*}
Furthermore, because there are no big jumps (larger than $\kappa$) in the interval $[t- \ell_{n}^{ \psi} \Delta_{n},t]$ on the event $\Omega_{n}( \psi, \kappa)$, we can also conclude that:
\begin{align*}
\sum_{i \in \mathcal{D}_{t-}^{n}} (tn- \ell_{n}-i)^{2} \mathbb{E} \left[ \left( \chi_{i}^{n, \prime \prime}( \kappa) \right)^{2} \mid \Omega_{n}( \psi, \kappa) \right] &= \sum_{i \in \mathcal{D}_{t-}^{n}, i \Delta_{n} \leq t - \ell_{n}^{ \psi} \Delta_{n}} (tn- \ell_{n}-i)^{2} \mathbb{E} \left[ \left( \chi_{i}^{n, \prime \prime}( \kappa) \right)^{2} \right] \\
&\leq C \frac{ \rho_{n}}{ \ell_{n}^{2}} \Delta_{n} \sum_{i \in \mathcal{D}_{t-}^{n}, i \Delta_{n} \leq t - \ell_{n}^{ \psi} \Delta_{n}} (tn- \ell_{n}-i)^{2},
\end{align*}
where the sum on the right-hand side has the following form:
\begin{equation}
\sum_{i \in \mathcal{D}_{t-}^{n}, i \Delta_{n} \leq t - \ell_{n}^{ \psi} \Delta_{n}} (tn- \ell_{n}-i)^{2} = \sum_{j=1}^{ \ell_{n}^{ \psi}} j^{2} = \frac{ \ell^{ \psi}_{n}( \ell^{ \psi}_{n}+1)(2 \ell^{ \psi}_{n}+1)}{6} \sim \ell_{n}^{3 \psi}.
\end{equation}
This produces
\begin{equation}
\sum_{i \in \mathcal{D}_{t-}^{n}} (tn- \ell_{n}-i)^{2} \mathbb{E} \left[ \left( \chi_{i}^{n, \prime \prime}( \kappa) \right)^{2} \mid \Omega_{n}( \psi, \kappa) \right] \sim \rho_{n} \Delta_{n} \ell_{n}^{3 \psi-2} \rightarrow 0,
\end{equation}
and it thus follows that
\begin{equation}
\limsup_{n \rightarrow \infty} \mathbb{P} \big(|R^{n}(t)| > c \big) \leq \limsup_{n \rightarrow \infty} \mathbb{P} \big(|R^{n}(t) > c \mid \bar{ \Omega}_{n}( \psi, \kappa) \big) + \limsup_{n \rightarrow \infty} A_{n},
\end{equation}
where
\begin{equation}
A_{n} \sim \rho_{n} \delta_{n} \int_{ \left\{x: \bar{ \Gamma}(x) \leq \kappa \right\}} \bar{ \Gamma}(x)^{2} F( \mathrm{d}x) + \rho_{n} \Delta_{n} \ell_{n}^{3 \psi-2}.
\end{equation}
As $\kappa \rightarrow 0$, $\int_{ \left\{x: \bar{ \Gamma}(x) \leq \kappa \right\}} \bar{ \Gamma}(x)^{2} F( \mathrm{d}x) \rightarrow 0$, so that as $\kappa \rightarrow 0$ and $n \rightarrow \infty$, $A_{n} \rightarrow 0$. Moreover, as $n \rightarrow \infty$, $\Omega_{n}^{ \complement}( \psi, \kappa) \rightarrow \emptyset$, so $\limsup_{n \rightarrow \infty} \mathbb{P} \big(|R^{n}(t)| > c \mid \Omega_{n}^{ \complement}( \psi, \kappa) \big) = 0$. To conclude, $\limsup_{n \rightarrow \infty} \mathbb{P} \big(|R^{n}(t)| > c \big) = 0$ and, hence, $\lim_{n \rightarrow \infty} \mathbb{P} \big(|R^{n}(t)| > c \big) = 0$, as was to be demonstrated. \qed

\subsection{Proof of Theorem \ref{theorem:clt-smoothness}}

As in proof of Theorem \ref{theorem:clt}, we start with a decomposition of $\widehat{ \lambda}_{t} - \mu_{t-}$:
\begin{equation*}
\sqrt{ \ell_{n}} ( \widehat{ \lambda}_{t} - \mu_{t-}) = \Lambda_{1}^{n}(t)  + \Lambda_{2}^{n}(t)  + \Lambda_{3}^{n}(t) + \mathrm{(III)},
\end{equation*}
where $\Lambda_{1}^{n}(t)$ -- $\Lambda_{3}^{n}(t)$ are defined as above, and
\begin{align*}
\mathrm{(III)} &= \frac{1}{ \sqrt{ \ell_{n}}} \sum_{i \in \mathcal{D}_{t-}^{n}} (tn- \ell_{n}-i)( \mu_{i \Delta_{n}} - \mu_{(i-1) \Delta_{n}}).
\end{align*}
The treatment of the first triplet does not change from before, so it follows that $\Lambda_{1}^{n}(t)$ and $\Lambda_{2}^{n}(t)$ are asymptotically negligible, whereas $\Lambda_{3}^{n}(t)$ converges stably in law:
\begin{equation}
\Lambda_{3}^{n}(t) \overset{ \mathfrak{D}_{s}}{ \longrightarrow} \sqrt{ \mu_{t-}} \mathcal{Z}.
\end{equation}
The difference is that we do not split (III) into a drift and martingale part, $\Lambda_{4}^{n}(t)$ and $\Lambda_{5}^{n}(t)$, as there is no such structural assumption to work with now. So, our sole task is to prove (III) is asymptotically negligible under Assumption \ref{assumption:holder}.

We set $\zeta_{i}^{n} = \frac{1}{ \sqrt{ \ell_{n}}} (tn- \ell_{n}-i)( \mu_{i \Delta_{n}} - \mu_{(i-1) \Delta_{n}})$, such that $\mathrm{(III)} = \sum_{i \in \mathcal{D}_{t-}^{n}} \zeta_{i}^{n}$. It is then straightforward to show that:
\begin{align*}
\sum_{i \in \mathcal{D}_{t-}^{n}} \mathbb{E} \big[| \zeta_{i}^{n}| \big] &\leq C \Delta_{n}^{H^{ \mu}} \frac{1}{ \sqrt{ \ell_{n}}} \sum_{i \in \mathcal{D}_{t-}^{n}}  (tn- \ell_{n}-i) \\
&= C \Delta_{n}^{H^{ \mu}} \frac{1}{ \sqrt{ \ell_{n}}} \sum_{j=1}^{ \ell_{n}} j \\
&= C \Delta_{n}^{H^{ \mu}} \sqrt{ \ell_{n}}( \ell_{n}+1) \\[0.10cm]
&\sim \ell_{n}^{3/2} \Delta_{n}^{H^{ \mu}},
\end{align*}
which converges to zero so long as we select $\ell_{n} = O \big( n^{ \frac{2H^{ \mu}}{3} - \epsilon} \big)$, for some $0 < \epsilon < \frac{2H^{ \mu}}{3}$. This implies that (III) is asymptotically negligible. The upper bound on $\epsilon$ ensures that $\ell_{n} \rightarrow \infty$. \qed

\subsection{Proof of Theorem \ref{theorem:observed-asymptotic-variance}}

The observed asymptotic local variance can be expressed as
\begin{equation*}
\widetilde{ \mathsf{avar}} ( \nabla \widehat{ \lambda}_{t}) = \frac{ \rho_{n}}{K_{n}} \sum_{j=0}^{K_{n}-1} \big( \widehat{ \lambda}_{t_{j}} - \widehat{ \lambda}_{t_{j}- \ell_{n} \Delta_{n}} \big)^{2}.
\end{equation*}
where $t_{j} = t - 2j \ell_{n} \Delta_{n}$.

We denote by $\mathcal{D}_{t_{j}}^{n}$ the union of $\mathcal{D}_{t_{j}-}^{n}$ and $\mathcal{D}_{(t_{j}- \ell_{n} \Delta_{n})-}^{n}$. Then, for every $t_{j}$, as in the proof of Theorem \ref{theorem:clt} the local intensity estimator can be decomposed as:
\begin{equation*}
\sqrt{ \rho_{n}} \big( \widehat{ \lambda}_{t_{j}} - \widehat{ \lambda}_{t_{j}- \ell_{n} \Delta_{n}} \big) = \Lambda_{1}^{n}(t_{j}) + \Lambda_{2}^{n}(t_{j}) + \Lambda_{3}^{n}(t_{j}) + \Lambda_{4}^{n}(t_{j}) + \Lambda_{5}^{n}(t_{j}),
\end{equation*}
where
\begin{align*}
\Lambda_{1}^{n}(t_{j}) &= \frac{ \sqrt{ \rho_{n}}}{ \ell_{n}} \sum_{i \in \mathcal{D}_{t_{j}}^{n}} (-1)^{ \mathds{1}_{ \left\{ i<t_{j}- \ell_{n} \Delta_{n} \right\}}} \big( \Delta_{i} N^{n} - \Delta_{i} \widetilde{N}^{n} - \mathbb{E} \big[ \Delta_{i} N^{n} - \Delta_{i} \widetilde{N}^{n} \mid \mathcal{F}_{(i-1) \Delta_{n}} \big] \big), \\[0.10cm]
\Lambda_{2}^{n}(t_{j}) &= \frac{ \sqrt{ \rho_{n}}}{ \ell_{n}} \sum_{i \in \mathcal{D}_{t_{j}}^{n}} (-1)^{ \mathds{1}_{ \left\{ i<t_{j}- \ell_{n} \Delta_{n} \right\}}}\mathbb{E} \big[ \Delta_{i} N^{n} - \Delta_{i} \widetilde{N}^{n} \mid \mathcal{F}_{(i-1) \Delta_{n}} \big], \\[0.10cm]
\Lambda_{3}^{n}(t_{j}) &= \frac{ \sqrt{ \rho_{n}}}{ \ell_{n}} \sum_{i \in \mathcal{D}_{t_{j}}^{n}} (-1)^{ \mathds{1}_{ \left\{ i<t_{j}- \ell_{n} \Delta_{n} \right\}}}\big( \Delta_{i} \widetilde{N}^{n} - \mu_{(i-1) \Delta_{n}} \big), \\[0.10cm]
\Lambda_{4}^{n}(t_{j}) &= \frac{ \sqrt{ \rho_{n}}}{ \ell_{n}} \sum_{i \in \mathcal{D}_{t_{j}}^{n}} (-1)^{ \mathds{1}_{ \left\{ i<t_{j}- \ell_{n} \Delta_{n} \right\}}}(t_{j}n- 2 \ell_{n}-i)(A_{i \Delta_{n}} - A_{(i-1) \Delta_{n}}).
\\[0.10cm]
\Lambda_{5}^{n}(t_{j}) &= \frac{ \sqrt{ \rho_{n}}}{ \ell_{n}} \sum_{i \in \mathcal{D}_{t_{j}}^{n}} (-1)^{ \mathds{1}_{ \left\{ i<t_{j}- \ell_{n} \Delta_{n} \right\}}}(t_{j}n- 2 \ell_{n}-i)(M_{i \Delta_{n}} - M_{(i-1) \Delta_{n}}),
\end{align*}
with $N_{t}$, $M_{t}$, and $A_{t}$ defined as in Theorem \ref{theorem:clt}.

Then,
\begin{equation*}
\widetilde{ \mathsf{avar}}( \nabla \widehat{ \lambda}_{t})  = \mathcal{E}_{1}^{n}(t) + \mathcal{E}_{2}^{n}(t) + \mathcal{R}^{n}(t),
\end{equation*}
where
\begin{align*}
\mathcal{E}_{1}^{n}(t) &= \frac{1}{K_{n}} \sum^{K_{n}-1}_{j=0} \big( \Lambda_{3}^{n}(t_{j}) \big)^{2}, \\[0.10cm]
\mathcal{E}_{2}^{n}(t) &= \frac{1}{K_{n}} \sum^{K_{n}-1}_{j=0} \big( \Lambda_{5}^{n}(t_{j}) \big)^{2}, \\[0.10cm]
\mathcal{R}^{n}(t) &= \widetilde{ \mathsf{avar}}( \nabla \widehat{ \lambda}_{t})- \mathcal{E}_{1}^{n}(t)- \mathcal{E}_{2}^{n}(t).
\end{align*}
What is left amounts to showing that $\mathcal{E}_{1}^{n}(t)$ and $\mathcal{E}_{2}^{n}(t)$ converge to the first and the second term in the true asymptotic variance, whereas $\mathcal{R}^{n}(t) \overset{p}{ \longrightarrow} 0$.

For the first term, we observe that
\begin{equation*}
\big( \Lambda_{3}^{n}(t_{j}) \big)^{2} = \sum_{i \in \mathcal{D}_{t_{j}}^{n}} \big( \zeta_{i}^{n}(3) \big)^{2} + 2 \sum_{s,i \in \mathcal{D}_{t_{j}}^{n}: s>i} \zeta_{i}^{n}(3) \zeta_{s}^{n}(3) (-1)^{  \mathds{1}_{ \left\{i<t_{j}- \ell_{n} \Delta_{n} \right\}} +  \mathds{1}_{ \left\{s<t_{j}- \ell_{n} \Delta_{n} \right\}}},
\end{equation*}
where $\zeta_{i}^{n}(3) = \displaystyle \frac{ \sqrt{ \rho_{n}}}{ \ell_{n}} \big( \Delta_{i} \widetilde{N}^{n} - \mu_{(i-1) \Delta_{n}} \big)$ as above. Since $\zeta_{i}^{n}(3)$ is a $\mathcal{F}_{i \Delta_{n}}$-martingale and
\begin{equation*}
\mathbb{E} \Big[ \big( \zeta_{i}^{n}(3) \big)^{2} \mid \mathcal{F}_{(i-1) \Delta_{n}} \Big] = \frac{ \rho_{n}}{ \ell_{n}^{2}} \mu_{(i-1) \Delta_{n}},
\end{equation*}
it follows that
\begin{equation*}
\mathbb{E} \bigg[ \big( \Lambda_{3}^{n}(t_{j}) \big)^{2} - \frac{ \rho_{n}}{ \ell_{n}^{2}} \sum_{i \in \mathcal{D}_{t_{j}}^{n}} \mu_{(i-1) \Delta_{n}} \bigg] = 0.
\end{equation*}
Then, we decompose $\mathcal{E}_{1}^{n}(t)$ into
\begin{equation*}
\mathcal{E}_{1}^{n}(t) = \mathcal{E}_{1}^{ \prime n}(t)+ \mathcal{E}_{1}^{ \prime \prime n}(t),
\end{equation*}
where
\begin{align*}
\mathcal{E}_{1}^{ \prime n}(t) &= \frac{1}{K_{n}} \sum_{j=0}^{K_{n}-1} \bigg( \big( \Lambda_{3}^{n}(t_{j}) \big)^{2} - \frac{ \rho_{n}}{ \ell_{n}^{2}} \sum_{i \in \mathcal{D}_{t_{j}}^{n}} \mu_{(i-1) \Delta_{n}} \bigg), \\[0.10cm]
\mathcal{E}_{1}^{ \prime \prime n}(t) &= \frac{1}{K_{n}} \frac{ \rho_{n}}{ \ell_{n}^{2}} \sum_{j=0}^{K_{n}-1} \sum_{i \in \mathcal{D}_{t_{j}}^{n}} \mu_{(i-1) \Delta_{n}}.
\end{align*}
So $\mathbb{E} \big[ \mathcal{E}_{1}^{ \prime n}(t) \big] = 0$, and because the $\Lambda_{3}^{n}(t_{j})$'s are based on non-overlapping blocks of observations, the variance of $\mathcal{E}_{1}^{ \prime n}(t)$ has the following form:
\begin{equation*}
\mathbb{E} \Big[ \big( \mathcal{E}_{1}^{ \prime n}(t) \big)^{2} \Big] = \frac{1}{K_{n}^{2}} \sum_{j=0}^{K_{n}-1} \mathbb{E} \bigg[ \Big( \big( \Lambda_{3}^{n}(t_{j}) \big)^{2} - \frac{ \rho_{n}}{ \ell_{n}^{2}} \sum_{i \in \mathcal{D}_{t_{j}}^{n}} \mu_{(i-1) \Delta_{n}} \Big)^{2} \bigg] \leq \frac{C}{K_{n}^{2}} \sum_{j=0}^{K_{n}-1} \mathbb{E} \Big[ \big( \Lambda_{3}^{n}(t_{j}) \big)^{4} \Big].
\end{equation*}
Next,
\begin{equation*}
\mathbb{E} \Big[ \big( \Lambda_{3}^{n}(t_{j}) \big)^{4} \Big] = \sum_{i \in \mathcal{D}_{t_{j}}^{n}} \mathbb{E} \Big[ \big( \zeta_{i}^{n}(3) \big)^{4} \Big] +  4 \sum_{s,i \in \mathcal{D}_{t_{j}}^{n}:s>i} \mathbb{E} \Big[ \big( \zeta_{i}^{n}(3) \zeta_{s}^{n}(3) \big)^{2} \Big].
\end{equation*}
Alluding to the boundedness of $\mu$, we get the estimates:
\begin{equation*}
\mathbb{E} \Big[ \big( \zeta_{i}^{n}(3) \big)^{2} \Big] \leq C \frac{ \rho_{n}}{ \ell_{n}^{2}} \quad \text{and} \quad \mathbb{E} \Big[ \big( \zeta_{i}^{n}(3) \big)^{4} \Big] \leq C \frac{ \rho_{n}^{2}}{ \ell_{n}^{4}}.
\end{equation*}
Based on this, we deduce that:
\begin{equation} \label{equation:estimate}
\mathbb{E} \Big[ \big( \Lambda_{3}^{n}(t_{j}) \big)^{4} \Big] \leq \frac{ \rho_{n}^{2}}{ \ell_{n}^{4}} \bigg( \sum_{i =1}^{2 \ell_{n}}C + 4 \sum_{i=1}^{2 \ell_{n}-1} \sum_{s=i+1}^{2 \ell_{n}} C \bigg) = C \frac{ \rho_{n}^{2}(2 \ell_{n} + 4 \ell_{n}(2 \ell_{n}-1))}{ \ell_{n}^{4}}.
\end{equation}
Ergo,
\begin{equation*}
\mathbb{E} \Big[ \big( \mathcal{E}_{1}^{ \prime n}(t) \big)^{2} \Big] \leq \frac{C}{K_{n}^{2}} \sum_{j=0}^{K_{n}-1} \frac{ \rho_{n}^{2}(2 \ell_{n} + 4 \ell_{n}(2 \ell_{n}-1))}{ \ell_{n}^{4}} = O \Big(K_{n}^{-1} \ell_{n}^{-2} \rho_{n}^{2} \Big),
\end{equation*}
so that $\mathcal{E}^{ \prime n}_{1}(t) \overset{p}{ \longrightarrow} 0$.

The second term, $\mathcal{E}_{1}^{ \prime \prime n}(t)$, can be represented as:
\begin{equation*}
\mathcal{E}_{1}^{ \prime \prime n}(t) = \frac{ \rho_{n}}{ \ell_{n}^{2}K_{n}} \sum_{s=1}^{2K_{n} \ell_{n}} \mu_{t-s \Delta_{n}} = 2 \frac{ \rho_{n}}{ \ell_{n}} \frac{1}{2K_{n} \ell_{n}} \sum_{s=1}^{2K_{n} \ell_{n}} \mu_{t-s \Delta_{n}}.
\end{equation*}
Since $K_{n} \ell_{n} \Delta_{n} \rightarrow 0$, following the train of thought in the proof of Theorem \ref{theorem:clt} implies that
\begin{equation*}
\mathcal{E}_{1}^{n}(t) \overset{p}{ \longrightarrow}
\begin{cases}
2 \mu_{t-}, & \text{if } \ell_{n} \sqrt{ \Delta_{n}} \rightarrow 0, \\[0.10cm]
\displaystyle \frac{2}{ \theta^{2}} \mu_{t-}, & \text{if } \ell_{n} \sqrt{ \Delta_{n}} \rightarrow \theta, \\[0.10cm]
0, & \text{if } \ell_{n} \sqrt{ \Delta_{n}} \rightarrow \infty.
\end{cases}
\end{equation*}
Now, moving on to $\mathcal{E}_{2}^{n}(t)$, we again observe that
\begin{equation*}
\big( \Lambda_{5}^{n}(t_{j}) \big)^{2} = \sum_{i \in \mathcal{D}_{t_{j}}^{n}} \big( \zeta_{i}^{n}(5) \big)^{2} + 2 \sum_{s,i \in \mathcal{D}_{t_{j}}^{n}: s>i} \zeta_{i}^{n}(5) \zeta_{s}^{n}(5) (-1)^{  \mathds{1}_{ \left\{i<t_{j}- \ell_{n} \Delta_{n} \right\}} +  \mathds{1}_{ \left\{s<t_{j}- \ell_{n} \Delta_{n} \right\}}},
\end{equation*}
where $\zeta_{i}^{n}(5) = \displaystyle \frac{ \sqrt{ \rho_{n}}}{ \ell_{n}} (t_{j}n-2 \ell_{n}-i)( M_{i \Delta_{n}} - M_{(i-1) \Delta_{n}})$. Since $\zeta_{i}^{n}(5)$ is $\mathcal{F}_{i \Delta_{n}}$-martingale and
\begin{equation*}
\mathbb{E} \Big[ \big( \zeta_{i}^{n}(5) \big)^{2} \mid \mathcal{F}_{(i-1) \Delta_{n}} \Big] = \frac{ \rho_{n}}{ \ell_{n}^{2}} (t_{j}n-2 \ell_{n}-i)^{2} \int_{(i-1) \Delta_{n}}^{i \Delta_{n}} \mathbb{E} \big[ \nu_{s}^{2} \mid \mathcal{F}_{(i-1) \Delta_{n}} \big] \mathrm{d}s,
\end{equation*}
we have:
\begin{equation*}
\mathbb{E} \bigg[ \big( \Lambda_{5}^{n}(t_{j}) \big)^{2} - \sum_{i \in \mathcal{D}_{t_{j}}^{n}} \frac{ \rho_{n}}{ \ell_{n}^{2}} (t_{j}n-2 \ell_{n}-i)^{2} \int_{(i-1) \Delta_{n}}^{i \Delta_{n}} \mathbb{E} \big[ \nu_{s}^{2} \mid \mathcal{F}_{(i-1) \Delta_{n}} \big] \mathrm{d}s \bigg] = 0.
\end{equation*}
Then, we decompose $\mathcal{E}_{2}^{n}(t)$ as the sum of four terms:
\begin{equation*}
\mathcal{E}_{2}^{n}(t) = \mathcal{E}_{2}^{ \prime n}(t) + \mathcal{E}_{2}^{ \prime \prime n}(t) + \mathcal{E}_{2}^{ \prime \prime \prime n}(t) + \mathcal{E}_{2}^{ \prime \prime \prime \prime n}(t),
\end{equation*}
where
\begin{align*}
\mathcal{E}^{ \prime n}_{2}(t) &= \frac{1}{K_{n}} \sum_{j=0}^{K_{n}-1} \bigg( \big( \Lambda_{5}^{n}(t_{j}) \big)^{2} - \frac{ \rho_{n}}{ \ell_{n}^{2}} \sum_{i \in \mathcal{D}_{t_{j}}^{n}} (t_{j}n-2 \ell_{n}-i)^{2} \int_{(i-1) \Delta_{n}}^{i \Delta_{n}} \mathbb{E} \big[ \nu_{s}^{2} \mid \mathcal{F}_{(i-1) \Delta_{n}} \big] \mathrm{d}s \bigg), \\[0.10cm]
\mathcal{E}^{ \prime \prime n}_{2}(t) &= \frac{ \rho_{n}}{ \ell_{n}^{2} K_{n}} \sum_{j=0}^{K_{n}-1} \sum_{i \in \mathcal{D}_{t_{j}}^{n}} (t_{j}n-2 \ell_{n}-i)^{2} \int_{(i-1) \Delta_{n}}^{i \Delta_{n}} \mathbb{E} \big[ \nu_{s}^{2} - \nu_{(i-1) \Delta_{n}}^{2} \mid \mathcal{F}_{(i-1) \Delta_{n}} \big] \mathrm{d}s, \\[0.10cm]
\mathcal{E}^{ \prime \prime \prime n}_{2}(t) &= \frac{ \rho_{n}}{ \ell_{n}^{2} K_{n}} \sum_{j=0}^{K_{n}-1} \sum_{i \in \mathcal{D}_{t_{j}}^{n}} (t_{j}n-2 \ell_{n}-i)^{2} \big( \nu_{(i-1) \Delta_{n}}^{2} - \nu_{t}^{2} \big) \Delta_{n}, \\[0.10cm]
\mathcal{E}^{ \prime \prime \prime \prime n}_{2}(t) &= \frac{ \rho_{n}}{ \ell_{n}^{2} K_{n}} \sum_{j=0}^{K_{n}-1} \sum_{i \in \mathcal{D}_{t_{j}}^{n}} (t_{j}n-2 \ell_{n}-i)^{2} \nu_{t}^{2} \Delta_{n}.
\end{align*}
By construction, $\mathbb{E}[ \mathcal{E}_{2}^{ \prime n}(t)] = 0$ and, as in the proof for $\mathcal{E}_{1}^{ \prime n}(t)$,
\begin{equation*}
\mathbb{E} \Big[ \big( \mathcal{E}_{2}^{ \prime n}(t) \big)^{2} \Big] \leq \frac{C}{K_{n}} \sum_{j=0}^{K_{n}-1} \mathbb{E} \Big[ \big( \Lambda_{5}^{n}(t_{j}) \big)^{4} \Big],
\end{equation*}
where, since $\zeta_{i}^{n}(5)$ in the definition of $\Lambda_{5}^{n}(t_{j})$ is an $\mathcal{F}_{i \Delta_{n}}$-martingale,
\begin{equation*}
\mathbb{E} \Big[ \big( \Lambda_{5}^{n}(t_{j}) \big)^{4} \Big] = \sum_{i \in \mathcal{D}_{t_{j}}^{n}} \mathbb{E} \Big[ \big( \zeta_{i}^{n}(5) \big)^{4} \Big] +  4 \sum_{s,i \in \mathcal{D}_{t_{j}}^{n}:s>i} \mathbb{E} \Big[ \big( \zeta_{i}^{n}(5) \zeta_{s}^{n}(5) \big)^{2} \Big].
\end{equation*}
From the proof of Theorem \ref{theorem:clt}:
\begin{equation*}
\mathbb{E} \Big[ \big( \zeta_{i}^{n}(5) \big)^{4} \Big] \leq \frac{C \rho_{n}^{2} \Delta_{n}^{2}}{ \ell_{n}^{4}} \sum_{i \in \mathcal{D}_{t_{j}}^{n}} (t_{j}n-2 \ell_{n}-i)^{4} \quad \text{and} \quad \mathbb{E} \Big[ \big( \zeta_{i}^{n}(5) \big)^{2} \Big] \leq \frac{C \rho_{n} \Delta_{n}}{ \ell_{n}^{2}} \sum_{i \in \mathcal{D}_{t_{j}}^{n}} (t_{j}n-2 \ell_{n}-i)^{2}.
\end{equation*}
Standard formulas for calculating the sum of powers of integers yield
\begin{equation*}
\sum_{i \in \mathcal{D}_{t_{j}}^{n}} (t_{j}n-2 \ell_{n}-i)^{4} = \frac{96 \ell_{n}^{5} + 120 \ell_{n}^{4} +40 \ell_{n}^{3} - \ell_{n}}{15},
\end{equation*}
and
\begin{equation} \label{equation:sum}
\sum_{i \in \mathcal{D}_{t_{j}}^{n}} (t_{j}n-2 \ell_{n}-i)^{2} = \frac{8 \ell_{n}^{3} + 6 \ell_{n}^{2} + \ell_{n}}{3}.
\end{equation}
Hence,
\begin{equation*}
\mathbb{E} \Big[ \big( \zeta_{i}^{n}(5) \big)^{4} \Big] = O( \rho_{n}^{2} \Delta_{n}^{2} \ell_{n}),
\end{equation*}
and it follows that $\mathcal{E}_{2}^{ \prime n}(t) = o_{p}(1)$.

For the second term,
\begin{equation*}
\mathbb{E} \big[ | \mathcal{E}^{\prime \prime n}_{2}(t) | \big] \leq \overline{M} \frac{ \rho_{n} \Delta_{n}}{ \ell_{n}^{2}K_{n}} \sum_{j=0}^{K_{n}-1} \sum_{i \in \mathcal{D}_{t_{j}}^{n}} (t_{j}n-2 \ell_{n}-i)^{2},
\end{equation*}
where $\overline{M} = \mathbb{E} \Big[ \sup_{s \in[t-K_{n} \ell_{n} \Delta_{n},t]} | \nu_{s}^{2} - \nu_{t}^{2}| \Big]$. Since $\nu$ is c\`{a}dl\`{a}g and $K_{n} \ell_{n} \Delta_{n} \rightarrow 0$, $\overline{M} = o_{p}(1)$. The sum was computed in \eqref{equation:sum}, so combining the terms shows that
\begin{equation*}
\frac{\rho_{n} \Delta_{n}}{ \ell_{n}^{2}K_{n}} \sum_{j=0}^{K_{n}-1} \sum_{i \in \mathcal{D}_{t_{j}}^{n}} (t_{j}n-2 \ell_{n}-i)^{2} = \frac{8}{3} \rho_{n} \ell_{n} \Delta_{n} + o(1),
\end{equation*}
where
\begin{equation*}
\frac{8}{3} \rho_{n} \ell_{n} \Delta_{n} \rightarrow
\begin{cases}
0, & \text{if } \ell_{n} \sqrt{ \Delta_{n}} \rightarrow 0, \\[0.10cm]
\displaystyle \frac{8}{3}, & \text{if } \ell_{n} \sqrt{ \Delta_{n}} \rightarrow \theta > 0 \text{ or } \ell_{n} \sqrt{ \Delta_{n}} \rightarrow \infty.
\end{cases}
\end{equation*}
As such, $\mathcal{E}_{2}^{ \prime \prime n}(t) = o_{p}(1)$.

Following the above steps, we also deduce that
\begin{equation*}
\mathcal{E}_{2}^{ \prime \prime \prime n}(t) \overset{p}{ \longrightarrow} 0.
\end{equation*}
The last term can be written as
\begin{equation*}
\mathcal{E}_{2}^{ \prime \prime \prime \prime n}(t) = \nu_{t}^{2} \Bigg( \frac{ \rho_{n} \Delta_{n}}{ \ell_{n}^{2}K_{n}} \sum_{j=0}^{K_{n}-1} \sum_{i \in \mathcal{D}_{t_{j}}^{n}} (t_{j}n-2 \ell_{n}-i)^{2} \Bigg),
\end{equation*}
where the sum on the right-hand side is given by \eqref{equation:sum}. As a consequence,
\begin{equation*}
\mathcal{E}_{2}^{ \prime  \prime \prime \prime n}(t) \overset{p}{ \longrightarrow}
\begin{cases}
0, & \text{if } \ell_{n} \sqrt{ \Delta_{n}} \rightarrow 0, \\[0.10cm]
\displaystyle \frac{8}{3} \nu_{t}^{2}, & \text{if } \ell_{n} \sqrt{ \Delta_{n}} \rightarrow \theta>0 \text{ or } \ell_{n} \sqrt{ \Delta_{n}} \rightarrow \infty,
\end{cases}
\end{equation*}
which handles the analysis of $\mathcal{E}_{2}^{n}(t)$.

To finalize the proof up, we notice that $\mathcal{R}^{n}(t)$ is an average of terms that all converge in probability to zero, and therefore it also converges in probability to zero. \qed

\subsection{Proof of Theorem \ref{observed_variance_LLN2}}

We largely copy from the proof of Theorem \ref{theorem:observed-asymptotic-variance}. Therefore, many repeated details are omitted and we concentrate on explaining the main differences.

The observed asymptotic local variance can again be expressed as follows:
\begin{equation*}
\widehat{ \mathsf{avar}}( \nabla \widehat{ \lambda}_{t}) = \frac{ \rho_{n}}{K_{n}} \sum_{j=0}^{K_{n}-1} \big( \widehat{ \lambda}_{t_{j}} - \widehat{ \lambda}_{t_{j}- \ell_{n} \Delta_{n}} \big)^{2},
\end{equation*}
where $t_{j} = t - j \Delta_{n}$.

This can further be split into
\begin{equation*}
\widehat{ \mathsf{avar}}( \nabla \widehat{ \lambda}_{t})  = \mathcal{E}_{1}^{n}(t) + \mathcal{E}_{2}^{n}(t) + \mathcal{R}^{n}(t),
\end{equation*}
where
\begin{align*}
\mathcal{E}_{1}^{n}(t) &= \frac{1}{K_{n}} \sum^{K_{n}-1}_{j=0} \big( \Lambda_{3}^{n}(t_{j}) \big)^{2}, \\[0.10cm]
\mathcal{E}_{2}^{n}(t) &= \frac{1}{K_{n}} \sum^{K_{n}-1}_{j=0} \big( \Lambda_{5}^{n}(t_{j}) \big)^{2}, \\[0.10cm]
\mathcal{R}^{n}(t) &= \widehat{ \mathsf{avar}}( \nabla \widehat{ \lambda}_{t})- \mathcal{E}_{1}^{n}(t)- \mathcal{E}_{2}^{n}(t).
\end{align*}

The only difference here compared to the proof of Theorem \ref{theorem:observed-asymptotic-variance} is the definition of $t_{j}$. It therefore follows that $\mathcal{R}^{n}(t) = o_{p}(1)$. So, to complete the proof it suffices to establish the convergence of $\mathcal{E}_{1}^{n}(t)$ and  $\mathcal{E}_{2}^{n}(t)$.

As above, we add and substract $\mu_{(i-1) \Delta_{n}}$ terms to write
\begin{equation*}
\mathcal{E}_{1}^{n}(t) = \mathcal{E}_{1}^{ \prime n}(t)+ \mathcal{E}_{1}^{ \prime \prime n}(t),
\end{equation*}
with
\begin{align*}
\mathcal{E}_{1}^{ \prime n}(t) &= \frac{1}{K_{n}} \sum_{j=0}^{K_{n}-1} \bigg( \big( \Lambda_{3}^{n}(t_{j}) \big)^{2} - \frac{ \rho_{n}}{ \ell_{n}^{2}} \sum_{i \in \mathcal{D}_{t_{j}}^{n}} \mu_{(i-1) \Delta_{n}} \bigg) \equiv \frac{1}{K_{n}} \sum_{j=0}^{K_{n}-1} \epsilon(t_{j}), \\[0.10cm]
\mathcal{E}_{1}^{ \prime \prime n}(t) &= \frac{1}{K_{n}} \frac{ \rho_{n}}{ \ell_{n}^{2}} \sum_{j=0}^{K_{n}-1} \sum_{i \in \mathcal{D}_{t_{j}}^{n}} \mu_{(i-1) \Delta_{n}} = 2 \frac{\rho_{n}}{ \ell_{n}} \frac{1}{2 K_{n} \ell_{n}} \sum_{j=0}^{K_{n}-1} \sum_{i=1}^{2 \ell_{n}} \mu_{t -(i+j) \Delta_{n}}.
\end{align*}
where $\epsilon(t_{j}) = \Big( \big( \Lambda_{3}^{n}(t_{j}) \big)^{2} - \frac{ \rho_{n}}{ \ell_{n}^{2}} \sum_{i \in \mathcal{D}_{t_{j}}^{n}} \mu_{(i-1) \Delta_{n}} \Big)$. As before, $\mathbb{E} \big[ \mathcal{E}_{1}^{ \prime n}(t) \big] = 0$, but now the variance of $\mathcal{E}_{1}^{ \prime n}(t)$ has a more complicated structure due to the overlapping sampling:
\begin{equation*}
\mathbb{E} \Big[ \big( \mathcal{E}_{1}^{ \prime n}(t) \big)^{2} \Big] = \frac{1}{K_{n}^{2}} \sum_{j=0}^{K_{n}-1} \mathbb{E} \Big[ \big( \epsilon(t_{j}) \big)^{2} \Big] + \frac{2}{K_{n}^{2}} \sum_{j=0}^{K_{n}-2} \sum_{s=j+1}^{K_{n}-1} \mathbb{E} \big[ \epsilon(t_{j}) \epsilon(t_{s}) \big].
\end{equation*}
To show $\mathbb{E} \Big[ \big( \mathcal{E}_{1}^{ \prime n}(t) \big)^{2} \Big]$ converges to zero, we look at each part in the above equation separately. First, by Theorem \ref{theorem:observed-asymptotic-variance}---employing the estimate in \eqref{equation:estimate}---we deduce that:
\begin{equation*}
\frac{1}{K_{n}^{2}} \sum_{j=0}^{K_{n}-1} \mathbb{E} \Big[ \big( \epsilon(t_{j}) \big)^{2} \Big]  \leq C \frac{ \ell_{n}}{K_{n}^{2}} \sum_{j=0}^{K_{n}-1} \mathbb{E} \Big[ \big( \Lambda_{3}^{n}(t_{j}) \big)^{4} \Big] = O \Big(K_{n}^{-1} \ell_{n}^{-2} \rho_{n}^{2} \Big).
\end{equation*}
Next, to deal with the cross-product term we express $\epsilon(t_{j})$ as
\begin{equation*}
\epsilon(t_{j}) = \sum_{i \in \mathcal{D}_{t_{j}}^{n}} \chi_{i} + \mathcal{R}_{t_{j}}(t),
\end{equation*}
where $\chi_{i} = \Big( \big( \zeta_{i}(3) \big)^{2} - \frac{ \rho_{n}}{ \ell_{n}} \mu_{(i-1) \Delta_{n}} \Big)$ is an $\mathcal{F}_{i \Delta_{n}}$-martingale difference sequence and $\mathcal{R}_{t_{j}}(t)$ is a reminder term of the form:
\begin{equation*}
\mathcal{R}_{t_{j}}(t) = 2 \sum_{s,i \in \mathcal{D}_{t_{j}}^{n}: s>i} \zeta_{i}^{n}(3) \zeta_{s}^{n}(3) (-1)^{ \mathds{1}_{ \left\{i<t_{j}- \ell_{n} \Delta_{n} \right\}} + \mathds{1}_{ \left\{s<t_{j}- \ell_{n} \Delta_{n} \right\}}}.
\end{equation*}
We note that for $s > 2 \ell_{n} j$, $\mathcal{D}_{t_{j}}^{n} \cap \mathcal{D}_{t_{s}}^{n} = \emptyset$, which implies that $\mathbb{E} \big[ \epsilon(t_{j}) \epsilon(t_{s}) \big] = 0$. In contrast for $s \leq 2 \ell_{n} j$, $\mathcal{D}_{t_{j}}^{n} \cap \mathcal{D}_{t_{s}}^{n} \neq \emptyset$, but nevertheless
\begin{equation*}
\mathbb{E} \big[ \mathcal{R}_{t_{j}}(t) \mathcal{R}_{t_{s}}(t) \big] = \mathbb{E} \big[ \epsilon(t_{j}) \mathcal{R}_{t_{s}}(t) \big] =  \mathbb{E} \big[ \epsilon(t_{s}) \mathcal{R}_{t_{j}}(t) \big] = 0,
\end{equation*}
since $\zeta_{i}(3)$ is also a $\mathcal{F}_{i \Delta_{n}}$-martingale difference.

Thus,
\begin{equation*}
\frac{2}{K_{n}^{2}} \sum_{j=0}^{K_{n}-2} \sum_{s=j+1}^{K_{n}-1} \mathbb{E} \big[ \epsilon(t_{j}) \epsilon(t_{s}) \big] =  \frac{2}{K_{n}^{2}} \sum_{j=0}^{K_{n}-2} \sum_{s=j+1}^{j + 2 \ell_{n}} \mathbb{E} \bigg[ \sum_{i \in \mathcal{D}_{t_{j}}^{n}} \chi_{i} \sum_{i \in \mathcal{D}_{t_{s}}^{n}} \chi_{i} \bigg].
\end{equation*}
For each summand on the right-hand side of the above equation,
\begin{equation*}
\mathbb{E} \bigg[ \sum_{i \in \mathcal{D}_{t_{j}}^{n}} \chi_{i} \sum_{i \in \mathcal{D}_{t_{s}}^{n}} \chi_{i} \bigg] = \sum_{i \in \mathcal{D}_{t_{j}}^{n} \cap \mathcal{D}_{t_{s}}^{n}}  \mathbb{E} \big[ \chi_{i}^{2} \big].
\end{equation*}
By the boundedness of $\mu$, $\mathbb{E} \Big[ \big( \zeta_{i}^{n}(3) \big)^{4} \Big] \leq C \frac{ \rho_{n}^{2}}{ \ell_{n}^{4}}$ and this bound also applies to $\mathbb{E} \big[ \chi_{i}^{2} \big]$. As the intersection $\mathcal{D}_{t_{j}}^{n} \cap \mathcal{D}_{t_{s}}^{n}$ contains at most $2\ell_n$ terms, it therefore shows that:
\begin{equation*}
\sum_{i \in \mathcal{D}_{t_{j}}^{n} \cap \mathcal{D}_{t_{s}}^{n}}  \mathbb{E} \big[ \chi_{i}^{2} \big] \leq C \ell_{n}^{-3} \rho_{n}^{2}.
\end{equation*}
Consequently,
\begin{equation*}
\frac{2}{K_{n}^{2}} \sum_{j=0}^{K_{n}-2} \sum_{s=j+1}^{j + 2 \ell_{n}} \mathbb{E} \Big[ \sum_{i \in \mathcal{D}_{t_{j}}^{n}} \chi_{i} \sum_{i \in \mathcal{D}_{t_{s}}^{n}} \chi_{i} \Big] \leq C K_{n}^{-1} \ell_{n}^{-2} \rho_{n}^{2}.
\end{equation*}
Combining the orders of the terms that make up $\mathbb{E} \Big[ \big( \mathcal{E}_{1}^{ \prime n}(t) \big)^{2} \Big]$, we deduce that
\begin{equation*}
\mathbb{E} \Big[ \big( \mathcal{E}_{1}^{ \prime n}(t) \big)^{2} \Big] = O(K_{n}^{-1} \ell_{n}^{-1} \rho_{n}^{2}).
\end{equation*}
Now, by the assumptions made in the theorem, it follows that $\mathbb{E} \Big[ \big( \mathcal{E}_{1}^{ \prime n}(t) \big)^{2} \Big] = o(1)$. Hence, the above inequality implies $\mathcal{E}_{1}^{ \prime n}(t) \overset{p}{ \longrightarrow} 0$.

In the $\mathcal{E}_{1}^{ \prime \prime n}(t)$ term, since by assumption $\ell_{n} / K_{n} \rightarrow 0$ and $K_{n} \Delta_{n} \rightarrow 0$, $(K_{n} + 2 \ell_{n}) \Delta_{n} \rightarrow 0$. This shows that $t - (i+j) \Delta_{n} \rightarrow t$ for every $i$ and $j$ in the sum. Since $\mu$ is c\`{a}dl\`{a}g, as $\Delta_{n} \rightarrow 0$,
\begin{equation*}
\mathcal{E}_{1}^{ \prime \prime n}(t) \overset{p}{ \longrightarrow}
\begin{cases}
2 \mu_{t-}, & \text{if } \rho_{n} / \ell_{n} \rightarrow 1, \\[0.10cm]
\displaystyle \frac{2}{ \theta^{2}} \mu_{t-}, & \text{if } \rho_{n} / \ell_{n} \rightarrow \displaystyle \frac{1}{ \theta^{2}}, \\[0.10cm]
0, & \text{if } \rho_{n} / \ell_{n} \rightarrow 0.
\end{cases}
\end{equation*}
Thus, alluding to the definition of $\rho_{n}$, we see that
\begin{equation*}
\mathcal{E}_{1}^{n}(t) \overset{p}{ \longrightarrow}
\begin{cases}
2 \mu_{t-}, & \text{if } \ell_{n} \sqrt{ \Delta_{n}} \rightarrow 0, \\[0.10cm]
\displaystyle \frac{2}{ \theta^{2}} \mu_{t-}, & \text{if } \ell_{n} \sqrt{ \Delta_{n}} \rightarrow \theta, \\[0.10cm]
0, & \text{if } \ell_{n} \sqrt{ \Delta_{n}} \rightarrow \infty.
\end{cases}
\end{equation*} \qed

\subsection{Proof of Theorem \ref{theorem:test-statistic}}

The statement of the theorem under $\mathscr{H}_{0}$ follows immediately by combining Theorem \ref{theorem:clt} and Theorem \ref{observed_variance_LLN2} with Slutsky's theorem.

We next look at $\mathscr{H}_{1}$. We use the version of the test statistic based on the non-overlapping observed asymptotic local variance in \eqref{equation:observed-variance}, $\widetilde{ \mathsf{avar}} \big( \nabla \widehat{ \lambda}_{t} \big)$. In that instance,
\begin{equation*}
\phi_{ \tau}^{ \text{ib}} = \frac{ \widetilde{D}_{ \tau,0}}{\displaystyle \sqrt{ \frac{1}{K_{n}} \sum_{j=0}^{K_n-1} \widetilde{D}_{ \tau,j}^{2}}},
\end{equation*}
where
\begin{equation*}
\widetilde{D}_{ \tau,j} = \frac{N^{n}( \tau-(1+2j) \delta_{n}, \tau-2j \delta_{n}) - N^{n}( \tau-(2+2j) \delta_{n}, \tau-(1+2j) \delta_{n})}{n \delta_{n}},
\end{equation*}
for $j = 0, 1, \dots, K_{n}-1$.

We next derive a lower (upper) bound on the stochastic order of the numerator (denominator). We show that the numerator diverges \textit{at least} as fast as the numerator. Note that it again suffices to explore the diverging part:
\begin{equation*}
D_{ \tau,j} =  \frac{N^{n, \beta}( \tau-(1+2j) \delta_{n}, \tau-2j \delta_{n}) - N^{n, \beta}( \tau-(2+2j) \delta_{n}, \tau-(1+2j) \delta_{n})}{n \delta_{n}},
\end{equation*}
for which
\begin{align} \label{equation:D}
\begin{split}
\mathbb{E} \big[|D_{ \tau,j}| \big] &\leq \mathbb{E} \left[ \frac{N^{n, \beta}( \tau-(1+2j) \delta_{n}, \tau-2j \delta_{n}) + N^{n, \beta}( \tau-(2+2j) \delta_{n}, \tau-(1+2j) \delta_{n})}{n \delta_{n}} \right] \\[0.10cm]
&= \mathbb{E} \left[ \frac{N^{n, \beta}( \tau-(2+2j) \delta_{n}, \tau-2j \delta_{n})}{n \delta_{n}} \right] \\[0.10cm]
&= \mathbb{E} \left[ \frac{1}{ \delta_{n}} \int_{ \tau-(2+2j) \delta_{n}}^{ \tau-2j \delta_{n}} \sigma_{u} | \tau-u|^{- \alpha} \mathrm{d}u \right] \\[0.10cm]
&\leq C \delta_{n}^{- \alpha} \big((j+1)^{1- \alpha} - j^{1- \alpha} \big),
\end{split}
\end{align}
by the proof of Lemma \ref{lemma:consistency}.

By the mean-value theorem there exists a $j^{*} \in (j,j+1)$, such that
\begin{equation*}
(j+1)^{1- \alpha} - j^{1- \alpha} = (j^*)^{-\alpha} \leq j^{- \alpha}.
\end{equation*}
Therefore,
\begin{equation*}
\mathbb{E} \big[|D_{ \tau,j}| \big] \leq C \delta_{n}^{- \alpha} j^{- \alpha}.
\end{equation*}
Thus, for $j$ fixed:
\begin{equation*}
D_{ \tau,j} = O_{p} ( \delta_{n}^{- \alpha}).
\end{equation*}
as $\delta_{n} \rightarrow 0$.

Regarding the denominator,
\begin{equation*}
\mathbb{E} \bigg[ \frac{1}{K_{n}} \sum_{j=0}^{K_{n}-1} D_{ \tau,j}^{2} \bigg] = \frac{1}{K_{n}} \sum_{j=0}^{K_{n}-1} \mathbb{E} \big[ D_{ \tau,j}^{2} \big], \end{equation*}
where
\begin{align*}
\mathbb{E} \big[ D_{ \tau,j}^{2} \big] &= \mathbb{E} \Bigg[ \bigg( \frac{N^{n, \beta}( \tau-(1+2j) \delta_{n}, \tau-2j \delta_{n}) - N^{n, \beta}( \tau-(2+2j) \delta_{n}, \tau-(1+2j) \delta_{n})}{n \delta_{n}} \bigg)^{2} \Bigg] \\[0.10cm]
&\leq \frac{1}{n^{2} \delta_{n}^{2}} \mathbb{E} \left[ N^{n, \beta}( \tau-(2+2j) \delta_{n}, \tau-2j \delta_{n})^{2} \right] \\[0.10cm]
&= \frac{1}{n^{2} \delta_{n}^{2}} \left( \mathbb{E} \left[ N^{n, \beta}( \tau-(2+2j) \delta_{n}, \tau-2j \delta_{n}) \right] + \mathbb{E} \left[ N^{n, \beta}( \tau-(2+2j) \delta_{n}, \tau-2j \delta_{n}) \right]^{2} \right),
\end{align*}
due to the properties of the Poisson distribution.

It follows from \eqref{equation:D} that
\begin{equation*}
\mathbb{E} \left[ N^{n, \beta}( \tau-(2+2j) \delta_{n}, \tau-2j \delta_{n}) \right] \leq Cn \delta_{n}^{1-\alpha}j^{- \alpha},
\end{equation*}
Hence, by the rate conditions imposed in Theorem \ref{theorem:test-statistic}, $n\delta_{n} \rightarrow \infty$ and $\delta_{}^{- \alpha} \rightarrow \infty$, so the above expectation diverges. This implies that the square of the expectation is the leading term, so it suffices to look at the second summand in the above decomposition:
\begin{equation*}
\frac{1}{K_{n}} \sum_{j=0}^{K_{n}-1} \frac{1}{n^{2} \delta_{n}^{2}} \mathbb{E} \left[ N^{n, \beta}( \tau-(2+2j) \delta_{n}, \tau-2j \delta_{n}) \right]^{2} \leq C \delta_{n}^{-2 \alpha} \frac{1}{K_{n}} \sum_{j=0}^{K_{n}-1} j^{- \alpha} = O \Big( \big[ \delta_{n} K_{n} \big]^{- 2 \alpha} \Big)
\end{equation*}
since $\sum_{j=0}^{K_{n}-1} j^{- 2\alpha} = O(K_{n}^{1- 2 \alpha})$.
Thus, the denominator is \textit{at most} $O_{p} \Big( \big[ \delta_{n} K_{n} \big]^{- \alpha} \Big)$.

Now, we look at the numerator:
\begin{equation*}
D_{ \tau,0} =  \frac{N^{n, \beta}( \tau- \delta_{n}, \tau) - N^{n, \beta}( \tau -  2\delta_{n}, \tau - \delta_{n})}{n \delta_{n}}.
\end{equation*}
Let
\begin{equation*}
\mathfrak{B}_{1, \tau}^{n} = \frac{N^{n, \beta}( \tau- \delta_{n}, \tau)}{n \delta_{n}} - \frac{1}{ \delta_{n}} \int^{ \tau}_{ \tau- \delta_{n}} \frac{ \sigma_{s}}{| \tau - s|^{ \alpha}} \mathrm{d}s.
\end{equation*}
As in the proof of Lemma \ref{lemma:consistency}, we can condition on $\mathcal{F}^{ \sigma}$, the $\sigma$-algebra generated by $\sigma$, and then employ the law of iterated expectations to obtain:
\begin{equation*}
\mathbb{E} \left[ \mathfrak{B}_{1, \tau}^{n} \right] = \mathbb{E} \left[ \mathbb{E} \left[ \mathfrak{B}_{1, \tau}^{n} \mid \mathcal{F}^{ \sigma} \right] \right] = 0,
\end{equation*}
and, by the law of total variance,
\begin{equation*}
\mathrm{var} \left( \mathfrak{B}_{1, \tau}^{n} \right) = \mathbb{E} \left[ \mathrm{var} \left( \mathfrak{B}_{1, \tau}^{n} \mid \mathcal{F}^{ \sigma} \right) \right] + \underbrace{ \mathrm{var} \left( \mathbb{E} \left[ \mathfrak{B}_{1, \tau}^{n} \mid \mathcal{F}^{ \sigma} \right] \right)}_{=0},
\end{equation*}
where
\begin{align*}
\mathrm{var} \left( \mathfrak{B}_{1, \tau}^{n} \mid \mathcal{F}^{ \sigma} \right) &= \frac{1}{n^{2} \delta_{n}^{2}} n \int^{ \tau}_{ \tau- \delta_{n}} \frac{ \sigma_{s}}{| \tau - s|^{ \alpha}} \mathrm{d}s \\[0.10cm]
&\leq \frac{C}{n \delta_{n}^{2}} \int^{ \tau}_{ \tau- \delta_{n}} \frac{1}{| \tau - s|^{ \alpha}} \mathrm{d}s = \frac{C}{n \delta_{n}^{2}} \delta_{n}^{1- \alpha} = C \frac{1}{n \delta_{n}^{(1+ \alpha)}} = O \left(n^{-1} \delta_{n}^{-(1+ \alpha)} \right).
\end{align*}
Thus,
\begin{equation*}
\mathrm{var} \left( \delta_{n}^{ \alpha/2} \mathfrak{B}_{1, \tau}^{n} \right) \leq \frac{C}{n \delta_{n}} \rightarrow 0,
\end{equation*}
which has the implication that
\begin{equation*}
\delta_{n}^{ \alpha/2} \left( \frac{N^{n, \beta}( \tau- \delta_{n}, \tau)}{n \delta_{n}} - \frac{1}{ \delta_{n}} \int_{ \tau- \delta_{n}}^{ \tau} \frac{ \sigma_{s}}{| \tau - s|^{ \alpha}} \mathrm{d}s \right) \overset{p}{ \longrightarrow} 0.
\end{equation*}
Analogously,
\begin{equation*}
\delta_{n}^{ \alpha/2} \left( \frac{N^{n, \beta}( \tau- 2\delta_{n}, \tau- \delta_{n})}{n \delta_{n}} - \frac{1}{ \delta_{n}} \int_{ \tau- 2\delta_{n}}^{ \tau- \delta_{n}} \frac{ \sigma_{s}}{| \tau - s|^{ \alpha}} \mathrm{d}s \right) \overset{p}{ \longrightarrow} 0,
\end{equation*}
such that
\begin{equation*}
\delta_{n}^{ \alpha/2} \left( D_{ \tau,0} - \left[ \frac{1}{ \delta_{n}} \int^{ \tau}_{ \tau- \delta_{n}} \frac{ \sigma_{s}}{| \tau - s|^{ \alpha}} \mathrm{d}s - \frac{1}{ \delta_{n}} \int^{ \tau - \delta_{n}}_{ \tau-2 \delta_{n}} \frac{ \sigma_{s}}{| \tau - s|^{ \alpha}} \mathrm{d}s \right] \right) \overset{p}{ \longrightarrow} 0.
\end{equation*}

Next, we define $\underline{ \sigma} = \inf_{s \in [ \tau-2 \delta_{n}]} \sigma_{s}$ and $\overline{ \sigma} = \sup_{s \in[ \tau-2 \delta_{n}]} \sigma_{s}$. Then, by construction
\begin{align*}
\frac{1}{ \delta_{n}} \int^{ \tau}_{ \tau- \delta_{n}} \frac{ \sigma_{s}}{| \tau - s|^{ \alpha}} \mathrm{d}s - \frac{1}{ \delta_{n}} \int_{ \tau-2 \delta_{n}}^{ \tau- \delta_{n}} \frac{ \sigma_{s}}{| \tau - s|^{ \alpha}} \mathrm{d}s &\geq \frac{1}{ \delta_{n}} \left( \int_{ \tau- \delta_{n}}^{ \tau} \frac{ \underline{ \sigma}}{| \tau - s|^{ \alpha}} \mathrm{d}s - \int_{ \tau-2 \delta_{n}}^{ \tau- \delta_{n}} \frac{ \overline{ \sigma}}{| \tau - s|^{ \alpha}} \mathrm{d}s \right) \\[0.10cm]
&= \frac{1}{ \delta_{n}} \left( \frac{ \underline{ \sigma}}{1- \alpha} \delta_{n}^{1- \alpha} - \frac{ \overline{ \sigma}}{1- \alpha} (2^{1- \alpha} - 1) \delta_{n}^{1- \alpha} \right) \\[0.10cm]
&= \frac{ \underline{ \sigma} - \overline{ \sigma} (2^{1- \alpha} - 1)}{1- \alpha} \delta_{n}^{- \alpha}.
\end{align*}
Since $\overline{ \sigma} - \underline{ \sigma} \geq | \sigma_{ \tau}- \sigma_{ \tau-2 \delta_{n}}|$, it follows that for every $\epsilon>0 : \{ \omega \in \Omega : \overline{ \sigma} - \underline{ \sigma} > \epsilon \} \subseteq \{ \omega \in \Omega : |\sigma_{ \tau}- \sigma_{ \tau-2 \delta_{n}}| > \epsilon \}$, and so
\begin{equation*}
\mathbb{P} \left(| \overline{ \sigma} - \underline{ \sigma}| > \epsilon \right) \leq \mathbb{P} \left(| \sigma_{ \tau} - \sigma_{ \tau- \delta_{n}}| > \varepsilon \right) \longrightarrow 0,
\end{equation*}
as $\delta_{n} \rightarrow 0$, where the right-hand side follows by stochastic continuity of $\sigma$. In other words, for a large enough $n$, $\underline{ \sigma}$ is arbitrarily close to $\overline{ \sigma}$ in probability. Consequently,
\begin{equation*}
\mathbb{P} \left( \frac{ \underline{ \sigma} - \overline{ \sigma}(2^{1- \alpha} - 1)}{1- \alpha} > 0 \right) \longrightarrow 1,
\end{equation*}
as $\delta_{n} \rightarrow 0$. Hence,
\begin{equation}
\frac{1}{ \delta_{n}} \int^{ \tau}_{ \tau- \delta_{n}} \frac{ \sigma_{s}}{| \tau - s|^{ \alpha}} \mathrm{d}s - \frac{1}{ \delta_{n}} \int_{ \tau-2 \delta_{n}}^{ \tau- \delta_{n}} \frac{ \sigma_{s}}{| \tau - s|^{ \alpha}} \mathrm{d}s \overset{p}{ \longrightarrow} \infty.
\end{equation}
so the explosion rate of the numerator is \textit{at least} $\delta^{- \alpha}$. This is faster than the explosion rate of the denominator (at most $\left( \delta_{n} K_{n} \right)^{- \alpha}$), so the test statistic diverges under the alternative. \qed

\subsection{Proof of Lemma \ref{lemma:cof}}

\noindent \textit{Part (i) --- Conditional on $\omega \in \Omega^{B}_{1}$}:

The observed process can be broken into the parts
\begin{equation*}
N_{t}^{n} = N_{t}^{n, \mu} + N_{t}^{n, \beta},
\end{equation*}
where $N_{t}^{n, \mu}$ and $N_{t}^{n, \beta}$ are inhomogeneous Poisson processes with rates $n \mu_{t}$ and $n \beta_{t}$. This yields the following decomposition for $\widetilde{ \lambda}_{ \theta}(k \delta_{n})$:
\begin{equation*}
\widetilde{ \lambda}_{ \theta}(k \delta_{n}) = \widetilde{ \mu}_{ \theta}(k \delta_{n}) + \widetilde{ \beta}_{ \theta}(k \delta_{n}),
\end{equation*}
where $\widetilde{ \mu}_{ \theta}(k \delta_{n}) \overset{p}{ \longrightarrow} \mu_{ \theta-}$ and
\begin{equation*}
\widetilde{ \beta}_{ \theta}(k \delta_{n}) = \frac{N_{ \theta+k \delta_{n}}^{n, \beta} - N_{ \theta-k \delta_{n}}^{n, \beta}}{2nk \delta_{n}}.
\end{equation*}
Since $\delta_{n} \rightarrow 0$ and $\alpha>0$, the convergence $\widetilde{ \mu}_{ \theta}(k \delta_{n}) \overset{p}{ \longrightarrow} \mu_{ \theta-}$ further implies that $ \delta_{n}^{ \alpha/2} \widetilde{ \mu}_{ \theta}(k \delta_{n}) \overset{p}{ \longrightarrow} 0$. It immediately follows that
\begin{equation*}
\delta_{n}^{ \alpha/2} \left( \widetilde{ \lambda}_{ \theta}(k \delta_{n}) - \sigma_{ \theta-} (2k \delta_{n})^{- \alpha} \right) - \delta_{n}^{ \alpha/2} \left( \widetilde{ \beta}_{ \theta}(k \delta_{n}) - \sigma_{ \theta-} (2k \delta_{n})^{- \alpha} \right) \overset{p}{ \longrightarrow} 0.
\end{equation*}
Note that $\sigma_{ \theta} (2k \delta_{n})^{- \alpha} = \dfrac{1}{2k \delta_{n}} \int_{ \theta-k \delta_{n}}^{ \theta+k \delta_{n}} \dfrac{ \sigma_{ \theta}}{| \theta - s|^{ \alpha}} \mathrm{d}s$. This implies that
\begin{equation} \label{equation:beta-decomposed}
\delta_{n}^{ \alpha/2} \left( \widetilde{ \beta}_{ \theta}(k \delta_{n}) - \sigma_{ \theta}(2k \delta_{n})^{- \alpha} \right) = \delta_{n}^{ \alpha/2} \left( \mathfrak{B}_{1, \theta}^{n} + \mathfrak{B}_{2, \theta}^{n} \right),
\end{equation}
where
\begin{align*}
\mathfrak{B}_{1, \theta}^{n} &= \widetilde{ \beta}_{ \theta}(k \delta_{n}) - \frac{1}{2k \delta_{n}} \int_{ \theta-k \delta_{n}}^{ \theta+k \delta_{n}} \frac{ \sigma_{s}}{| \theta - s|^{ \alpha}} \mathrm{d}s, \\[0.10cm]
\mathfrak{B}_{2, \theta}^{n} &= \frac{1}{2k \delta_{n}} \int_{ \theta-k \delta_{n}}^{ \theta+k \delta_{n}} \frac{ \sigma_{s}}{| \theta - s|^{ \alpha}} \mathrm{d}s - \frac{1}{2k \delta_{n}} \int_{ \theta-k \delta_{n}}^{ \theta+k \delta_{n}} \frac{ \sigma_{ \theta}}{| \theta - s|^{ \alpha}} \mathrm{d}s.
\end{align*}
Following the footsteps in the proof of Theorem \ref{theorem:test-statistic} by conditioning on $\mathcal{F}^{ \sigma}$ and employing the laws of iterated expectation and total variance, we find that
\begin{equation*}
\mathbb{E} \left[ \mathfrak{B}_{1, \theta}^{n} \right] = 0 \quad \text{and} \quad \mathrm{var} \left( \mathfrak{B}_{1, \theta}^{n} \right) = \mathbb{E} \left[ \mathrm{var} \left( \mathfrak{B}_{1, \theta}^{n} \mid \mathcal{F}^{ \sigma} \right) \right],
\end{equation*}
where $\mathrm{var} \left( \mathfrak{B}_{1, \theta}^{n} \mid \mathcal{F}^{ \sigma} \right) = O_{p} \left(n^{-1} \delta_{n}^{-(1+ \alpha)} \right)$, so that $\delta_{n}^{ \alpha/2} \mathfrak{B}_{1, \theta}^{n} \overset{p}{ \longrightarrow} 0$.

$\mathfrak{B}_{2, \theta}^{n}$ can be bounded by
\begin{align*}
| \mathfrak{B}_{2, \theta}^{n}| &= \left| \frac{1}{2k \delta_{n}} \int_{ \theta-k \delta_{n}}^{ \theta+k \delta_{n}} \frac{ \sigma_{s} - \sigma_{ \theta}}{| \theta - s|^{ \alpha}} \mathrm{d}s \right| \\[0.10cm]
&\leq \frac{1}{2k \delta_{n}} \int_{ \theta-k \delta_{n}}^{ \theta+k \delta_{n}} \frac{| \sigma_{s} - \sigma_{ \theta}|}{| \theta - s|^{ \alpha}} \mathrm{d}s \\[0.10cm]
&\leq \frac{(2k \delta_{n})^{H^{ \sigma}}}{2k \delta_{n}} \int_{ \theta-k \delta_{n}}^{ \theta+k \delta_{n}} \frac{L}{| \theta - s|^{ \alpha}} \mathrm{d}s = (2k \delta_{n})^{H^{ \sigma} - \alpha} L.
\end{align*}
Hence,
\begin{equation*}
\delta_{n}^{ \alpha/2}| \mathfrak{B}_{2, \theta}^{n}| \leq C \delta_{n}^{H^{ \sigma} - \alpha/2} L \overset{p}{ \longrightarrow} 0,
\end{equation*}
because $L$ is a bounded random variable and $H^{ \sigma} - \alpha/2>0$. Thus, $\delta_{n}^{ \alpha/2}| \mathfrak{B}_{1, \theta}^{n}|$ and $\delta_{n}^{ \alpha/2}| \mathfrak{B}_{2, \theta}^{n}|$ converge to zero in probability.

Combining this fact with \eqref{equation:beta-decomposed},
\begin{equation*}
\delta_{n}^{ \alpha/2} \left( \widetilde{ \lambda}_{ \theta}(k \delta_{n}) - \sigma_{ \theta} (2k \delta_{n})^{- \alpha} \right) \overset{p}{ \longrightarrow} 0.
\end{equation*}
for any $k \geq 1$. This means that:
\begin{equation*}
\frac{ \widetilde{ \lambda}_{ \theta}(k \delta_{n})}{ \widetilde{ \lambda}_{ \theta}( \delta_{n})} = \frac{ \delta_{n}^{ \alpha/2} \left( \widetilde{ \lambda}_{ \theta}(k \delta_{n}) - \sigma_{ \theta}(2k \delta_{n})^{- \alpha} \right) + \delta_{n}^{ \alpha/2} \left( \sigma_{ \theta}(2k \delta_{n})^{- \alpha} \right)}{ \delta_{n}^{ \alpha/2} \left( \widetilde{ \lambda}_{ \theta}(k \delta_{n}) - \sigma_{ \theta}(2k \delta_{n})^{- \alpha} \right) + \delta_{n}^{ \alpha/2} \left( \sigma_{ \theta}(2 \delta_{n})^{- \alpha} \right)} \overset{p}{ \longrightarrow}  \frac{ \delta_{n}^{ \alpha/2} \left( \sigma_{ \theta}(2k \delta_{n})^{- \alpha} \right)}{ \delta_{n}^{ \alpha/2} \left( \sigma_{ \theta}(2 \delta_{n})^{- \alpha} \right)} = k^{- \alpha},
\end{equation*}
which completes the proof of part (i). \\[-0.25cm]

\noindent \textit{Part (ii) --- Conditional on $\omega \in \Omega^{J}_{1}$}:

We start by observing that $\widetilde{ \lambda}_{ \theta}(k \delta_{n})$ can be represented as the average of a forward- and backward-looking intensity estimator:
\begin{equation*}
\widetilde{ \lambda}_{ \theta}(k \delta_{n}) = \frac{1}{2} \left( \widehat{ \lambda}_{ \theta}^{(-)}(k \delta_{n}) + \widehat{ \lambda}_{ \theta}^{(+)}(k \delta_{n}) \right).
\end{equation*}
The proof of Theorem \ref{theorem:clt-smoothness} implies that $\widehat{ \lambda}_{ \theta}^{(-)}(k \delta_{n}) \overset{p}{ \longrightarrow} \mu_{ \theta-}$. As the forward-looking estimator $\widehat{ \lambda}^{(+)}_{ \theta}(k \delta_{n})$ is the mirror image of $\widehat{ \lambda}_{ \theta}^{(-)}(k \delta_{n})$, we also deduce that $\widehat{ \lambda}_{ \theta}^{(+)}(k \delta_{n}) \overset{p}{ \longrightarrow} \mu_{ \theta+}$, where $\mu_{ \theta+} = \mu_{ \theta-} + \Delta \mu_{ \theta}$. In summary, by the continuous mapping theorem
\begin{equation*}
\widetilde{ \lambda}_{ \theta}(k \delta_{n}) \overset{p}{ \longrightarrow} \frac{1}{2} \left( \mu_{ \theta-} + \mu_{ \theta-} + \Delta \mu_{ \theta} \right) = \mu_{ \theta-} + \frac{ \Delta \mu_{ \theta}}{2}.
\end{equation*}
As this convergence holds for every $k \geq 1$,
\begin{equation*}
\frac{ \widetilde{ \lambda}_{ \theta}(k \delta_{n})}{ \widetilde{ \lambda}_{ \theta}( \delta_{n})} \overset{p}{ \longrightarrow} 1.
\end{equation*} \qed

\subsection{Proof of Theorem \ref{theorem:cof-jump}}

The proof of the theorem consists of deriving the bivariate asymptotic distribution of the change-of-frequency intensity estimator computed with different $k$:
\begin{equation} \label{equation:bivariate-limit}
\sqrt{n \delta_{n}}
\begin{pmatrix}
\widetilde{ \lambda}_{ \theta}(k \delta_{n}) - \mu_{ \theta-} - \dfrac{ \Delta \mu_{ \theta}}{2} \\
\widetilde{ \lambda}_{ \theta}( \delta_{n}) - \mu_{ \theta-} - \dfrac{ \Delta \mu_{ \theta}}{2}
\end{pmatrix}
\overset{ \mathfrak{D}_{s}}{ \longrightarrow} \Sigma^{1/2} \mathcal{Z}_{2},
\end{equation}
where $\mathcal{Z}_{2} \sim N(0, I_{2})$ and $\Sigma$ is the asymptotic covariance matrix. The statement of the theorem then follows by application of the delta method.

To derive \eqref{equation:bivariate-limit}, by the Cram\'{e}r-Wold device it suffices to show that every fixed linear combination
\begin{equation*}
\mathbf{L}_{n} = c_{1} \sqrt{n \delta_{n}} \left( \widetilde{ \lambda}_{ \theta}(k \delta_{n}) - \mu_{ \theta-} - \frac{ \Delta \mu_{ \theta}}{2} \right) + c_{2} \sqrt{n \delta_{n}} \left( \widetilde{ \lambda}_{ \theta}( \delta_{n}) - \mu_{ \theta-}- \frac{ \Delta \mu_{ \theta}}{2} \right),
\end{equation*}
converges stably in law, i.e.
\begin{equation*}
\mathbf{L}_{n} \overset{ \mathfrak{D}_{s}}{ \longrightarrow} \sqrt{ c_{1}^{2} \Sigma_{1,1} + c_{2}^{2} \Sigma_{2,2} + 2c_{1}c_{2} \Sigma_{1,2}} \mathcal{Z},
\end{equation*}
where $(c_{1}, c_{2}) \in \mathbb{R}^{2}$ and $\mathcal{Z} \sim N(0,1)$. To prove this, we split $\widetilde{ \lambda}_{ \theta}(k \delta_{n})$ into the average of the forward- and backward-looking intensity estimator:
\begin{equation*}
\sqrt{n \delta_{n}} \left( \widetilde{ \lambda}_{ \theta}(k \delta_{n}) - \mu_{ \theta-} + \frac{ \Delta \mu_{ \theta}}{2} \right) = \frac{1}{2} \left[ \sqrt{n \delta_{n}} \left( \widetilde{ \lambda}_{ \theta}^{(-)}(k \delta_{n}) - \mu_{ \theta-} \right) + \sqrt{n \delta_{n}} \left( \widetilde{ \lambda}_{ \theta}^{(+)}(k \delta_{n}) - \mu_{ \theta+} \right) \right].
\end{equation*}
Next, both $\widetilde{ \lambda}_{ \theta}^{(-)}(k \delta_{n})$ and $\widetilde{ \lambda}_{ \theta}^{(+)}(k \delta_{n})$ are decomposed as in Theorem \ref{theorem:clt-smoothness}. For example,
\begin{equation*}
\sqrt{n \delta_n } \left( \widetilde{ \lambda}_{ \theta}^{(-)}(k \delta_{n}) - \mu_{ \theta-} \right) = \sum_{j=1}^{4} \Lambda_{j}^{n,(-)}( \theta;k),
\end{equation*}
where
\begin{align*}
\Lambda_{1}^{n,(-)}( \theta;k) &= \frac{1}{k \sqrt{ \ell_{n}}} \sum_{i=n \theta-k \ell_{n}+1}^{n \theta} \big( \Delta_{i} N^{n} - \Delta_{i} \widetilde{N}^{n} - \mathbb{E} \big[ \Delta_{i} N^{n} - \Delta_{i} \widetilde{N}^{n} \mid \mathcal{F}_{(i-1) \Delta_{n}} \big] \big), \\[0.10cm]
\Lambda_{2}^{n,(-)}( \theta;k) &= \frac{1}{k \sqrt{ \ell_{n}}} \sum_{i=n \theta-k \ell_{n}+1}^{n \theta} \mathbb{E} \big[ \Delta_{i} N^{n} - \Delta_{i} \widetilde{N}^{n} \mid \mathcal{F}_{(i-1) \Delta_{n}} \big], \\[0.10cm]
\Lambda_{3}^{n,(-)}( \theta;k) &= \frac{1}{k \sqrt{ \ell_{n}}} \sum_{i=n \theta-k \ell_{n}+1}^{n \theta} \big( \Delta_{i} \widetilde{N}^{n} - \mu_{(i-1) \Delta_{n}} \big), \\[0.10cm]
\Lambda_{4}^{n,(-)}( \theta;k) &= \frac{1}{k \sqrt{ \ell_{n}}} \sum_{i=n \theta-k \ell_{n}+1}^{n \theta} (n \theta- \ell_{n}-i)( \mu_{i \Delta_{n}} - \mu_{(i-1) \Delta_{n}}).
\end{align*}
Proceeding as above,
\begin{equation*}
\mathbf{L}_{n} = L_{n}^{(-)} + L_{n}^{(+)} + R_{n},
\end{equation*}
where
\begin{equation*}
L_{n}^{(-)} = \frac{1}{2} \left(c_{1} \Lambda_{3}^{n,(-)}( \theta;k) + c_{2} \Lambda_{3}^{n,(-)}( \theta;1) \right), \qquad L_{n}^{(+)} = \frac{1}{2} \left(c_{1} \Lambda_{3}^{n,(+)}( \theta;k) + c_{2} \Lambda_{3}^{n,(+)}( \theta;1) \right),
\end{equation*}
are the leading terms and the remainder is
 \begin{equation*}
R_{n} = \sum_{j=1,j \neq 3}^{4} \left(c_{1} \Lambda_{1}^{n,(+)}( \theta;k) + c_{2} \Lambda_{1}^{n,(+)}( \theta;1) \right) + \sum_{j=1,j \neq 3}^{4} \left(c_{1} \Lambda_{1}^{n,(-)}( \theta;k) + c_{2} \Lambda_{1}^{n,(-)}( \theta;1) \right).
\end{equation*}
Notice that since $R_{n}$ is a sum of a finite number of asymptotically negligible terms, it is itself asymptotically negligible. Thus, it remains to inspect the behavior of $L_{n}^{(-)}$ and $L_{n}^{(+)}$. We do this in detail for the former only by writing
\begin{equation*}
L_{n}^{(-)} = \sum_{i=n \theta-k \ell_{n}+1}^{n \theta} \xi_{i}^{n},
\end{equation*}
where
\begin{equation*}
\xi_{i}^{n} = \frac{c_{1}}{2k \sqrt{ \ell_{n}}} \big( \Delta_{i} \widetilde{N}^{n} - \mu_{(i-1) \Delta_{n}} \big) + \frac{c_{2}}{2 \sqrt{ \ell_{n}}} \big( \Delta_{i} \widetilde{N}^{n} - \mu_{(i-1) \Delta_{n}} \big) \mathbf{1}_{ \left\{i \geq n \theta- \ell_{n}+1 \right\}}.
\end{equation*}
It holds that $\mathbb{E} \left[ \xi_{i}^{n} \mid \mathcal{F}_{(i-1) \Delta_{n}} \right] = 0$, and
\begin{equation*}
\mathbb{E} \left[ \left( \xi_{i}^{2} \right)^{2} \mid \mathcal{F}_{(i-1) \Delta_{n}} \right] = \frac{ \mu_{(i-1) \Delta_{n}}}{4 \ell_{n}} \left[ \frac{c_{1}^{2}}{k^{2}} + \left(c_{2}^{2} + \frac{2c_{1}c_{2}}{k} \right) \mathbf{1}_{ \left\{i \geq n \theta- \ell_{n}+1 \right\}} \right].
\end{equation*}
Therefore,
\begin{align*}
\sum_{i=n \theta-k \ell_{n}+1}^{n \theta} \mathbb{E} \left[ \left( \xi_{i}^{2} \right)^{2} \mid \mathcal{F}_{(i-1) \Delta_{n}} \right] &= c_{1}^{2} \frac{1}{4k} \sum_{i = n \theta - k \ell_{n}+1}^{n \theta} \frac{ \mu_{(i-1) \Delta_{n}}}{k \ell_{n}} + \left(c_{2}^{2}+ \frac{2c_{1}c_{2}}{k} \right) \frac{1}{4} \sum_{i=n \theta- \ell_{n}+1}^{n \theta} \frac{ \mu_{(i-1) \Delta_{n}}}{ \ell_{n}} \\
&\overset{p}{ \longrightarrow} c_{1}^{2} \frac{1}{4k} \mu_{ \theta-} + c_{2}^{2} \frac{1}{4} \mu_{ \theta-} + 2c_{1}c_{2} \frac{1}{4k} \mu_{ \theta-}.
\end{align*}
Furthermore,
\begin{equation*}
\sum_{i=n \theta-k \ell_{n}+1}^{n \theta} \mathbb{E} \left[ \left( \xi_{i}^{n} \right)^{4} | \mathcal{F}_{(i-1) \Delta_{n}} \right] \leq C \sum_{i = n \theta-k \ell_{n}+1}^{n \theta} \frac{1}{ \ell_{n}^{2}} \mathbb{E} \left[ \left( \Delta_{i} \widetilde{N}^{n} - \mu_{(i-1) \Delta_{n}} \right)^{4} \mid \mathcal{F}_{(i-1) \Delta_{n}} \right] \overset{p}{ \longrightarrow} 0,
\end{equation*}
where the convergence was already established in the proof of Theorem \ref{theorem:clt} (for the analogous term, which was denoted by $\zeta_{i}^{n}(3)$ back then). Thus, by Lemma \ref{lemma:stable-convergernce}
\begin{equation*}
L_{n}^{(-)} \overset{ \mathfrak{D}_{s}}{ \longrightarrow} \sqrt{ c_1^2 \frac{1}{4k}\mu_{\theta-}  + c_2^2 \frac{1}{4}\mu_{\theta-}  + 2 c_1 c_2 \frac{1}{4k}\mu_{\theta-} } \mathcal{Z}^{(-)},
\end{equation*}
where $\mathcal{Z}^{(-)} \sim N(0,1)$.

Analogously,
\begin{equation*}
L_{n}^{(+)} \overset{ \mathfrak{D}_{s}}{ \longrightarrow} \sqrt{c_{1}^{2} \frac{1}{4k} \mu_{ \theta+} + c_{2}^{2} \frac{1}{4} \mu_{ \theta+} + 2 c_{1}c_{2} \frac{1}{4k} \mu_{ \theta+}} \mathcal{Z}^{(+)},
\end{equation*}
where $\mathcal{Z}^{(+)} \sim N(0,1)$ independently from $\mathcal{Z}^{(-)}$. Since $\mu_{ \theta+} = \mu_{ \theta-} + \Delta \mu_{ \theta}$:
\begin{equation*}
\mathbf{L}_{n} \overset{ \mathfrak{D}_{s}}{ \longrightarrow} \sqrt{c_{1}^{2} \frac{1}{2k} \left( \mu_{ \theta-} + \frac{ \Delta \mu_{ \theta}}{2} \right) + c_{2}^{2} \frac{1}{2} \left( \mu_{ \theta-} + \frac{ \Delta \mu_{ \theta}}{2} \right) + 2c_{1}c_{2} \frac{1}{2k} \left( \mu_{ \theta-} + \frac{ \Delta \mu_{ \theta}}{2} \right)} \mathcal{Z}.
\end{equation*}
So the convergence in \eqref{equation:bivariate-limit} holds with
\begin{equation*}
\Sigma =
\begin{bmatrix}
\frac{1}{2k} \left( \mu_{ \theta-} + \frac{ \Delta \mu_{ \theta}}{2} \right) & \frac{1}{2k} \left( \mu_{ \theta-} + \frac{ \Delta \mu_{ \theta}}{2} \right) \\
\frac{1}{2k} \left( \mu_{ \theta-} + \frac{ \Delta \mu_{ \theta}}{2} \right) & \frac{1}{2 } \left( \mu_{ \theta-} + \frac{ \Delta \mu_{ \theta}}{2} \right).
\end{bmatrix}
\end{equation*} \qed

\clearpage


\renewcommand{\baselinestretch}{1.0}
\small
\bibliography{userref}

\end{document}